\documentclass[a4paper,12pt]{article}
\usepackage[dvipdfmx]{graphicx}
\usepackage{amsfonts,amsthm,amsmath,amssymb,graphicx,float,mathtools}
\numberwithin{equation}{section}
\usepackage{subfigure}
\usepackage{color}
\usepackage{bbm} 
\usepackage{tensor} 
\usepackage{verbatim} 
\usepackage[
pagebackref=false,
colorlinks=true,
linkcolor=blue,
urlcolor=blue,
filecolor=black,
citecolor=red,
pdfstartview=FitV,
pdftitle={},
pdfauthor={},
pdfsubject={},
pdfkeywords={},
pdfpagemode=None,
bookmarksopen=true
]{hyperref}
\usepackage{ascmac}
\usepackage{ulem}

\begin{document}

\newtheorem{definition}{Definition}[section]
\newcommand{\be}{\begin{equation}}
\newcommand{\ee}{\end{equation}}
\newcommand{\bea}{\begin{eqnarray}}
\newcommand{\eea}{\end{eqnarray}}
\newcommand{\LE}{\left[}
\newcommand{\R}{\right]}
\newcommand{\nn}{\nonumber}
\newcommand{\Tr}{\text{Tr}}
\newcommand{\N}{\mathcal{N}}
\newcommand{\G}{\Gamma}
\newcommand{\vf}{\varphi}
\newcommand{\LL}{\mathcal{L}}
\newcommand{\Op}{\mathcal{O}}
\newcommand{\HH}{\mathcal{H}}
\newcommand{\arctanh}{\text{arctanh}}
\newcommand{\up}{\uparrow}
\newcommand{\down}{\downarrow}
\newcommand{\ket}[1]{\left| #1 \right>}
\newcommand{\bra}[1]{\left< #1 \right|}
\newcommand{\ketbra}[1]{\left|#1\right>\left<#1\right|}
\newcommand{\rd}{\partial}
\newcommand{\de}{\partial}
\newcommand{\ba}{\begin{eqnarray}}
\newcommand{\ea}{\end{eqnarray}}
\newcommand{\db}{\bar{\partial}}
\newcommand{\we}{\wedge}
\newcommand{\ca}{\mathcal}
\newcommand{\lr}{\leftrightarrow}
\newcommand{\f}{\frac}
\newcommand{\s}{\sqrt}
\newcommand{\vp}{\varphi}
\newcommand{\hvp}{\hat{\varphi}}
\newcommand{\tvp}{\tilde{\varphi}}
\newcommand{\tp}{\tilde{\phi}}
\newcommand{\ti}{\tilde}
\newcommand{\ap}{\alpha}
\newcommand{\pr}{\propto}
\newcommand{\mb}{\mathbf}
\newcommand{\ddd}{\cdot\cdot\cdot}
\newcommand{\no}{\nonumber \\}
\newcommand{\la}{\langle}
\newcommand{\lb}{\rangle}
\newcommand{\ep}{\epsilon}
 \def\we{\wedge}
 \def\lr{\leftrightarrow}
 \def\f {\frac}
 \def\ti{\tilde}
 \def\ap{\alpha}
 \def\pr{\propto}
 \def\mb{\mathbf}
 \def\ddd{\cdot\cdot\cdot}
 \def\no{\nonumber \\}
 \def\la{\langle}
 \def\lb{\rangle}
 \def\ep{\epsilon}
\newcommand{\mcl}{\mathcal}
 \def\g{\gamma}
\def\Tr{\text{tr}}

\begin{titlepage}
\thispagestyle{empty}

\begin{flushright}

\end{flushright}
\bigskip

\begin{center}
  \noindent{\large \textbf{
  Scrambling and Recovery of Quantum Information in Inhomogeneous 
  Quenches in Two-dimensional Conformal Field Theories
  }}\\
  
\vspace{0cm}

\renewcommand\thefootnote{\mbox{$\fnsymbol{footnote}$}}
Kanato Goto\footnote{kanato.goto@yukawa.kyoto-u.ac.jp}${}^{1,2,3}$,
Masahiro Nozaki\footnote{mnozaki@ucas.ac.cn}${}^{4,3}$,
Shinsei Ryu\footnote{shinseir@princeton.edu}${}^{1}$,\\
Kotaro Tamaoka\footnote{tamaoka.kotaro@nihon-u.ac.jp}${}^{5}$
and Mao Tian Tan\footnote{maotian.tan@apctp.org}${}^{6}$\\

\vspace{0cm}

${}^{1}${\small \sl Department of Physics, Princeton University, Princeton, New Jersey, 08544, USA}\\
${}^{2}${\small \sl Center for Gravitational Physics and Quantum Information (CGPQI),\\ Yukawa Institute for Theoretical Physics,
Kyoto University, Kyoto 606-8501, Japan}\\
${}^{3}${\small \sl RIKEN Interdisciplinary Theoretical and Mathematical Sciences (iTHEMS), \\Wako, Saitama 351-0198, Japan}\\
${}^{4}${\small \sl Kavli Institute for Theoretical Sciences, University of Chinese Academy of Sciences,
Beijing 100190, China}\\
${}^{5}${\small \sl Department of Physics, College of Humanities and Sciences, Nihon University, \\Sakura-josui, Tokyo 156-8550, Japan}\\
${}^{6}${\small \sl Asia Pacific Center for Theoretical Physics, Pohang, Gyeongbuk, 37673, Korea}\\

\end{center}
\begin{abstract}
  We study
  various quantum quench processes induced by the M\"obius/sine-square
  deformation of the Hamiltonian in two-dimensional conformal field theories
  starting from the thermofield double state in the two copies of the Hilbert space.
  These quantum quenches, some of which are directly related to
  the operator entanglement of the time-evolution operators,
  allow us to study scrambling and recovery of quantum information.
  In particular, under the SSD time-evolution,
  we show from the time-dependence of mutual information that the Bell pairs,
  initially shared by the subsystems of the two Hilbert spaces,
  may revive even after the mutual information for small subsystems is
  completely destroyed by quantum information scrambling dynamics.
  This mutual information is robust against the strong scrambling dynamics.
  As a consequence, the steady state has a non-local correlation
  shared not by any of two parties but by three parties.
  In the holographic dual description, a wormhole connecting the two Hilbert spaces
  may non-linearly grow with time during the quantum quenches.
  We also propose effective pictures that describe the dynamics of mutual
  information during the time-evolution by inhomogeneous Hamiltonians.
\end{abstract}
\end{titlepage} 
\tableofcontents

\section{Introduction and summary}

Non-equilibrium dynamics in quantum many-body systems is a subject of intense research.
One of the recurrent themes is how quantum entanglement is generated 
and propagates during non-equilibrium processes.
It has been shown that complex (“chaotic”) quantum many-body systems can scramble quantum information non-locally.
Quantum information scrambling entails the loss of the information of initial states at least locally
and results in thermalization
\cite{PhysRevA.43.2046,PhysRevE.50.888,2008Natur.452..854R,2011RvMP...83..863P,2005JSMTE..04..010C}.
Experimental techniques to measure scrambling in laboratories have rapidly been
developed in the past few years (e.g., \cite{2019Natur.567...61L,2020PhRvL.124x0505J,2021PhRvX..11b1010B,2016PhRvA..94d0302S,PhysRevA.94.062329,2016arXiv160701801Y,2017PhRvA..95a2120Y,2018PhRvA..97d2105Y,2017PhRvE..95f2127C,2017arXiv171003363Y,2017NatPh..13..781G,2016arXiv161205249W,PhysRevX.7.031011,2017arXiv170506714M}).

Non-equilibrium dynamics in the context of (1+1)-dimensional conformal field theory (CFT)
have been widely studied in recent years 
\cite{2005JSMTE..04..010C,2016JSMTE..06.4003C,2007JSMTE..10....4C,2013JHEP...05..080N,2014PhRvL.112v0401C,Nozaki:2014hna,He:2014mwa,2014arXiv1405.5946C,2015JHEP...02..171A,Asplund_2015}.
In particular, recent works constructed a series of solvable models of quantum quench and Floquet dynamics in (1+1)-dimensional CFT using a class of inhomogeneous Hamiltonians. 
These works provide rare examples where the dynamics of interacting many-body quantum systems can be solved exactly. 
The inhomogeneous Hamiltonians used in these works include, in particular, the so-called sine-square deformation (SSD) and M\"obius deformation of (1+1)-dimensional quantum many-body systems. 
In these deformations, evolution operators are given as a linear superposition of three Virasoro generators
($L_0$, $L_{\pm 1}$),
which form an ${\it sl}(2, \mathbb{R})$ subalgebra of the Virasoro algebra. 
Not only being exactly solvable, these quantum quench and Floquet dynamics exhibit
rich behaviors, such as dynamical “phase transitions”
that separate heating and non-heating behaviors during time evolution
\cite{PhysRevB.97.184309,2019JPhA...52X5401M,Goto:2021sqx, PhysRevLett.118.260602,2018arXiv180500031W,2020PhRvX..10c1036F,2020PhRvB.102t5125H,2021PhRvR...3b3044W,2020arXiv201109491F,2021arXiv210910923W}.

Ref.\ \cite{Goto:2021sqx} studied quantum quench problems in 2d CFT 
using these inhomogeneous Hamiltonians starting from the Gibbs state as an initial state.
One of the main findings of Ref.\ \cite{Goto:2021sqx} is that the time evolution 
generates an inhomogenous temperature profile.
In particular, when the inhomogeneous post-quench Hamiltonian is the SSD Hamiltonian,
it heats up a spatial sub-region near the point where the Hamiltonian density vanishes, 
while it cools down the rest of the system. 
(The idea of using inhomogeneous Hamiltonians to prepare low-temperature states 
has been explored also outside the M\"obius/sine-square deformation -- see, for example,
\cite{2016arXiv161104591Z,2020PhRvR...2c3347R,HM,2018PhRvL.120u0604A,2019PhRvB..99j4308M,2022arXiv221100040W}.)
This heating process results in a local excitation that carries the (almost) entire entropy of the system,
which we call a black-hole-like excitation (B.H.-like excitation).
On the other hand, for the cooled region, non-local quantum correlations emerge 
under the inhomogeneous time evolution.
The SSD Hamiltonian can thus be used to ``simulate'' the formation of a black hole.

In this paper, we further study inhomogeneous deformations of the CFT Hamiltonian 
and the associated non-equilibrium dynamics. 
To be concrete, we will discuss three setups presented in Section \ref{Section:Preliminary}.
All these processes are quantum quenches starting from the thermofield double (TFD) state
defined on two copies of the Hilbert space, $\mathcal{H}_1$ and $\mathcal{H}_2$.

There are three motivations for studying 
these setups (roughly one for each setup).
First, in the previous works almost all properties discussed
(cooling/heating, the formation of B.H.-like excitations) are universal 
in the sense that they depend only on conformal symmetry.
Little is known about the effects of the inhomogeneous deformations on the details of theories
and quantum information scrambling.\footnote{We however note that
  the dynamics of mutual information was studied in 
  \cite{Goto:2021sqx}.
  Out-of-time-order correlators during the Floquet dynamics
  using M\"obius Hamiltonians were also studied
  in \cite{2022JHEP...08..221D}.
}
Different CFTs can exhibit different kinds of dynamics, e.g., integrable,
chaotic, or something in between.
For these dynamics, effective descriptions of dynamics have been developed
– the quasi-particle picture for integrable dynamics and the membrane picture
(line-tension picture) for chaotic and holographic dynamics. 
As discussed in \cite{Hosur:2015ylk,Nie:2018dfe}, quantum information scrambling 
can be detected by studying operator entanglement.
In particular, the operator entanglement for undeformed CFT time-evolution 
operators was previously discussed in \cite{Nie:2018dfe}.
In this work, we study the effect of the inhomogeneous temperature profile and B.H.-like excitations on 
quantum information scrambling.
%
In the quantum quench setups starting from the TFD state, we will study 
bipartite and tripartite mutual information between subsystems
in $\mathcal{H}_1$ and $\mathcal{H}_2$ which measures operator entanglement in disguise.

Second, by considering two-step time-evolution  operators, we discuss the recovery of quantum information. 
In the past decades, information retrieval from a black hole has received
considerable attention
\cite{2007JHEP...09..120H,2017arXiv171003363Y,2020arXiv200700895N,Tajima2021}.
In the setups considered in these works, quantum information is thrown into a
black hole, scrambled in its interior, and then emitted as the Hawking radiation.
These works investigated efficient ways of retrieving the quantum state from the emitted Hawking radiations.
Investigating the information retrieval from typical states, 
i.e., states in which information is scrambled, 
should lead to a deep understanding of quantum thermalization and black hole dynamics.
Our setups using inhomogeneous time evolution operators in 2d CFT that are rather different from 
those considered in the above works, where quantum information theoretical models were considered. 
Nevertheless, we will demonstrate the recovery of quantum information in our setups: 
If we start from the TFD state or a typical state and then evolve the system
with the SSD Hamiltonian acting on the single Hilbert space, 
then the mutual information between the subsystems on the different Hilbert spaces,
$\mathcal{H}_1$ and $\mathcal{H}_2$,
locally returns to its initial value. (Here, in our setups, the time evolution operator acts solely 
on $\mathcal{H}_1$.)
From this mutual information recovery, we can see 
the Bell pairs initially shared by the subsystems of $\mathcal{H}_1$ and
$\mathcal{H}_2$ may be revived during the SSD time evolution.
This recovered correlation may be robust against the scrambling effect of $2$d holographic CFTs.
Furthermore, under the evolution induced by the uniform holographic Hamiltonian, 
when the subsystems do not include the so-called fixed points, 
the system can develop a genuine tripartite correlation, 
i.e., 
a non-local correlation shared by three parties, but not by two parties only.
Finally, we are also interested in the dynamics of B.H.-like excitations.
In Setup 3 presented in Section \ref{Section:Preliminary},
we once again consider two-step time-evolution where the first step creates
a pair of B.H.-like excitations while the second step induces non-trivial dynamics thereof. 

We back up the above analyses for the specific setups by developing 
an effective description of the entanglement dynamics.
In particular, we develop the line-tension picture for inhomogeneous time evolution.
We also develop the holographic bulk description of these inhomogeneous quenches
by keeping track of the spatiotemporal deformations  
of the bulk black hole horizon. 
Finally, we also discuss the wormhole connecting the two Hilbert spaces.
Due to the non-trivial dynamics of the B.H.-like excitations,
the size of the wormhole exhibits an oscillatory growth.

The rest of the paper is organized as follows:
In Section \ref{Section:Preliminary},
we will describe the inhomogeneously-deformed Hamiltonians in 2d CFT, 
the three setups considered in this paper, 
and the measures of entanglement of our interest.
In Sections \ref{Section:thesystem1} and \ref{Section:thesystem2}, 
we will present the time-dependence of mutual information under the evolution by the inhomogeneous Hamiltonians, starting from the thermofield double and typical states.
In the following three sections, we report
the time-dependence of the entanglement measures in 
the three setups:
In Section \ref{Section:thesystem3}, we will report the time dependence of entanglement entropy and mutual information when we start from the thermofield double state, evolve the system with the SSD Hamiltonian, and then subsequently evolve it with the uniform Hamiltonian.
In Section \ref{Sec:Line-tension-picture},
we will propose an effective model that describes the operator entanglement hydrodynamics of the M\"obius/SSD 
time evolution operators.
In Section \ref{Section:gravitational-description}, we will report the dual geometries of the systems considered in this paper, and also present the growth of wormholes.
Finally, in Section \ref{Section:Discussions-and-future-directions }, we will discuss the possible applications of our results to experiments, and comment on a few future directions.
\section{Preliminaries\label{Section:Preliminary}}

In this section, we describe the inhomogeneously-deformed Hamiltonian,
the setups of our interest, and the measures of entanglement considered in this paper.

\subsection{Inhomogeneously-deformed Hamiltonians \label{Section:IDHs}}

Inhomogeneously-deformed Hamiltonians considered in this paper are defined by modulating 
the Hamiltonian density. Let us start from a homogeneous Hamiltonian 
$H_0 = \int^L_0 h(x) dx$
defined on a one-dimensional circle of circumference $L$,
i.e.,  the periodic spatial boundary condition is imposed.
The integrand of the homogeneous Hamiltonian, $h(x)$, is the Hamiltonian density.
An inhomogeneous deformation of $H_0$ can be introduced as 
\begin{align}
H_{\text{Inho}}= \int^{L}_0 dx\, f(x) h(x), 
\end{align}
where
$f(x)$ is an envelope function.
The envelope functions considered in this paper are 
\be
\begin{split}
f_{\text{M\"obius}}(x)&=1-\tanh{2\theta}\cos{\left(\f{2\pi x}{L}\right)},
\\
f_{\text{SSD}}(x)&=2\sin^2{\left(\f{\pi x}{L}\right)}, 
\\
f_{\text{CSD}} (x)&=2\cos^2{\left(\f{\pi x}{L}\right)}.
\end{split}
\ee
Here, $f_{\text{M\"obius}}(x)$ reduces to $f_{\text{SSD}}(x)$
and $f_{\text{CSD}}(x)$ in the $\theta\to \pm \infty$ limits respectively.
The inhomogeneously-deformed Hamiltonians with the envelope functions
$f(x)=f_{\text{M\"obius}}(x)$, $f_{\text{SSD}}(x)$, $f_{\text{CSD}}(x)$ are called M\"obius, sine-square (SS) and cosine-square (CS) deformed Hamiltonians respectively.

The SSD Hamiltonian was originally proposed as a simple way of removing the
boundary effect in finite size systems
\cite{PhysRevB.83.060414,PhysRevB.84.165132,2012JPhA...45k5003K,2009PThPh.122..953G,2011PhRvA..83e2118G,2011PhRvB..84k5116S,2011JPhA...44y2001K,PhysRevB.86.041108,PhysRevB.87.115128}.
Subsequently, the SSD in 2d CFTs and its one-parameter deformation (M\"obius deformation) 
were also discussed
\cite{2015JPhA...48E5402I,2016IJMPA..3150170I,2016arXiv160309543O,PhysRevB.93.235119,2017arXiv170906238T,2018PTEP.2018f1B01T}.
Recently, these deformations have been used to study thermalization and
non-thermalization
\cite{PhysRevB.97.184309,2019JPhA...52X5401M,Goto:2021sqx}
and Floquet dynamics
\cite{PhysRevLett.118.260602,2018arXiv180500031W,2020PhRvX..10c1036F,2020PhRvB.102t5125H,2021PhRvR...3b3044W,2020arXiv201109491F,2021arXiv210910923W}.

In these deformations, we naturally identify two special locations on the spatial circle, 
$x=0 \equiv X_f^1$ and $x=L/2 \equiv X^2_f$.
Being the minimum or maximum of the envelope functions,
we expect the effect of the envelope functions on quantum dynamics is most significant 
around these points. 
We will soon show that these points play special roles under the inhomogeneous
time evolution
by looking at various quantities such as the Heisenberg time evolution of operators. 

For the bulk of the paper, we mainly focus on the M\"obius and SS deformations.
The details of the analysis and calculations of the entanglement dynamics under
the CSD time-evolution are presented in Appendix \ref{App:system4}.

\subsection{The systems evolved with the inhomogeneously-deformed Hamiltonians \label{Section:The systems}}

We consider the following three setups in this paper.
In all setups, we consider the thermofield double (TFD) state  
\begin{align}
  \label{eq:def_of_TFD}
  \ket{\text{TFD}} =\mathcal{N} e^{-\f{\epsilon (H^1_0+H^2_0)}{2}}\sum_{a} \ket{a}_1 \otimes \ket{a}_2,
\end{align}
as our initial state of time evolution.
Here, the TFD state is defined in the doubled Hilbert space, $\mathcal{H}=\mathcal{H}_1\otimes \mathcal{H}_2$,
and $H^{i=1,2}_0$ and $\ket{a}_{i=1,2}$ denote the un-deformed $2$d CFT
Hamiltonian, and its eigenstates respectively.
The regulator $\epsilon$ is half of the inverse temperature, $\epsilon=\beta/2$. 
The square of the normalization factor $\mathcal{N}$ guarantees that $\left \langle \text{TFD}| \text{TFD} \right \rangle=1$.
We will mainly work with holographic CFTs, i.e., CFTs that admit holographic dual descriptions. 
However, we also study the 2d free fermion CFT as a representative of
non-chaotic (integrable) CFTs and make comparisons between the two.
The TFD state was previously used as a ``convenient'' initial condition
in quantum quench problems \cite{HM}. 
The TFD state is a short-range entangled state, 
and can be considered as a ground state of a gapped Hamiltonian
\cite{https://doi.org/10.48550/arxiv.1804.00491, Cottrell_2019}.
Our setups above are hence in a similar spirit to the seminal work 
by Calabrese and Cardy on quantum quench in 2d CFTs 
\cite{2005JSMTE..04..010C,2016JSMTE..06.4003C}.

{\bf Setup 1}: 
In the first setup, starting from the TFD state
we consider the time evolution under the unitary operator 
$
U_{\text{M\"obius/SSD}} 
=e^{-it_1 H^{1}_{\text{M\"obius/SSD}}}\otimes {\bf 1}_2$,
where $H^{1}_{\text{M\"obius/SSD}}$ and ${\bf 1}_2$ denote the M\"obius/SS
deformed Hamiltonian acting on $\mathcal{H}_1$, and identity operator on $\mathcal{H}_2$, respectively.
The evolved state is
\begin{align}
  \label{setup1}
  |\Psi_1(t_1)\rangle
  =
  \left(
  e^{-it_1 H^{1}_{\text{M\"obius/SSD}}}\otimes {\bf 1}_2
  \right)\ket{\text{TFD}}.
\end{align}

{\bf Setup 2}: 
In the second setup, we once again start from the TFD state,
and then consider the two-step time-evolution first by 
$e^{-it_0 H_{0}}$ 
and then
$e^{-it_1 H_{\text{M\"obius/SSD}}}$,
both acting on $\mathcal{H}_1$:
\begin{align}
  \label{setup 2}
  \ket{\Psi_2(t_1,t_0)} =\left( e^{-iH^1_{\text{SSD}}t_1} \otimes {\bf 1}_2 \right)
  \left(e^{-iH^1_0t_0}\otimes {\bf 1}_2\right) \ket{\text{TFD}}.
\end{align}
Here, the first time evolution can be interpreted as creating an excited state,
which is then time evolved during the second step of the time evolution.

{\bf Setup 3}: 
Finally, in the third setup, we exchange the ordering of the two
time evolution operators in Setup 2, and consider:
\begin{align}
  \label{eq:stata-in-system3}
  \ket{\Psi_3(t_1,t_0)} =\left( e^{-iH^1_{0}t_0} \otimes {\bf 1}_2 \right)
  \left(e^{-iH^1_{\text{SSD}} t_1}\otimes {\bf 1}_2\right) \ket{\text{TFD}}.
\end{align}

Let us now elaborate on the motivations for studying these setups and provide an overview of our results. 

-- We first note that, in addition to the interpretation 
as quantum quench, we can give an interpretation of 
these states (and entanglement measures for these states)
from the perspective of operator entanglement.
Consider, for a unitary time evolution operator $U_{\text{unitary}}$, 
an effective unitary time evolution operator 
$
U_{\text{effective}}=U_{\text{unitary}}e^{-\epsilon H_0}.
$
By using the channel-state map \cite{Hosur:2015ylk, nielsen_chuang_2010},
define the dual state of $U_{\text{effective}}$ as the state on the doubled Hilbert space, $\mathcal{H}_1\otimes \mathcal{H}_2$:
\begin{align}
  \label{dual_state}
  \ket{U_{\text{effective}}}=\mathcal{N}\sum_{a}U_{\text{unitary}}e^{-\f{\epsilon}{2} \left(H^1+H^2\right)}\ket{a}_1\otimes\ket{a^*}_2,
\end{align}
where $\ket{\cdot^*}$ is CPT conjugate of $\ket{\cdot}$, and $\ket{a}_i$
is an eigenstate of $H_i$.\footnote{The definition of dual state is not unique.}
The unitary time evolution operator acts only on  $\mathcal{H}_1$.
The dynamical properties of $U_{\text{effective}}$
are represented as the entanglement structure of the dual state.
Thus, the above states can be interpreted as the dual states of the effective unitary time evolution operators. 
In particular, by considering the state \eqref{setup1}
and its entanglement structure, we can discuss
the operator entanglement of the M\"obius/SS deformed time evolution operator
and the effect of the inhomogeneous deformation on 
quantum information scrambling. 
For the case of regular, homogeneous Hamiltonian $H_0$ of 2d CFTs,
the operator entanglement and quantum information scrambling were studied in \cite{Nie:2018dfe}.

-- In Setup 2 and 3, we have two-step time evolution operators. 
In Setup 2, the first time evolution under $H_0$ is expected to scramble quantum information
(for holographic CFTs).
Our interest here is the effect of the second time evolution
on the scrambled information.
As we will see, the SSD evolution recovers the non-local correlation between 
subsystems $A$ and $B$ in $\mathcal{H}_1$ and $\mathcal{H}_2$ when 
the subsystem $A$ includes $X^1_f$.
Namely, by the SSD evolution, we can retrieve the information from the typical state, the
state where the information is fully scrambled. 
The motivation for Setup 2 is thus in line with
information retrieval from a black hole
\cite{2007JHEP...09..120H,2017arXiv171003363Y,2020arXiv200700895N,Tajima2021}.
In these works, quantum information is first thrown into a black hole, 
scrambled in its interior, and then emitted as Hawking radiation. 
They investigated efficient ways of retrieving the quantum state from
the emitted Hawking radiations.

-- In Setup 3,
the first part of the two-step time evolution (with the SSD Hamiltonian on $\mathcal{H}_1$)
can be interpreted as preparing a pair of black-hole-like excitations (B.H.-like excitations)
\cite{Goto:2021sqx}.
The created B.H.-like excitations are then subject to the second step of the time evolution under 
the regular Hamiltonian $H_0^1$.
Setup 3 can thus be used to study the dynamics of the B.H.-like excitations.
As we will show, the second time evolution 
induces an interesting dynamics of 
the B.H.-like excitations that can be detected by monitoring various entanglement measures. 
Also, the wormhole growth measured by the geodesic length is described by the propagation of the B.H.-like excitations.
From this perspective,
it would also be interesting to 
consider a similar time evolution, 
\begin{align}
   \label{eq:evolvedstatein3-CSD}
  \ket{\Phi(t_1,t_2)}=\left(e^{-iH^1_{\text{CSD}}t_2}\otimes {\bf 1}_2\right) \left(e^{-iH^1_{\text{SSD}} t_1}\otimes {\bf 1}_2\right)\ket{\text{TFD}}.
\end{align}
Here, the first step of the time evolution is the same and still creates
a pair of B.H.-like excitations.
The second time evolution is however given by $H_{\text{CSD}}$, instead of $H_0$,
whose envelope function profile is complimentary to $H_{\text{SSD}}$.
The details of the entanglement dynamics for (\ref{eq:evolvedstatein3-CSD}) are presented in Appendix \ref{App:system4}.

\subsection{Entanglement entropies and the twist operator formalism}

\subsubsection{Entanglement entropies, bipartite and tripartite mutual information}

The main quantities of interest in this paper are entanglement entropies
for various subsystems as well as the bipartite and tripartite mutual information
(BMI and TMI, respectively).
Below, we consider a subsystem (sub-Hilbert space) of $\mathcal{H}_2$, which we call $A$.
Similarly, we consider a sub-Hilbert space $B$ of $\mathcal{H}_1$.
When discussing TMI, we consider two subsystems of $\mathcal{H}_2$, denoted as $B_1$ and $B_2$.
More specifically, subsystem $A$ is a spatial interval with its left and right ends located 
at $X_1$ and $X_2$, and, similarly, $B$ is an interval
with its left and right ends located at $Y_1$ and $Y_2$.
Here, $0<X_2<X_1$ and $0<Y_2<Y_1$.
($B_1$ and $B_2$ are also intervals -- their geometries are specified in the following.)
Starting from the total density matrix $|\Psi\rangle\langle\Psi|$,
we consider the reduced density matrix $\rho_{{\cal V}}$ 
(${\cal V}=A,B, A\cup B$, $\cdots$),
and the corresponding von Neumann and/or R\'enyi entropies.
They are denoted by $S_{{\cal V}}$ and $S^{(n)}_{{\cal V}}$, respectively.

Bipartite mutual information (BMI) for $A$ and $B$ is defined as the linear combination of the entanglement entropies:
\begin{align}
  I_{A,B}=S_A+S_B-S_{A\cup B}.
\end{align}
We note that $I_{A,B}$ is independent of the lattice spacing. 
Since the universal pieces of entanglement entropies cancel out,
$I_{A,B}$ depends on only the non-universal pieces of these entropies. 

To define tripartite mutual information (TMI) we consider 
three subsystems $A$, $B_1$ and $B_2$.
Then, the TMI for $A$, $B_1$ and $B_2$ is defined as a linear combination of  BMI:
\begin{align}
  \label{TOMI_for_scrambling}
  I_{A,B_1,B_2}=I_{A,B_1}+I_{A,B_2}-I_{A,B_1\cup B_2}.
\end{align}
As in
\cite{2016JHEP...02..004H,Nie:2018dfe,Kudler-Flam:2019wtv,Kudler-Flam:2019kxq,Kudler-Flam:2020yml,Mascot:2020qep,MacCormack:2020auw,Kudler-Flam:2021kfl,Goto:2021gve},
the TMI for operator entanglement can be a measure of scrambling.
The time-dependence of TMI may detect how the Bell pairs initially shared by $A$
and $\mathcal{H}_1$ are delocalized and become non-locally hidden in $\mathcal{H}_1$ under the time evolution.

\subsubsection{Parameter regimes of interest}

For the bulk of the paper, we are interested in the above entanglement quantities 
in the coarse-grained regime. This regime is defined as follows.
Let $\hat{\mathcal{V}}$ denote the subsystem consisting of the spatial intervals,
and then let $\hat{L}$, $\hat{l}_{\mathcal{V}}$, $\hat{a}$, $\hat{\epsilon}$,
and $\hat{t}$ denote a system size, a subsystem size, a lattice spacing,
a regularization parameter that guarantees the norm of states considered in this
paper is one, and the times associated to some Hamiltonian considered. 
Here, $\hat{*}$ denotes a dimensionful parameter, and $*$ is the dimensionless one defined as $\f{\hat{*}}{\hat{a}}$.
In the following, we will use only dimensionless parameters.
The parameter region considered is 
\begin{align}
  \label{eq:parameter}
  L \gg l_{\mathcal{V}}, 
  t \gg \epsilon \gg 1.
\end{align}
The interest in this regime comes from the expectation that in this regime 
we can potentially use effective descriptions of entropy propagation such as the quasiparticle picture
or the line-tension (membrane) picture.



\subsection{Path integral formulation and twist operators}

To develop the path-integral formalism, let us define Euclidean density operators as 
\begin{align}
  \label{eq:density-matrix-Eulidean}
  \rho_{E}=\mathcal{N}_E^2\sum_{a,b}e^{-\epsilon(E_a+E_b)}\left( U^1_{E}\ket{a}\bra{b}_1\tilde{U}^{1}_{E}\otimes \ket{a^*}\bra{b^*}_2\right),
\end{align}
where $\mathcal{N}^{-2}_{E}=\Tr e^{-2\epsilon H_0}$ guarantees that $\Tr \rho_{E}=1$.  
These density operators may be obtained from the ones defined in Section \ref{Section:The systems}
by analytically-continuing to imaginary time.
Here, the Euclidean evolution operator is given, depending on the setups above,
\be
\label{Euclidean time evolutions}
\begin{split}
  U^1_{E}=\begin{cases}
            e^{-H^1_{\text{M\"obius}}\tau_1} & 
            \\
            e^{-H^1_{\text{SSD}}\tau_1}e^{-H^1_0\tau_0} & 
            \\ 
            e^{-H^1_0\tau_0}e^{-H^1_{\text{SSD}}\tau_1} & 
          \end{cases},
    \quad
     \tilde{U}^1_{E}=\begin{cases}
    e^{H^1_{\text{M\"obius}}\tau_1} & 
    \\
    e^{H^1_0\tau_0}e^{H^1_{\text{SSD}}\tau_1} & 
    \\ 
    e^{H^1_{\text{SSD}}\tau_1}e^{H^1_0\tau_0} & 
    \end{cases}.
\end{split}
\ee
We now define the reduced Euclidean density operators
for $\mathcal{V}$ as $\rho_{E,\mathcal{V}}=\Tr_{\overline{\mathcal{V}}}\, \rho_{E}$.
They are given explicitly as 
\begin{align}
  \rho_{E,\mathcal{V}}
  =\begin{cases}
     \mathcal{N}^2_E\, \Tr_{\overline{A}}
     \left(e^{-2\epsilon H_0}\right) & \mathcal{V}=A\\
     \mathcal{N}^2_E\, \Tr_{\overline{B}}\left( U^1_{E}e^{-2\epsilon H_0}\tilde{U}^{1}_{E}\right) & \mathcal{V}=B,\\
     \mathcal{N}^2_E\, \sum_{a,b}e^{-\epsilon(E_a+E_b)}\Tr_{\tilde{B}}\left( U^1_{E}\ket{a}\bra{b}_1\tilde{U}^{1}_{E}\right)\otimes \Tr_{\tilde{A}}\left(\ket{a^*}\bra{b^*}_2\right) & \mathcal{V} =A\cup B
   \end{cases}, 
\end{align}
where $\bar{A}$ denotes the complement of $A$.
Let us define Euclidean entanglement entropy associated with
$\rho_{E,\mathcal{V}}$ as von Neumann entropy for this reduced density matrix:
\begin{align}
  \label{eq:def_of_ee}
  S_{E,\mathcal{V}}
  = -\Tr_{\mathcal{V}}\left( \rho_{E,\mathcal{V}} \log{\rho_{E,\mathcal{V}}}\right)
  = \lim_{n\rightarrow 1} \f{1}{1-n}\log{\Tr_{\mathcal{V}}\left(\rho_{E,\mathcal{V}}\right)^n}.
\end{align}
Thus, in the von Neumann limit $n\rightarrow$ 1, the $n$-th R\'enyi entropy,
$S^{(n)}_{E,{\cal V}}=\f{1}{1-n}\log{\Tr_{\mathcal{V}}\left(\rho_{E,\mathcal{V}}\right)^n}$,
reduces to the  Euclidean entanglement entropy.
In the path-integral formalism, $S^{(n)}_{E,{\cal V}}$ is given by
$S^{(n)}_{E,{\cal V}}=\f{1}{1-n}\log{\f{Z_n}{Z^n_1}}$,
where $Z_{n}$ is the partition function on an $n$-sheeted geometry defined by
sewing $\mathcal{V}$ together
in a cyclic fashion as in \cite{2004JSMTE..06..002C,2009JPhA...42X4005C}.
At the end of the calculations, we analytically continue $\tau_{i=0,1,2}$ to $it_{i=0,1,2}$
to obtain the time evolution of entanglement entropies.
With this procedure in mind, from now on, we drop the subscript ``$E$''
and simply write $S_{E,{\cal V}}\to S_{\cal V}$.

To compute $S_{\mathcal{V}}$, let us now employ the twist operator formalism
where $\Tr_{\mathcal{V}}\left(\rho_{\mathcal{V}}\right)^n$ is given by the
$2m_{\mathcal{V}}$-point functions
arising from insertion of the twist and anti-twist operators on the torus. Here, 
$\mathcal{V}$ is composed of $m_{\mathcal{V}}$ intervals.
Consequently, 
the R\'enyi entropies can be expressed as
\be \label{EE_euclidean}
\begin{split}
    &S^{(n)}_{A}=\f{1}{1-n}
    \log{\left\{
    \left\langle
    \overline{\sigma}_{n}(w_{X_1},\overline{w}_{X_1})\sigma_{n}(w_{X_2},\overline{w}_{X_2})\right\rangle_{2\epsilon}\right\}}, \\
    &S^{(n)}_{B}=\f{1}{1-n}\log{
      \Big\{
      {\cal N}_E^2\, \,
      \Tr
      \Big[
      \tilde{U}^1_{E}\,
      \sigma_{n}(w_{Y_1},\overline{w}_{Y_1})\overline{\sigma}_{n}(w_{Y_2},\overline{w}_{Y_2})U^1_{E}\,
      e^{-2\epsilon H_0}
      \Big]
      \Big\}
      },
  \\
    &S^{(n)}_{A\cup B}=\f{1}{1-n}\log
      \Big\{
      \mathcal{N}^2_E\, \, 
      \Tr\Big{[}
      e^{-\epsilon H_0}\,
      \tilde{U}^1_{E}\, 
      \sigma_{n}(w_{Y_1},\overline{w}_{Y_1})\overline{\sigma}_{n}(w_{Y_2},\overline{w}_{Y_2})
      \\
  &\qquad\qquad
    \qquad \qquad \qquad
    \times
    U^1_{E}\, e^{-\epsilon H_0}\,
    \overline{\sigma}_{n}(w_{X_1},\overline{w}_{X_1})\sigma_{n}(w_{X_2},\overline{w}_{X_2})
 \Big] 
  \Big\},
\end{split}
\ee
where $\left\langle \cdot \right \rangle_{2\epsilon}$ denotes the expectation
value on the thermal torus where thermal and spatial circumstances are $2\epsilon$ and $L$, respectively. 
The complex coordinate is defined as $(w_x,\overline{w}_x)=(ix,-ix)$, and
$h_n=\f{c(n^2-1)}{24n}$
denotes the conformal dimension of twist and anti-twist operators.
By using the identities, 
$\tilde{U}^1_{E}U^1_{E} =U^1_{E}\tilde{U}^1_{E}={\bf 1}$ and $e^{-\epsilon H_0} e^{\epsilon H_0}= e^{\epsilon H_0} e^{-\epsilon H_0}={\bf 1}$, we can rewrite $2m_{\mathcal{V}}$-point functions in (\ref{EE_euclidean}) as the ones in Heisenberg picture.
In the Heisenberg picture, the evolution of the twist and anti-twist operators in Euclidean time is given by 
\be \label{transformation_of_operator}
e^{\epsilon H_0}\,
\tilde{U}^1_{E}\,
\sigma_{n}\left(w_x,\overline{w}_x\right)\,
U^1_{E}\, e^{-\epsilon H_0}
=\left|\f{dw_{x,\epsilon}^{\text{New}}}{dw_{x}}\right|^{2h_n}
\sigma_{n}
\left(w^{\text{New}}_{x,\epsilon},\overline{w}^{\text{New}}_{x,\epsilon}\right).
\ee
Some details of $w_{x,\epsilon}^{\text{New}}$ and $\overline{w}_{x,\epsilon}^{\text{New}}$ 
are presented in Appendix \ref{App:Evo_o_Ope}. 
During the evolution by $U^1_{E}e^{-\epsilon H_0}$,
the location of the operators is mapped to $w^{\text{New}}_{x,\epsilon},\overline{w}^{\text{New}}_{x,\epsilon}$.
As a consequence, $S^{(n)}_{\mathcal{V}}$ is written as 
\be \label{eq:EE_euclidean_during_ev}
\begin{split}
    &S^{(n)}_{A}=\f{1}{1-n}\log{\left[\left\langle\overline{\sigma}_{n}(w_{X_1},\overline{w}_{X_1})\sigma_{n}(w_{X_2},\overline{w}_{X_2})\right\rangle_{2\epsilon}\right]}, \\
    &S^{(n)}_{B}=\f{1}{1-n}\log{\left[\Pi_{i=1,2}\left|\f{dw_{Y_i,\epsilon}^{\text{New}}}{dw_{Y_i}}\right|^{2h_n}\right]}
    +\f{1}{1-n}\log{\left\langle 
    \sigma_n\left(w^{\text{New}}_{Y_1,\epsilon}, \overline{w}^{\text{New}}_{Y_1,\epsilon}\right) 
    \overline{\sigma}_n\left(w^{\text{New}}_{Y_2,\epsilon}, \overline{w}^{\text{New}}_{Y_2,\epsilon}\right)\right \rangle_{2\epsilon}},\\
    &S^{(n)}_{A\cup B}=\f{1}{1-n}
    \log{\left[\Pi_{i=1,2}\left|\f{dw_{Y_i,\epsilon}^{\text{New}}}{dw_{Y_i}}\right|^{2h_n}\right]}\\
    &\qquad 
    +\f{1}{1-n}
    \log{\left\langle \sigma_n\left(w^{\text{New}}_{Y_1,\epsilon}, \overline{w}^{\text{New}}_{Y_1,\epsilon}\right) 
    \overline{\sigma}_n\left(w^{\text{New}}_{Y_2,\epsilon}, \overline{w}^{\text{New}}_{Y_2,\epsilon}\right)
    \overline{\sigma}_{n}(w_{X_1},\overline{w}_{X_1})\sigma_{n}(w_{X_2},\overline{w}_{X_2})\right \rangle_{2\epsilon}},
\end{split}
\ee
We note that $\left|\f{dw_{x,\epsilon}^{\text{New}}}{dw_{x,\epsilon}}\right|^{2h_n}$ is independent of the details of 2d CFTs. 
We hence call this factor the universal piece.
On the other hand, the two- and four-point functions of
the twist fields on the torus depend on the details of $2$d CFTs, and we call them the non-universal pieces.
These variables, $w^{\text{New}}_{x,\epsilon}$ and $\overline{w}^{\text{New}}_{x,\epsilon}$, 
depend on the imaginary times $\tau_{i=0,1,2}$. 
After we analytically continue $\tau_{i=0,1,2}$ to $it_{i=0,1,2}$,  only these imaginary parts of $w^{\text{New}}_{x,\epsilon}$ 
and $\overline{w}^{\text{New}}_{x,\epsilon}$ depend on these real times. 
In other words, during the evolution by $U^1_{E}e^{-\epsilon H_0}$, the twist and anti-twist operators spatially move with time as in Appendix \ref{App:Evo_o_Ope_ana}.
Under the evolution by $H_{\text{SSD/CSD}}$, the primary operators at $x=X^f_1=0$ or $x=X^f_2=\f{L}{2}$ does not spatially move.
We call $X^f_1$ and $X^f_2$ fixed points.

\subsubsection{Non-universal pieces in $2$d holographic CFTs}

Let us have a closer look at the non-universal pieces of the entanglement
entropy for the single and double intervals in $2$d holographic CFTs. 
To compare the results on 2d holographic CFTs with the ones in the $2$d free fermion CFT, 
we also calculated the non-universal pieces 
in the free fermion CFT. The results and calculations for the free fermion CFT are reported in Appendix \ref{App:NUP_in_free}.

\subsubsection*{Single interval}

Here, we present the non-universal piece of entanglement entropy for the single interval in the coarse-grained regime.
In this regime, the gravity dual of the system on the torus is the BTZ black hole \cite{Witten:1998zw}.
Therefore, in the von Neumann limit when $n\rightarrow 1$, the non-universal piece is given by the geodesic length in the BTZ black hole  \cite{Ryu:2006ef,Ryu:2006bv}.
Let $\mathcal{V}$ denote the subsystem, and also $v_1$ and $v_2$ denote the endpoints of $\mathcal{V}$. Here, we assume that $v_1>v_2>0$.
The non-universal piece of entanglement entropy for the reduced density matrix associated with $\mathcal{V}$ is holographically given by
\be \label{EE_on_torus}
\begin{split}
&\lim_{n\rightarrow 1}
\f{1}{1-n}
\log{\left\langle \sigma_n
\left(w^{\text{New}}_{v_1,\epsilon}, 
\overline{w}^{\text{New}}_{v_1,\epsilon}\right) 
\overline{\sigma}_n
\left(w^{\text{New}}_{v_2,\epsilon}, 
\overline{w}^{\text{New}}_{v_2,\epsilon}
\right)\right \rangle_{2\epsilon}} \approx \f{c}{3}\log{\left(\f{2\epsilon}{\pi}\right)}\\
&+\begin{cases}
&\text{Min}
\left[\f{c}{6}
\log{\left|\sin{\left(\f{\pi}{2\epsilon}
(w^{\text{New}}_{v_1, \epsilon}-w^{\text{New}}_{v_2, \epsilon}
\pm iL)\right)}\right|^2},
\f{c}{6}\log{\left|\sin{\left(
\f{\pi}{2\epsilon}
(w^{\text{New}}_{v_1, \epsilon}
-w^{\text{New}}_{v_2, \epsilon})\right)}
\right|^2}+\f{c\pi L}{6\epsilon}\right]\\
&
\qquad 
\text{if } x=X^1_f \in \mathcal{V} \\
&\text{Min}\left[\f{c}{6}
\log{\left|\sin{\left(\f{\pi}{2\epsilon}
(w^{\text{New}}_{v_1, \epsilon}
-w^{\text{New}}_{v_2, \epsilon}\pm iL)\right)}\right|^2}
+\f{c\pi L}{6\epsilon},
\f{c}{6}\log{\left|\sin{\left(\f{\pi}{2\epsilon}
(w^{\text{New}}_{v_1, \epsilon}-w^{\text{New}}_{v_2, \epsilon})\right)}\right|^2}\right]\\
&
\qquad \text{if } x=X^1_f \notin \mathcal{V} \\
\end{cases}.
\end{split}
\ee
\subsubsection*{Double intervals}
Let us turn to the non-universal piece of the entanglement entropy for a union of double intervals.
In $2$d holographic CFTs, the non-universal piece for a pair of intervals is given by  
\be \label{eq:non-uni-SAB-double}
\begin{split}
    &\lim_{n\rightarrow 1}\f{1}{1-n}
    \log{\left\langle \sigma_n\left(w^{\text{New}}_{Y_1,\epsilon}, 
    \overline{w}^{\text{New}}_{Y_1,\epsilon}\right) \overline{\sigma}_n
    \left(w^{\text{New}}_{Y_2,\epsilon}, \overline{w}^{\text{New}}_{Y_2,\epsilon}\right)\overline{\sigma}_{n}(w_{X_1},\overline{w}_{X_1})\sigma_{n}(w_{X_2},\overline{w}_{X_2})\right \rangle_{2\epsilon}}\\
    &
      \qquad
      \approx \f{2c}{3}\log{\left(\f{2\epsilon}{\pi}\right)}+\text{Min}\left[S_{\text{dis}},S_{\text{con}}\right],
  \end{split}
\ee  
where $S_{\text{dis}}$ is determined by the length of geodesic that connects the endpoints of intervals at the same Euclidean time slices, while $S_{\text{con}}$ is determined by that of geodesics connecting points on different Euclidean time-slices. Some details of $S_{\text{dis}}$ and $S_{\text{con}}$ are reported in Appendix \ref{App:NUP_hol_CFT}.
The Euclidean temporal and spatial locations, $\tau^{\text{New}}_{x,\epsilon}$ 
and $X^{\text{New}}_{x,\epsilon}$, 
of endpoints are defined as
\be
\tau^{\text{New}}_{x,\epsilon}
=\f{w^{\text{New}}_{x,\epsilon} +\overline{w}^{\text{New}}_{x,\epsilon}}{2}
,
\quad 
  X^{\text{New}}_{x,\epsilon}
  =\f{w^{\text{New}}_{x,\epsilon} 
  -\overline{w}^{\text{New}}_{x,\epsilon}}{2i}.
\ee

\section{Setup 1\label{Section:thesystem1}} 
Let us now turn to the analysis of the time-dependence of BMI and TMI in
Setup 1, 
\eqref{setup1}.
%
One of the main findings is Fig.\ \ref{Fig:t1dependece_IAB_theta}
where we plot BMI as a function of time for various choices of $\theta$.
This plot should be compared with,
e.g., Fig.\ 11 in Ref.\ \cite{Nie:2018dfe}
where BMI (or bipartite operator mutual information) of
the regular, homogeneous time evolution operator for holographic CFTs was
studied. 
Interestingly, we find a threshold value of $\theta$
that separates the two types of behaviors of BMI
presented in the left and right panels of 
Fig.\ \ref{Fig:t1dependece_IAB_theta}, respectively.
We also compare holographic CFTs and the free fermion CFT
described by the quasiparticle picture.

\subsection{Analysis of the geodesic length \label{Section:analysis_on_gl}}
We first discuss the time-dependence of geodesics
corresponding to the non-universal pieces 
of $S_A$, $S_B$, and $S_{A\cup B}$
in the Heisenberg picture.
For simplicity, let us suppose that the center of $B$ is at $x =X^{f}_1$.
The twist and anti-twist operators associated with $\rho_A$ are stationary, so that in the coarse-grained region, the entanglement entropy is approximated by a stationary constant,
\be \label{eq:EE-for-A}
S_{A}\approx\f{c \pi l_A}{6\epsilon},
\ee
where $l_A$ is the subsystem size of $A$.
Let us look closely at the time-dependence of the non-universal pieces of $S_B$ and $S_{A\cup B}$.
The twist and anti-twist operators associated with $B$ evolve
under $H_{\text{M\"obius/SSD}}$
and 
periodically move between 
the two fixed points
$x=X^{f}_1$ and $x=X^{f}_2$ with period 
$L \cosh{2\theta}$.  
In the SSD limit $\theta \rightarrow \infty$, 
the oscillation disappears,
and  these operators move 
asymptotically 
toward one of the fixed points, $x=X^{f}_2$.
The traveling speed of these operators depends on their locations and $\theta$.
According to the time evolution of the twist and anti-twist operators, 
the size of the subsystem 
associated with these operators 
grows and shrinks with time.
Consequently, the geodesic length associated with this subsystem increases and decreases. 

\begin{figure}[tbp]
    \begin{tabular}{cc}
      \begin{minipage}[t]{0.5\hsize}
        \centering
        \includegraphics[keepaspectratio, scale=0.03]{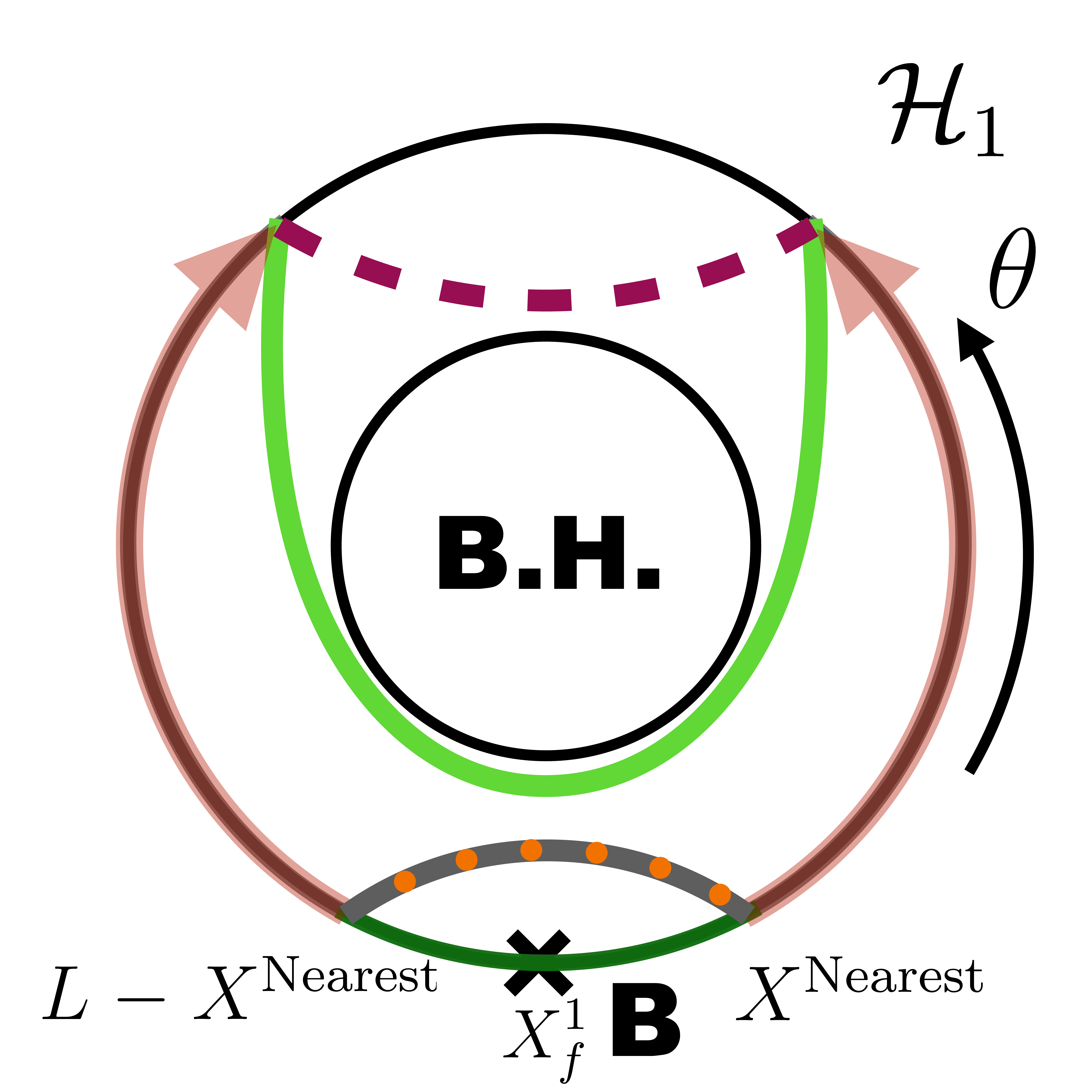}

    [a] The geodesics associated with $B$ in $\mathcal{H}_1$.

      \end{minipage} &
      \begin{minipage}[t]{0.5\hsize}
        \centering
        \includegraphics[keepaspectratio, scale=0.03]{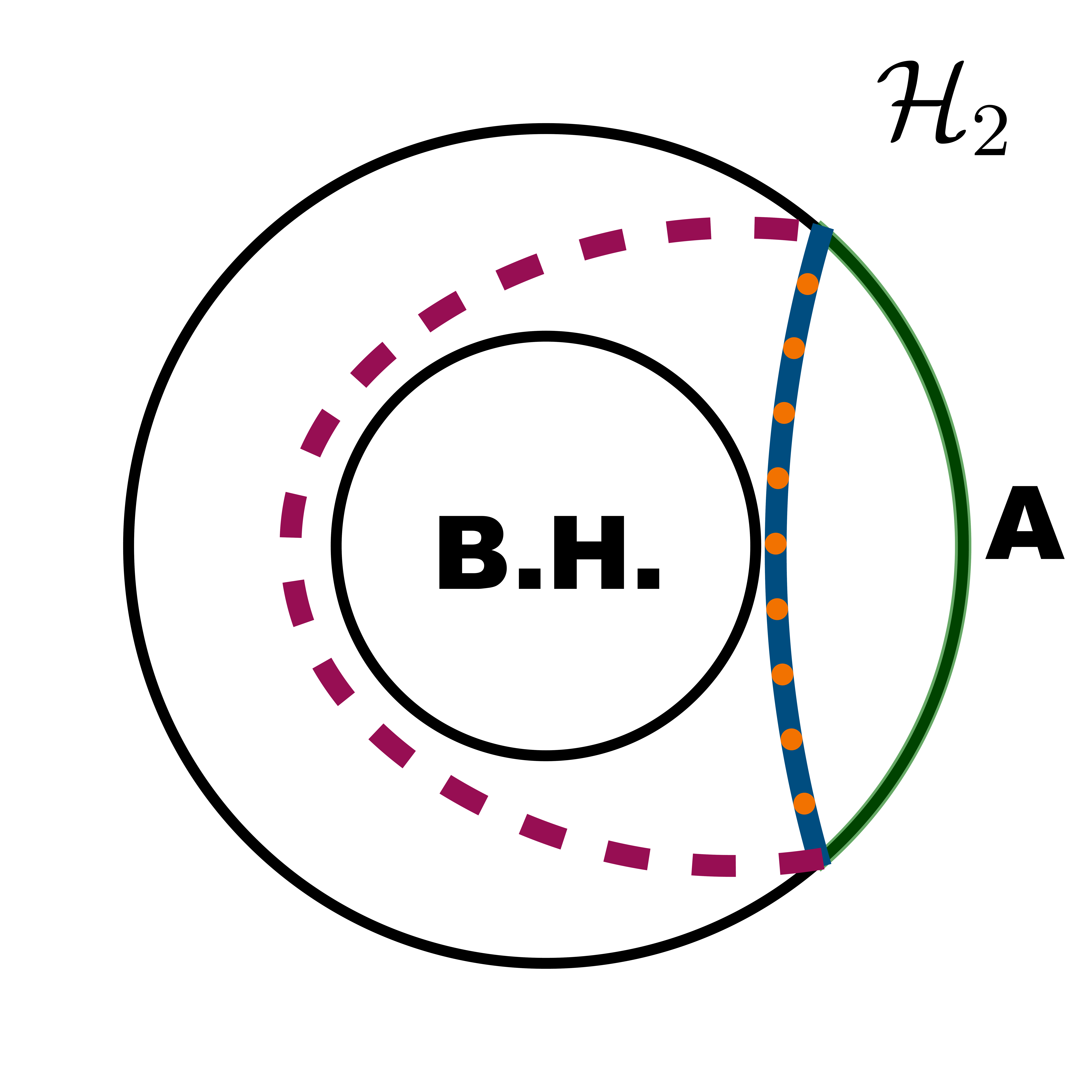}

     [b] The geodesics associated with $A$ in $\mathcal{H}_2$.

      \end{minipage} 
    \end{tabular}
    \caption{The M\"obius evolution of the subsystem in the Heisenberg picture. The green lines illustrate the subsystems, $A$ and $B$. 
    The blue solid line illustrates the non-universal piece of $S_A$.
    The gray solid line illustrates the non-universal pieces of $S_B$ for small $\theta$, while the light green solid line illustrates that of $S_B$ for large $\theta$. 
    For $0\le \theta \le \theta_C$, the non-universal piece of $S_{A \cup B}$ is the orange dashed line, while for $\theta_C <\theta$, it is given by the purple dotted lines. The red arrow illustrates the growth of $X^{\text{Nearest}}_{i=Y_1,Y_2}$ with the increase of $\theta$. 
    The details of $\theta_C$ is reported in Appendix
    \ref{Section:theta-critical}.
  }
    \label{geometry_correspond_thesystem1}
  \end{figure}

For the non-universal piece of $S_{A\cup B}$, in the small $t_1$-regime, the non-universal piece of $S_{A\cup B}$ may be given by the lengths of geodesics connecting the endpoints of $A$ and $B$, $S_{\text{con}}$, while in the large $t_1$-regime, it may be given by the ones connecting the endpoints on the same Euclidean time-slices, $S_{\text{dis}}$.
Therefore, for large $t_1$, the non-universal pieces of $S_B$ and $S_{A\cup B}$ may be determined by the lengths of the geodesics connecting the endpoints of the subsystems on the same Euclidean time-slices as in Fig.\ \ref{geometry_correspond_thesystem1}.
More specifically,
for the $t_1$-regime where 
$\left(w^{\text{New}}_{Y_1,\epsilon}-w^{\text{New}}_{Y_2,\epsilon}\right)/(i\epsilon) \gg 1$,
$\left(\overline{w}^{\text{New}}_{Y_1,\epsilon}-\overline{w}^{\text{New}}_{Y_2,\epsilon}\right)/(i\epsilon) \gg 1$,
$\left[iL-\left(w^{\text{New}}_{Y_1,\epsilon}-w^{\text{New}}_{Y_2,\epsilon}\right)\right]/(i\epsilon) \gg 1$ and $\left[iL+\left(\overline{w}^{\text{New}}_{Y_1,\epsilon}-\overline{w}^{\text{New}}_{Y_2,\epsilon}\right)\right]/(i\epsilon) \gg 1$, 
$S_{A\cup B}$ should be approximated by
\be
\begin{split}
    S_{A\cup B} \approx \text{Min}\left[\hat{S}_1,\hat{S}_2\right]\\
    =\f{c \pi}{6\epsilon}\times \text{Min}\bigg{[}&L+\left(X^{\text{New}}_{Y_1,\epsilon}-X^{\text{New}}_{Y_2,\epsilon}-l_A\right),
    L-\left(X^{\text{New}}_{Y_1,\epsilon}-X^{\text{New}}_{Y_2,\epsilon}\right)+l_A\bigg{]},
\end{split}
\ee
where $\hat{S}_2$ 
is the same as the non-universal piece of $S_B$ in this time regime.

Which of these contributions,
$\hat{S}_1$ and $\hat{S}_2$,
is dominant depends on $\theta$
and there is a threshold value 
$\theta_C$ separating the two cases.
In the small $\theta$ regime,
$0\le\theta\le \theta_C$, 
$S_{\text{dis}}$ is given by $\hat{S}_2$, so that for small $\theta$ but large $t_1$, $I_{A,B}$ is zero. 
On the other hand, in the large $\theta$-regime,
$\theta_C < \theta$, 
$S_{\text{dis}}$ is given by $\hat{S}_1$.
In this time regime, $S_A$ and $S_B$ are approximated by 
\eqref{eq:EE-for-A} 
and 
$\frac{c\pi  \left[ L- (X^{\text{New}}_{Y_1,\epsilon}-X^{\text{New}}_{Y_2,\epsilon}) \right]}{6\epsilon}$,
respectively, so that $I_{A,B}$ is approximated by
\be
I_{A,B} \approx\f{c \pi\left[l_A-(X^{\text{New}}_{Y_1,\epsilon}-X^{\text{New}}_{Y_2,\epsilon})\right]}{3\epsilon}.
\ee
The critical value $\theta_C$ separating these two cases 
depends on $Y_{i=1,2}$, $X_{i=1,2}$, and $L$
and can be determined as follows.
Let us suppose that $\overline{\sigma}_n(w^{\text{New}}_{Y_2,\epsilon},\overline{w}^{\text{New}}_{Y_2,\epsilon})$ moves with time between $x=X^{\text{Nearest}}_{Y_2}$ and $x=X^{\text{Furthest}}_{Y_2}$ where $0<X^{\text{Furthest}}_{Y_2}<X^{\text{Nearest}}_{Y_2}<L/2$, while $\sigma_n(w^{\text{New}}_{Y_1,\epsilon},\overline{w}^{\text{New}}_{Y_1,\epsilon})$ moves 
between $x=X^{\text{Nearest}}_{Y_1}$ and $x=X^{\text{Furthest}}_{Y_1}$ where $L/2<X^{\text{Nearest}}_{Y_1}<X^{\text{Furthest}}_{Y_1}<L$.
If $\theta$ becomes larger, then $X^{\text{Nearest}}_{i=Y_1,Y_2}$ gets closer to $X^f_2$.
Let $t_{1,\text{Max}}$ denote the time for the effective size of $B$ to reach its maximum. 
This time, $t_{1,\text{Max}}$, depends on $\theta$, $Y_1$ and $Y_2$.
Let $\theta_C$ denote the value of inhomogeneous parameter, for which $L- (X^{\text{New}}_{Y_1,\epsilon}
-X^{\text{New}}_{Y_2,\epsilon})$ is equal to $X^{\text{New}}_{Y_1,\epsilon}
-X^{\text{New}}_{Y_2,\epsilon}$ at $t_1=t_{1,\text{Max}}$.
The details of the analysis of $\theta_C$ are reported in Appendix \ref{Section:theta-critical}.

For $H_{\text{SSD}}$, 
in the time regime when the 
B.H.-like excitations,
with each of them having half of the thermal entropy on $\mathcal{H}_1$,
emerge around $x=X^f_1$ \cite{Goto:2021sqx}, $I_{A,B}$ is approximated by 
\be \label{eq:IAB_late}
I_{A,B}\approx \f{2c \pi l_A}{6\epsilon},
\ee
where $l_A=X_1-X_2$.
One possible interpretation for $I_{A,B}$ after the emergence of the B.H.-like excitations is that $I_{A,B}$ may measure the Bell pairs initially shared by $A$ and $\mathcal{H}_1$. 

If $B$ does not include $x=X^{1}_f$, then for the large $t_1$-regime under the SSD evolution, the non-universal piece of $S_{A\cup B}$ is given by that of $S_A+S_B$ where $S_B$ is approximated by 
the entanglement entropy of the vacuum state.
As a consequence, $I_{A,B}$ is zero
at late times.
This means 
that the reduced density matrix on $A \cup B$ 
approximately factorizes as 
\be
\rho_{A\cup B}(t_1 \gg 1) \approx \rho_{\text{Thermal},A} \otimes \rho_{\text{Vacuum},B},
\ee
where $\rho_{\text{Thermal},A}$ 
the reduced density matrix of a thermal state
at inverse temperature $2\epsilon$ for subsystem $A$,
and 
$\rho_{\text{Vacuum},B}$ 
is the reduced density matrix of 
the vacuum state for subsystem $B$.

\subsubsection{The $\theta$- and position-dependence of $I_{A,B}$ 
\label{Section:theta-position-dependence}}

\begin{figure}[tbp]
    \begin{tabular}{cc}
      \begin{minipage}[t]{0.45\hsize}
        \centering
        \includegraphics[keepaspectratio, scale=0.6]{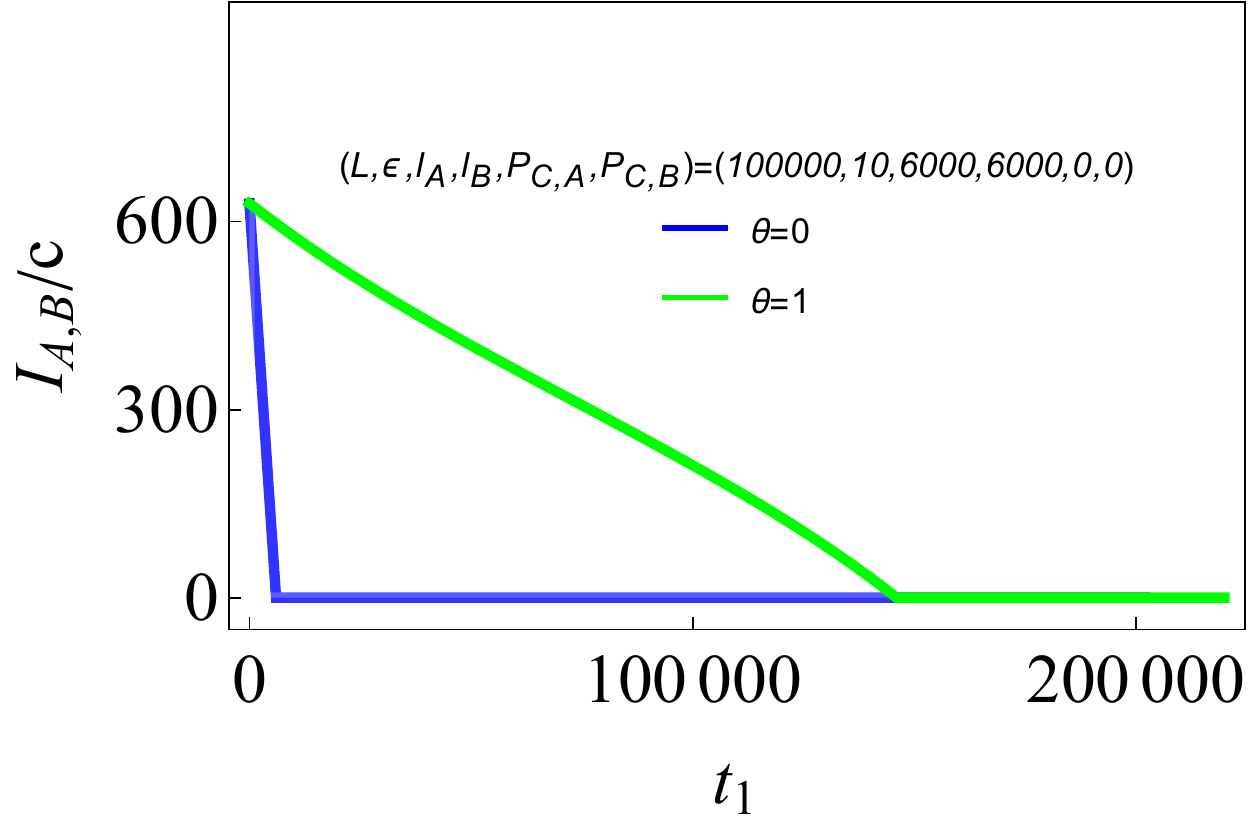}

    (a) $0\le \theta \le \theta_C$

      \end{minipage} &
      \begin{minipage}[t]{0.45\hsize}
        \centering
        \includegraphics[keepaspectratio, scale=0.6]{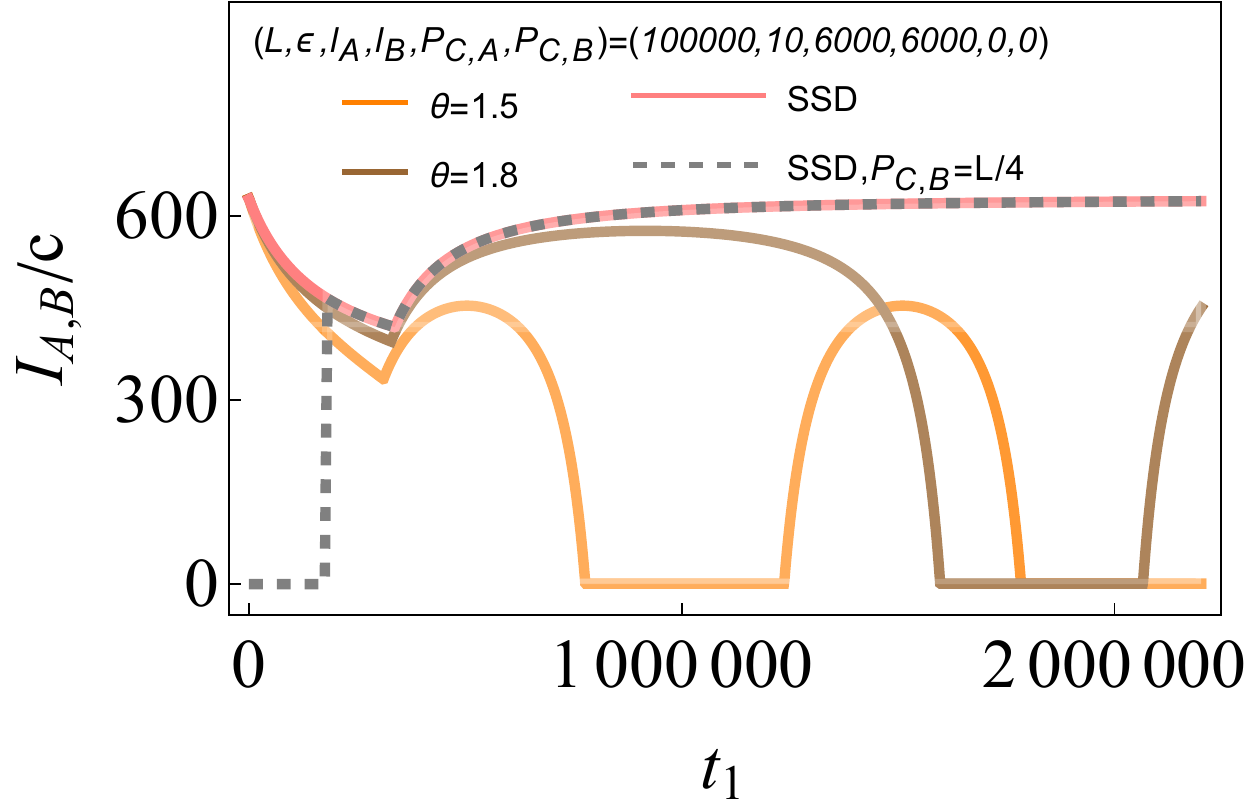}

    (b) $\theta_C< \theta$

      \end{minipage} 
    \end{tabular}
    \caption{The time-dependence of $I_{A,B}$ for 
    (a) $0 \le \theta \le \theta_C$
    and 
    (b) $\theta_C\le \theta$
    as a function of $t_1$.
    Here, $l_{\alpha=A,B}$ and $P_{C,\alpha=A,B}$ denote the size and the center
    of $\alpha$.
  }
    \label{Fig:t1dependece_IAB_theta}
  \end{figure}

The behavior of the geodesic
and the time-evolution of the subsystems in the Heisenberg picture
described above
is directly translated into 
the time-dependence of $I_{A,B}$. 
In Fig.\ \ref{Fig:t1dependece_IAB_theta}, 
we plot $I_{A,B}$ for various choices of $\theta$ as a function of $t_1$.
In this plot, the center of $B$ is $x=X^f_1$.
The solid lines illustrate the time-dependence of $I_{A,B}$ for $A$, the center of which is $x=X^f_1$, while the dashed line illustrates that for $A$, the center of which is $x=\f{L}{4}$.
In Fig.\ \ref{Fig:t1dependece_IAB_theta} (a), we show the time-dependence of $I_{A,B}$ for the small $\theta$-region where $0\le \theta \le \theta_C$, while in (b), we show that for the large $\theta$-region where $\theta_C <\theta$.
  As discussed in Section \ref{Section:analysis_on_gl},
  in the late time-regime, $I_{A,B}$ for $0\le\theta\le \theta_C$ is practically zero, while that for $\theta_C<\theta$ becomes positive.
For $0\le\theta\le \theta_C$, $I_{A,B}$ monotonically decreases with $t_1$ up to $t_{1,*}$, and then is practically zero.
Here, $t_{1,*}$ is the phase transition time where $S_{\text{con}}$ exchanges dominance with $S_{\text{dis}}$. 
The details of early-time decay
depends on $\theta$:
For larger 
$\theta$,
the early-time decay is slower
($t_{1,*}$ is bigger).
This behavior for $\theta < \theta_C$ is
similar to what was found 
for bipartite operator mutual information of
the regular homogeneous time-evolution operator
of holographic CFTs
in Ref.\ \cite{Nie:2018dfe}.

On the other hand,
the behvior 
for $\theta_C < \theta$ is markedly different.
Except for the SSD limit, 
$I_{A,B}$ first monotonically decreases with $t_1$ up to $t_{1,*}$, and then 
oscillates
with
periodicity $L \cosh 2\theta$.
For larger $\theta$ (closer to the SSD limit), 
the early-time decay is slower, and $I_{A,B}$ at $t_1=t_{1,*}$ is larger. 
In the SSD limit, after $t_1=t_{1,*}$ $I_{A,B}$ grows with $t_1$, and saturates to a value that is proportional to the size of $A$.
We will revisit this behavior
in Sec.\ \ref{Sec:Line-tension-picture}
by developing the line-tension picture (membrane picture)
for inhomogeneous chaotic time-evolution operators. 

\begin{figure}[htbp]
    \begin{tabular}{cc}
      \begin{minipage}[t]{1\hsize}
        \centering
        \includegraphics[keepaspectratio, scale=0.6]{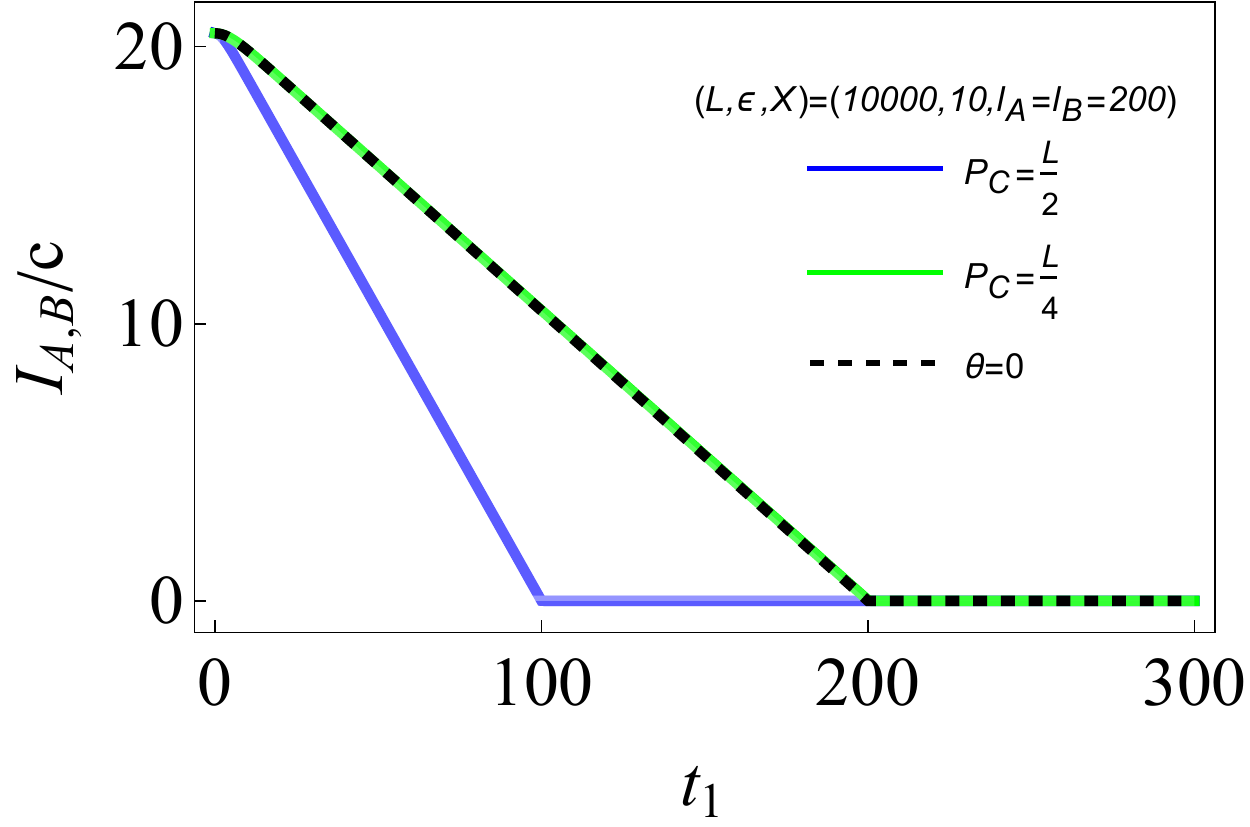}
      \end{minipage} &
     
    \end{tabular}
    \caption{
    The time-dependence of $I_{A,B}$ for $P_C=\f{L}{4}, \f{L}{2}$ for $\theta=0$ (dashed line) and in the SSD limit (solid lines)
    as a function of $t_1$.
    Here, we choose $l_A=l_B$ and $P_{C,A}=P_{C,B}=P_C$. The solid lines illustrate the $t_1$-dependence of $I_{A,B}$ for $P_C=\f{L}{4}, \f{L}{2}$ in the SSD limit. 
    For $\theta=0$, the time-dependence of $I_{A,B}$ is independent of $P_C$.}
    \label{Fig:t1dependece_IAB_position}
  \end{figure}

Let us turn to the analysis of the position-dependence of $I_{A,B}$.
For simplicity, let us consider the SSD limit, and $l_A=l_B$, and $P_{C,A}=P_{C,B}=P_C$.
We can see from the time-dependence of $I_{A,B}$ how scrambling may destroy the non-local correlation between $A$ and $B$.
Also, we can see the times when $\rho_{A\cup B}$ may approximately factorize into $\rho_A$ and $\rho_B$:
\be
\rho_{A \cup B} \approx \rho_A \otimes \rho_B.
\ee
In Fig.\ \ref{Fig:t1dependece_IAB_position}, we depict $I_{A,B}$ for various $P_C$ as the function of $t_1$.
In this figure, we take $P_C$ to be $\f{L}{4}$ and $\f{L}{2}$.
From the time-dependence of $I_{A,B}$ in Fig.\ \ref{Fig:t1dependece_IAB_position}, we can see that when $\theta$ becomes larger, the early-time decay of $I_{A,B}$ for $P_C=\f{L}{2}$ becomes faster and the time for $\rho_{A\cup B}$ to factorize into $\rho_A$ and $\rho_B$ may become smaller.
For $P_C=\f{L}{4}$, the $t_1$-dependence of $I_{A,B}$ may be independence of $\theta$.
One possible explanation for the $t_1$-dependence is that the inhomogeneous deformation may promote scrambling to destroy the non-local correlation around $P_C=X_f^2$, while this deformation may make scrambling destroy it around $P_C =X_f^1$ slower, and then prevent $\rho_{A\cup B}$ from factorizing into $\rho_A$ and $\rho_B$.

\subsubsection{Theory-dependence of $I_{A,B}$ under evolution 
\label{Section:theory-dependence}}

\begin{figure}[tbp]
    \begin{tabular}{cc}
      \begin{minipage}[t]{1\hsize}
        \centering
    \includegraphics[keepaspectratio, scale=0.6]{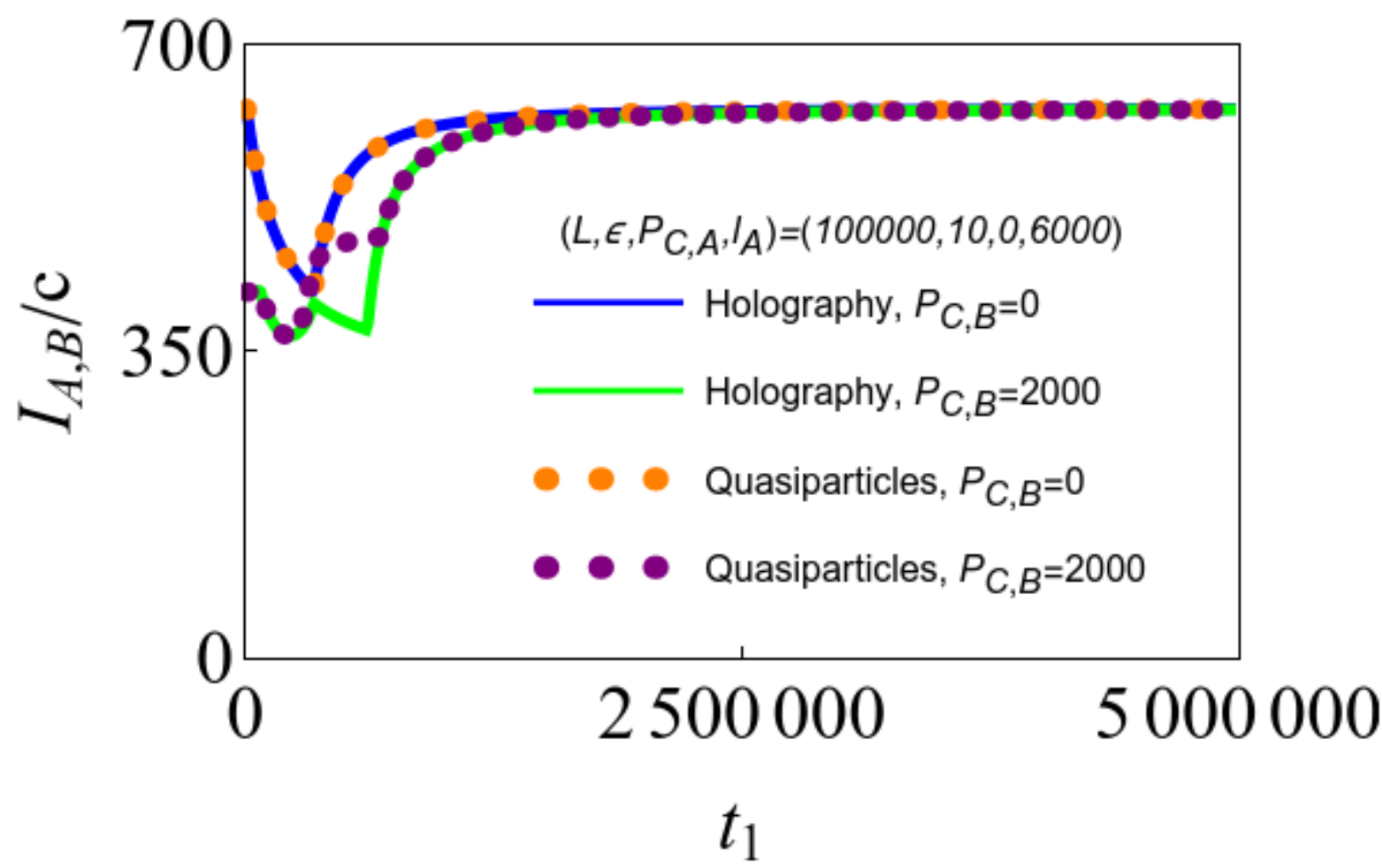}
      \end{minipage} &
     
    \end{tabular}
    \caption{
    The time-dependence of $I_{A,B}$ in the SSD limit as a function of $t_1$. Here, we compare the $t_1$-dependence of $I_{A,B}$ that follows from the propagation of quasiparticles 
    (dotted lines)
    with that of $I_{A,B}$ in $2$d holographic CFTs
    (solid lines). 
    In this figure, $P_{C,J=A,B}$ and $l_{J=A,B}$ denote the centers and sizes of $J=A$ and $B$, respectively. }
    \label{Fig:theory-dependence}
  \end{figure}

We have so far focused on holographic CFTs.
However, as one of our motivations is to understand
quantum information scrambling behaviors and their theory dependence,
we now make a comparison,
for the time-dependence of $I_{A,B}$,
between $2$d holographic CFTs with the $2$d free fermion CFT.
First,
as we show in Appendix \ref{App:free_fermion},
for the free fermion CFT with inhomogeneous time evolution, 
we can establish that its entanglement dynamics 
is described by the quasiparticle picture,
just like the standard case of homogeneous time evolution.
In this picture, the time-dependence of $I_{A,B}$ follows the propagation of quasiparticles at speeds given by ${H}_{\text{M\"obius/SSD}}$.
Some details of
the calculations of $I_{A,B}$
in the $2$d free fermion CFT
and a detailed description of the quasiparticle picture
can be found in Appendix \ref{App:free_fermion}. 
The upshot is that the BMI 
in the 2d free fermion CFT is carried separately by left and right-moving quasiparticles that move independently of one another. These quasiparticles are localized packets of information and their number is conserved. In Fig.\ \ref{Fig:theory-dependence}, 
we plot $I_{A,B}$ in the SSD limit as a function of $t_1$.
We see that if the size and center of $A$ are the same as $B$, 
then the time-dependence of $I_{A,B}$ in $2$d holographic CFTs follows the quasiparticle picture.
This is however not the case otherwise.
We will propose an effective picture that describes the $t_1$-dependence of $I_{A,B}$ in the $2$d holographic CFTs in Section \ref{Sec:Line-tension-picture}.

\subsection{Tripartite mutual information}
Let us turn to 
TMI.
Suppose that we divide $\mathcal{H}_{2}$ into $A$ and its complement, and also 
$\mathcal{H}_1$ into $B_1$ and $B_2$, and then define TMI as $I_{A,B_1,B_2}=I_{A,B_1}+I_{A,B_2}-I_{A,B_1\cup B_2}$. 
Here, we also assume that $l_A=l_{B_1}$ and $P_{C,A}=P_{C,B}=X^1_f$.
Then, the time-dependence of $I_{A,B_1}$ is the same as that of $I_{A,B}$ reported in Section \ref{Section:theta-position-dependence}, while $I_{A,B_2}$ is independent of $t_1$ and approximately zero.
In the coarse-grained regime, 
$I_{A,B_1\cup B_2}$ is also independent of $t_1$ and approximated by (\ref{eq:IAB_late}).
During the evolution by $H_{\text{M\"obius}}$ with $0\le \theta \le \theta_C$, $I_{A,B_1,B_2}$ is a stationary constant in (\ref{eq:IAB_late}), while for $\theta_C <\theta$, $I_{A,B_1,B_2}$ is a periodic function of $t_1$ with period $T=L \cosh{2\theta}$. The range of $I_{A,B}$ is between zero and (\ref{eq:IAB_late}).
In the SSD limit, the time dependence of $I_{A,B_1,B_2}$ is approximated by $I_{A,B_1,B_2} \approx I_{A,B}-\f{2c \pi l_A}{6\epsilon}$,
where $I_{A,B}$ is reported in Section \ref{Section:theta-position-dependence}.
For large $t_1$-regime, $I_{A,B_1,B_2}$ saturates
to zero.
One possible explanation for the late-time value of $I_{A,B_1,B_2}$ is that the correlation initially shared by $A$ and $B_1$ may not be scrambled in the whole $\mathcal{H}_1$, and this correlation between $A$ and $B_1$ may be revived.

\section{Setup 2 \label{Section:thesystem2}}
 
In this section, we present the time-dependence of BMI and TMI for 
Setup 2, \eqref{setup 2}.
Recall that in \eqref{setup 2}
the state is first time-evolved by the homogeneous Hamiltonian 
and then by the SSD Hamiltonian.
In holgraphic CFTs, the first step of the time-evolution
scrambles quantum information of the initial state
and produce a typical state (the Page state)
\cite{1993PhRvL..71.1291P,1996PhRvL..77....1S}.
Our focus here is the effect of the second step of the time-evolution
on the scrambled information. 

Let us focus on the analysis of the lengths of geodesics corresponding to $I_{A,B}$.
Let $\mathcal{V}_1$ and $\mathcal{V}_2$ denote the sub-regions on $\mathcal{H}_1$ and $\mathcal{H}_2$, respectively, and also let $l_{\mathcal{V}_{i=1,2}}$ denote the size of $\mathcal{V}_{i=1,2}$, respectively.
For large $t_0$, $t_0 \gg \mathcal{O}(L)$, the $2$d holographic CFTs Hamiltonian
evolves the system to the Page state,
so that for all $\mathcal{V}_{i=1,2}$ where $\sum_{i=1,2}l_{\mathcal{V}_i}<L$, $I_{\mathcal{V}_1, \mathcal{V}_2}$ should be completely destroyed.
Subsequently, we evolve it with $H_{\text{SSD}}$.
In the large-$t_1$ regime, $S_{\text{con}}$ should be larger than $S_{\text{dis}}$.
For simplicity 
let us assume that $A$ and $B$ include $x=X^1_f$.
In this case, 
$w^{\text{New}}_{x,\epsilon}$
and
$\bar{w}^{\text{New}}_{x,\epsilon}$
in this setup are obtained from 
those in Setup 1 by shifting by $i t_0$,
$w^{\text{New}}_{x,\epsilon}
\to it_0+w^{\text{New}}_{x,\epsilon}$
and 
$\overline{w}^{\text{New}}_{x,\epsilon}
\to it_0+\overline{w}^{\text{New}}_{x,\epsilon}$. 
For $S_{\text{dis}}$ and $S_B$, the shifts by $it_0$ are canceled, so that $S_{A\cup B}$ and $S_B$ 
in this setup is the same as those in Setup 1.
Since for small $t_1$, $S_{A\cup B}=S_A+S_B$, the early-time $I_{A,B}$ is zero. 
For large $t_1$,
excluding the $t_1$-regime when $X^{\text{New}}_{Y_1,\epsilon}-X^{\text{New}}_{Y_2,\epsilon} \ll \epsilon$, the $t_1$-dependence of $S_{A\cup B}$ should be given by
\be
S_{A\cup B} \approx\f{c \pi }{6\epsilon} \left[X^{\text{New}}_{Y_1,\epsilon}-X^{\text{New}}_{Y_2,\epsilon}+(X_1-X_2)\right].
\ee
The distance between 
$X^{\text{New}}_{Y_1,\epsilon}$ and $X^{\text{New}}_{Y_2,\epsilon}$ decreases with $t_1$, so that $I_{A,B}$ may grow with $t_1$ and saturate to (\ref{eq:IAB_late}).

\begin{figure}[tbp]
      \centering
      \includegraphics[keepaspectratio, scale=0.6]{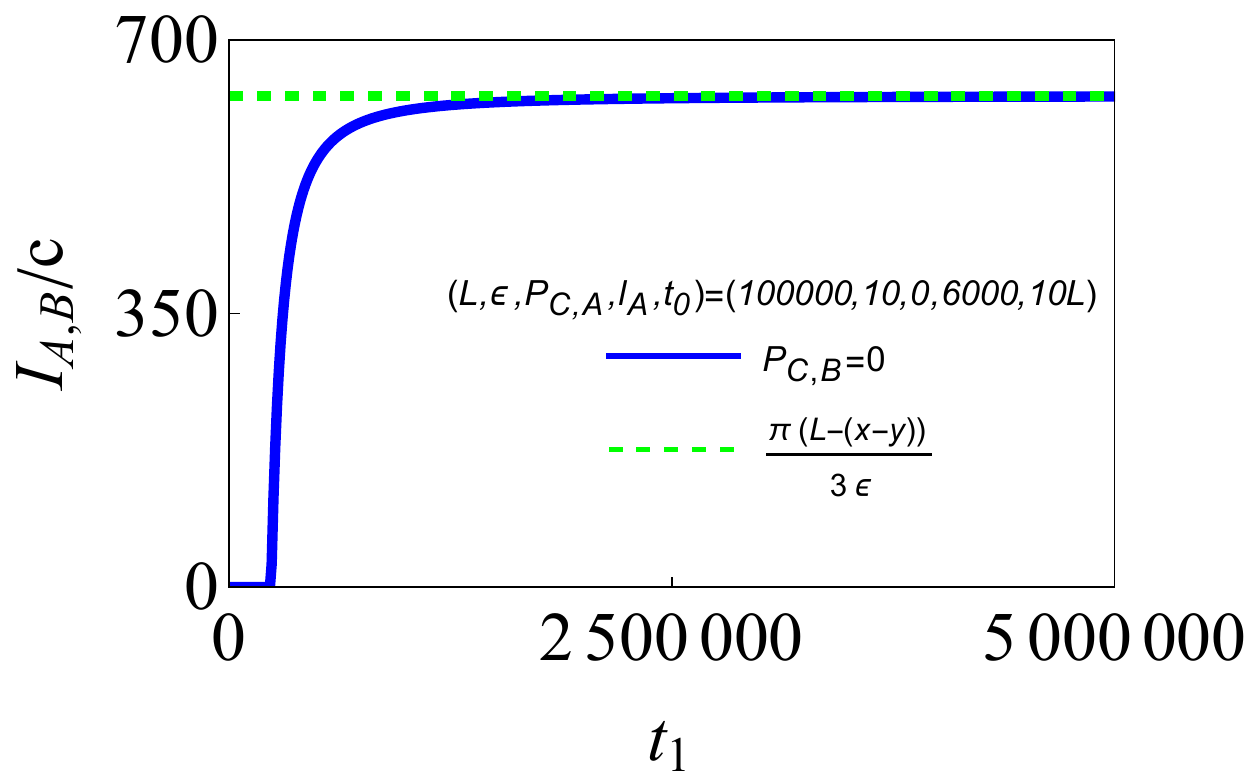}
      \hspace{0.7cm}
      \includegraphics[keepaspectratio, scale=0.14]{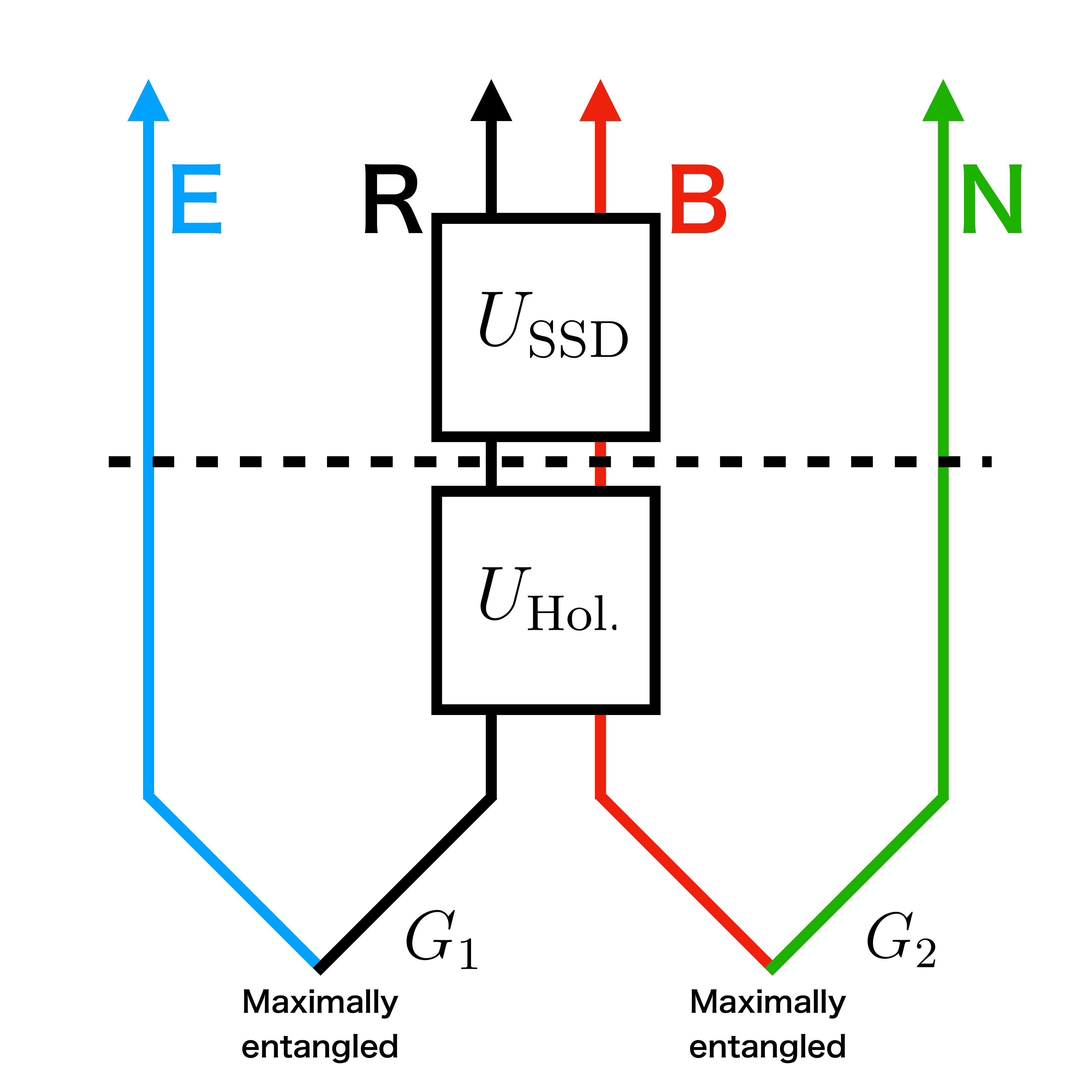}
  \caption{(Left)
    The time-dependence of $I_{A,B}$ in the SSD limit as a function of $t_1$. The solid line illustrates the $t_1$-dependence of $I_{A,B}$ for $t_0=10 L$. 
    The dashed line is the asymptotic value in (\ref{eq:IAB_late}).
    In this figure, $P_{C,J=A,B}$ and $l_{J=A,B}$ denote the centers and sizes
    of $J=A$ and $B$, respectively.
    (Right) Quantum circuit description of Setup 2. 
  }
  \label{Fig:MI_system2}
\end{figure}

In fact, as in Fig.\ \ref{Fig:MI_system2},
for $P_{C,A}=P_{C,B}=0$ and $l_A=l_B$,
even in 
the large $t_0$-regime, the $I_{A,B}$ grows with $t_1$, and then saturates at the value in (\ref{eq:IAB_late}).
One possible interpretation for the $t_1$-dependence of $I_{A,B}$ in this figure is that the SSD evolution
may recover the non-local correlation between $A$ and the subsystem including $x=X^{1}_f$ even 
when the system is in the typical state.
The SSD time-evolution is able to recover the mutual information from the typical state. 

The above recovery of quantum information is analogous to the one discussed in
quantum circuit models of quantum information scrambling and black holes,
e.g., 
the Hayden-Preskill thought experiment
\cite{Hayden_2007,2017arXiv171003363Y}
where 
the authors considered the retrieval of the quantum information from a
black hole.
To make a comparison, we can describe Setup 2 in the quantum circuit language
as in Fig.\ \ref{Fig:MI_system2}.
Here, in the parameter region considered in this paper (see \eqref{eq:parameter}),
the TFD state may be approximated by the product of Bell states,
$\ket{\text{TFD}}\approx \prod_{x=0}^{L}\ket{\text{Bell};x}$,
where $\ket{\text{Bell},x}$ denotes the Bell state at the spatial location $x$.
For example, if the dimension of the local Hilbert space at $x$ is $d$, then the definition of a single Bell state is given by
$\ket{\text{Bell};x}=\f{1}{\sqrt{d}}\sum_{q=1}^{d}\ket{q}_1\otimes\ket{q}_2$.
Let us divide these Bell pairs into two groups, $G_1$ and $G_2$.
Let $R$ and $E$ denote the sub-regions associated with $G_1$ of $\mathcal{H}_1$
and $\mathcal{H}_2$, respectively, while let $B$ and $N$ denote the sub-regions associated with $G_2$ of $\mathcal{H}_1$ and $\mathcal{H}_2$.
In Fig.\ \ref{Fig:MI_system2}, $U_{\text{SSD}}$ and $U_{\text{Hol.}}$ denote the time evolution induced by the holographic uniform Hamiltonian and $H_{\text{SSD}}$, respectively.
The process under the dashed line is the same as the one considered in the
Hayden-Preskill thought experiment.
If we interpret $U_{\text{SSD}}$ as a unitary decoder, the location where this decoder acts is different from that discussed in the Hayden-Preskill thought experiment.
Therefore, it would be interesting to consider the information retrieval in the system where the $U_{\text{SSD}}$ acts on $E$ and $R$. This is left for future work.

%
%

\section{Setup 3\label{Section:thesystem3}}

In this section, we study the entanglement dynamics of the state \eqref{eq:stata-in-system3}.
Here, the first part of the two-step time evolution
(with the SSD Hamiltonian on $\mathcal{H}_1$, $H^1_{{\rm SSD}}$)
can be interpreted as preparing a pair of B.H.-like excitations.
The created B.H.-like excitations are then subject to the second step of the time evolution under $H^1_0$
(Fig.\ \ref{Fig:emergence-and-propagation-of-B.H.-like-excitations}).
As we will show below, the propagating B.H.-like excitations
lead to periodic behaviors of entanglement quantities. 
Furthermore, in this setup, the system acquires genuine tripartite entanglement
due to the strong scrambling effect of the dynamics.
On the contrary, in the $2$d free fermion CFT, the B.H.-like excitations are just the clusters of 
quasi-particle, and no such tripartite entanglement arises.

\subsection{Entanglement entropy \label{Section:EE}}

\begin{figure}[t]
    \begin{tabular}{ccc}
      \begin{minipage}[t]{0.33\hsize}
        \centering
        \includegraphics[keepaspectratio, scale=0.26]{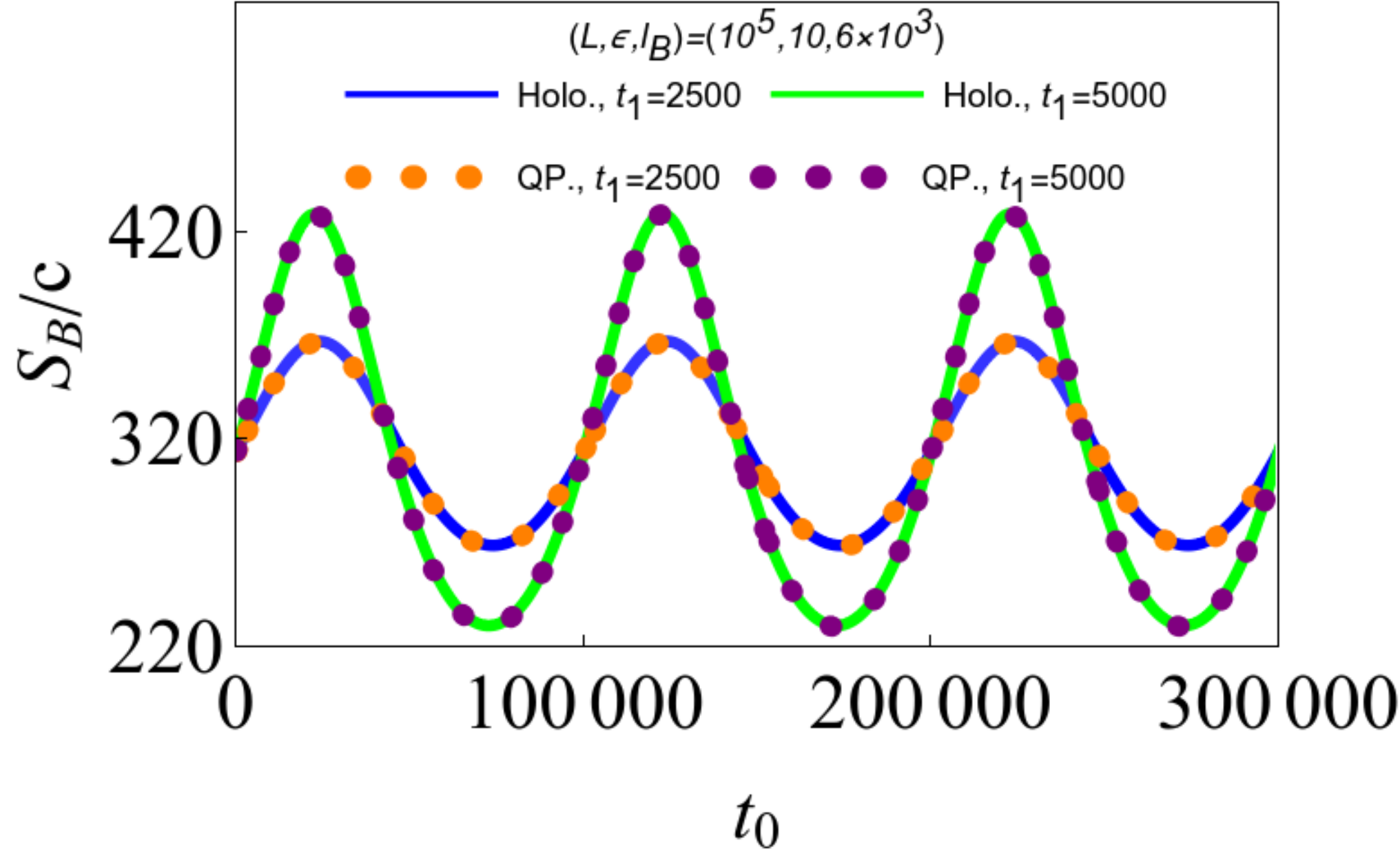}
    $(a_1)$ For the small $t_1$-regime. 
      \end{minipage} & 
     
     \begin{minipage}[t]{0.33\hsize}
        \centering
        \includegraphics[keepaspectratio, scale=0.24]{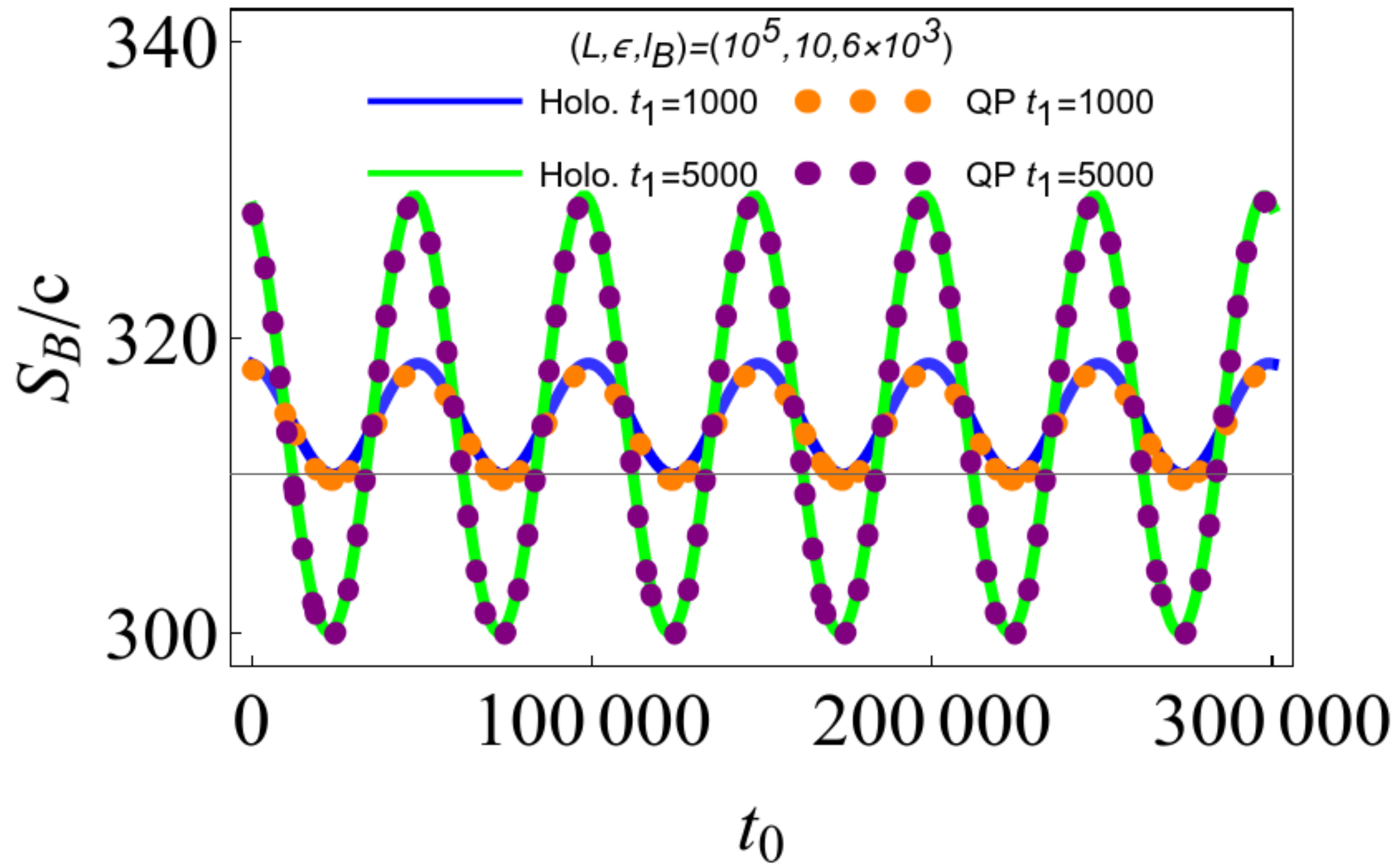}
     $(b_1)$ For the small $t_1$-regime. 
      \end{minipage} 
      
      \begin{minipage}[t]{0.33\hsize}
        \centering
        \includegraphics[keepaspectratio, scale=0.33]{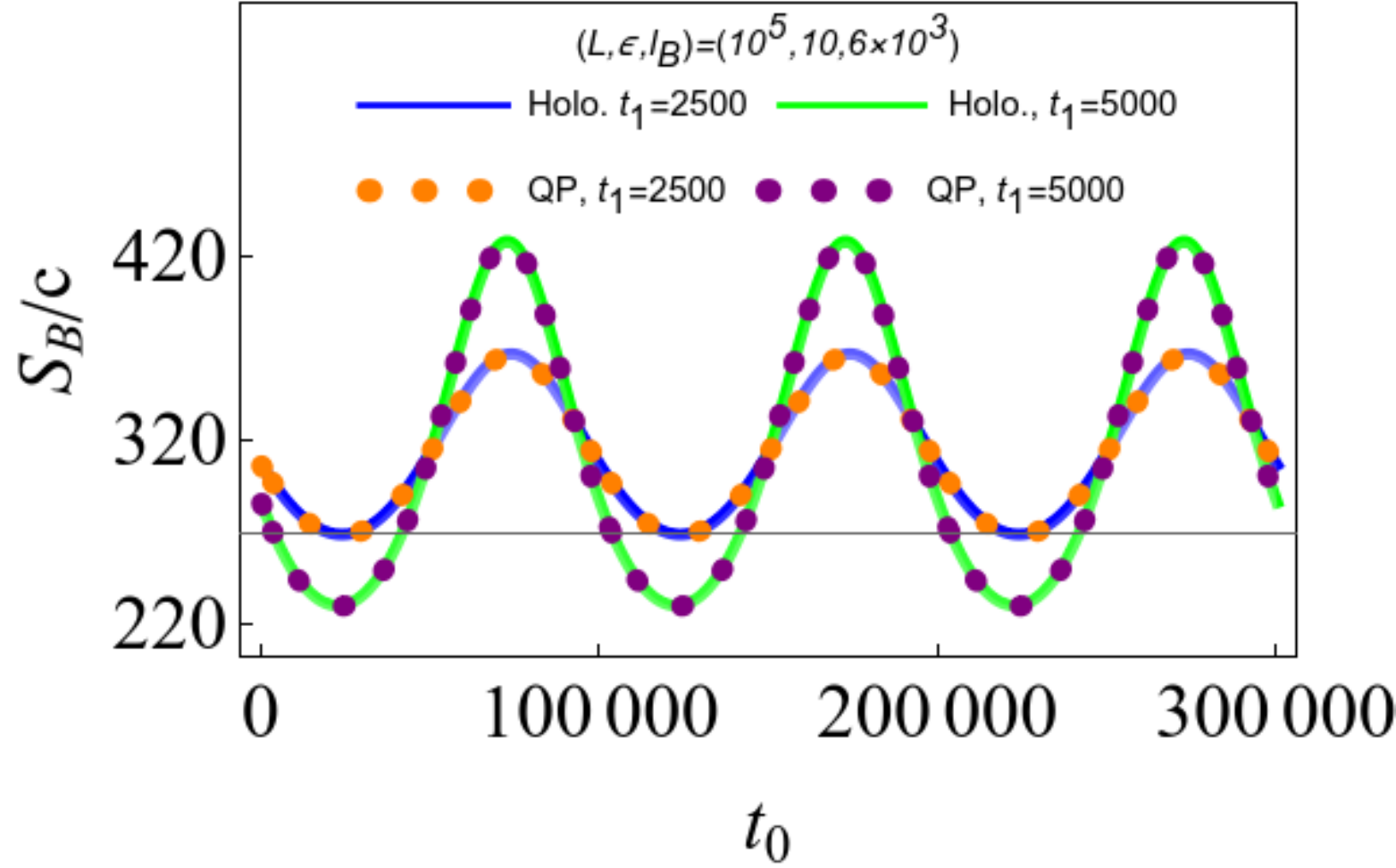}
      $(c_1)$ For the small $t_1$-regime. 
      \end{minipage}\\
      \begin{minipage}[t]{0.33\hsize}
        \centering
        \includegraphics[keepaspectratio, scale=0.27]{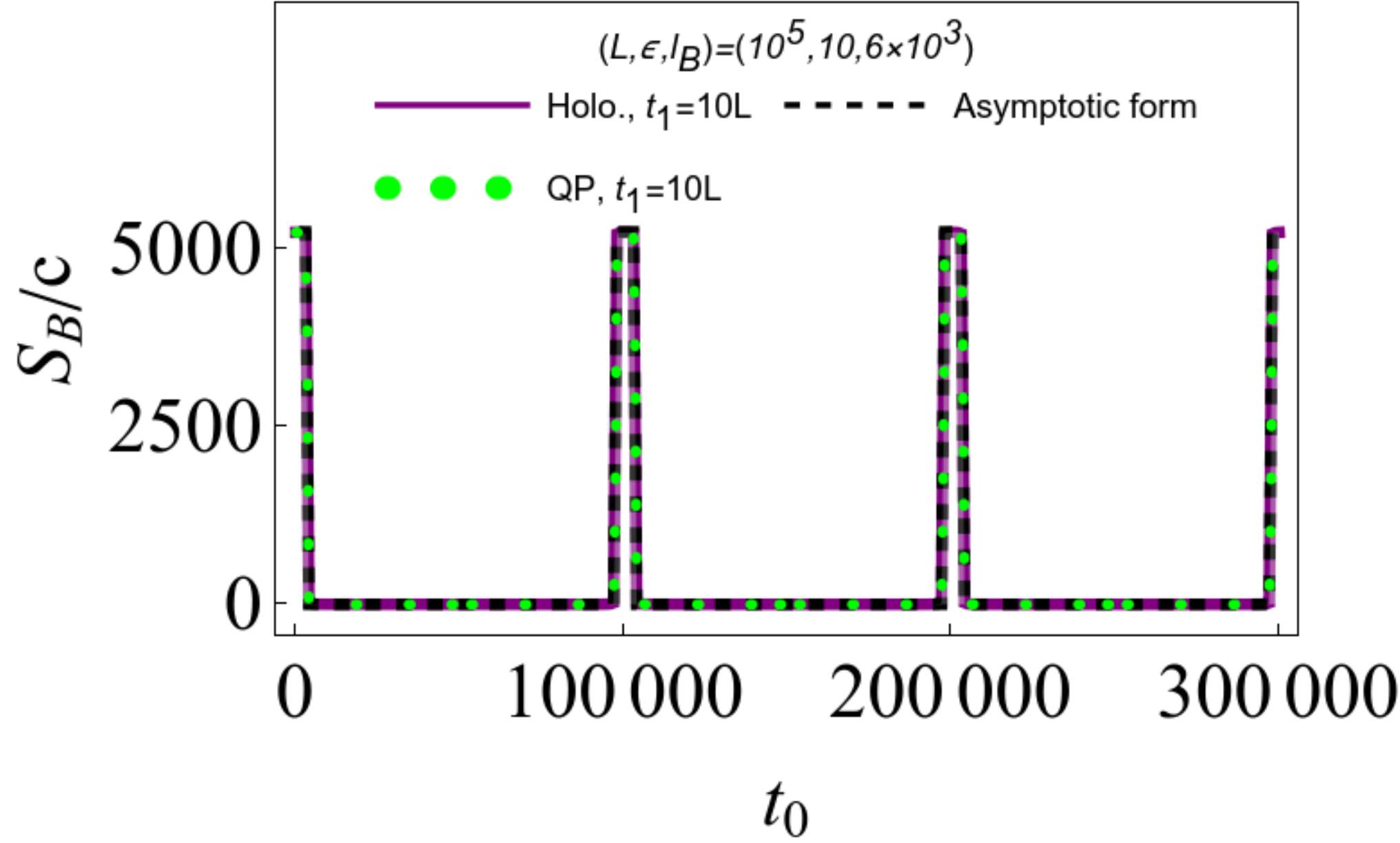}
    $(a_2)$ For the large $t_1$-regime. 
      \end{minipage} & 
     
     \begin{minipage}[t]{0.33\hsize}
        \centering
        \includegraphics[keepaspectratio, scale=0.26]{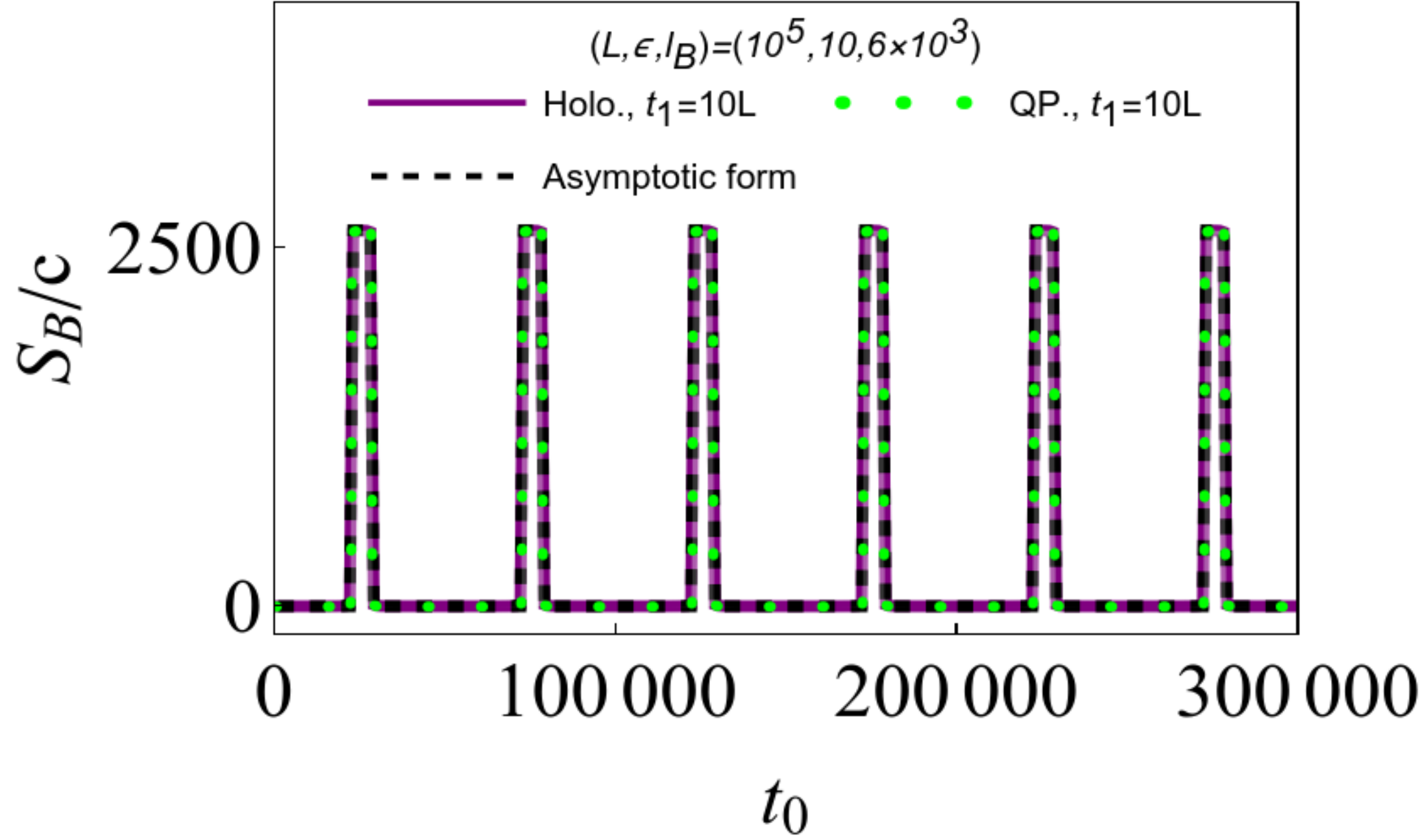}
     $(b_2)$ For the large $t_1$-regime. 
      \end{minipage} 
      
      \begin{minipage}[t]{0.33\hsize}
        \centering
        \includegraphics[keepaspectratio, scale=0.36]{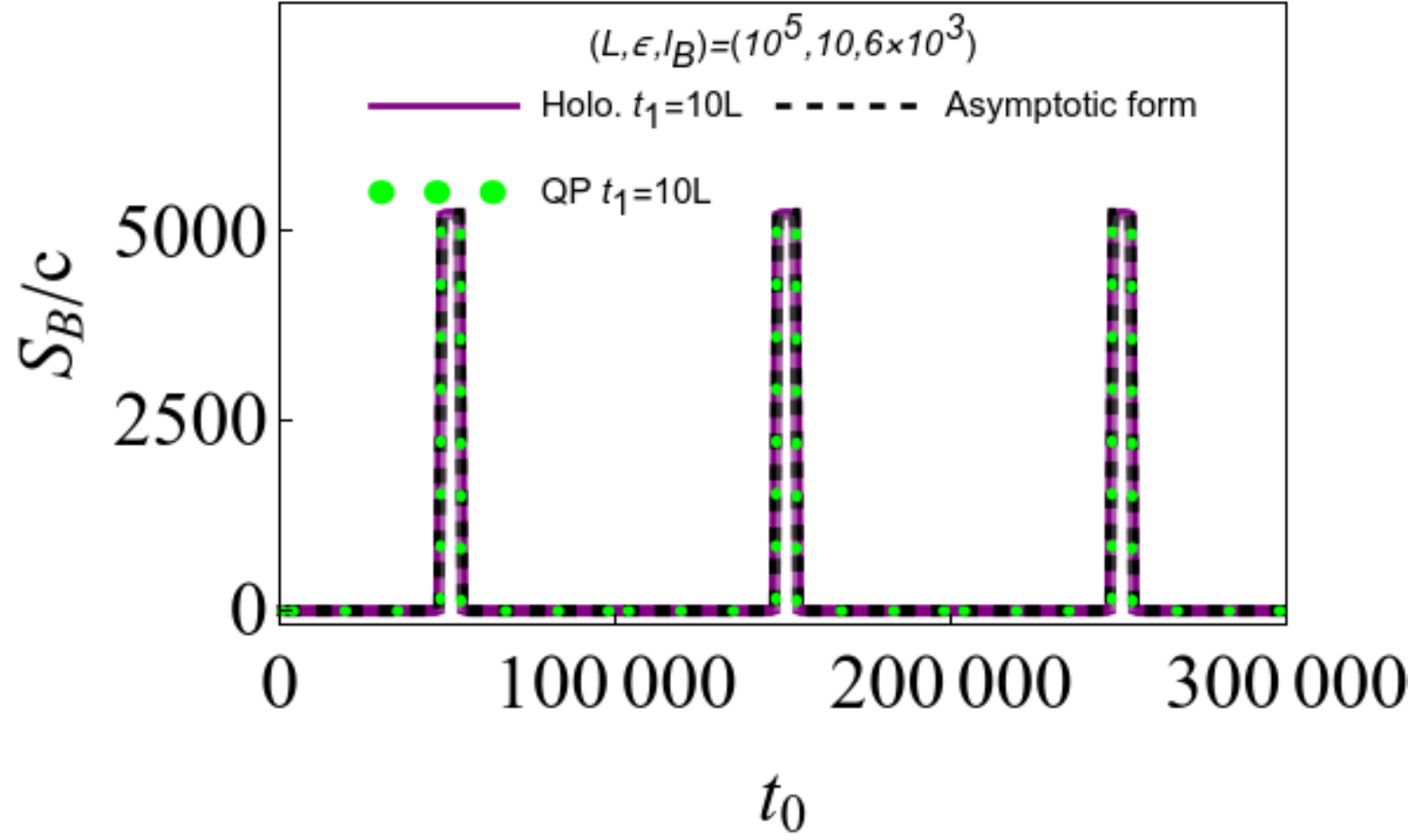}
      $(c_2)$ For the large $t_1$-regime. 
      \end{minipage}\\
    \end{tabular}
    \caption{
    The entanglement entropy $S_B$ 
    as a function of $t_0$
    for various choices of $t_1$ 
    for three configurations
     (a) $x=X^{1}_f \in B$,
     (b) $\f{L}{2}>Y_1>Y_2>0$,
     and (c) $x=X^{2}_f \in B$.
    The subscript in $(a_{i=1,2})$ distinguishes 
    the small and large $t_1$ regimes (top and bottom rows, respectively).
    The dashed line illustrates the asymptotic behavior of $S_B$ in (\ref{eq:Asym-beh-SB3}) in the large $t_1$ limit.}
    \label{Fig:t0-dependence-SB-in-3}
  \end{figure}

Let us first study $S_B$. In particular, we present the $t_0$-dependence of $S_B$ in three cases:
(a) $x=X^{1}_f \in B$; (b) $\f{L}{2}>Y_1>Y_2>0$; (c) $x=X^{2}_f \in B$.
In Fig.\ \ref{Fig:t0-dependence-SB-in-3}, we plot $S_B$ for various $t_1$ as a function of $t_0$.
The $t_0$-dependence of $S_B$ is periodic with period $L$. 
This periodic behavior follows from the evolution of twist and anti-twist operators reported in Appendix\ \ref{App:Evo_o_Ope_ana}.
The larger $t_1$ is, the larger the amplitude of the oscillation of $S_B$ is, 
and the system deviates further from the typical state.
The time-dependence of $S_B$ for the single interval can be understood by
the quasiparticle picture (we provide the details in Appendix \ref{App:free_fermion}).
For $t_1 \gg 1$, the $t_0$-dependence of $S_B$ is approximated by 
\begin{equation}
\begin{split}\label{eq:Asym-beh-SB3}
&\text{For~(a),}~~S_B\approx\begin{cases}
\f{c\pi L}{6\epsilon} & nL+Y_2>t_0>nL-Y_2\\
\f{c\pi L}{12\epsilon}& (n+1)L-Y_1>t_0>nL+Y_2\\
\f{c}{3}\log{\left[\f{L}{\pi}\sin{\left[\f{\pi(Y_1-Y_2)}{L}\right]}\right]}& nL+Y_1>t_0>(n+1)L-Y_1\\
\f{c\pi L}{12\epsilon}& (n+1)L-Y_2>t_0>nL+Y_1 \\
\end{cases},\\
   &\text{For~(b),}~~ S_B\approx\begin{cases}
\f{c}{3}\log{\left[\f{L}{\pi}\sin{\left[\f{\pi(Y_1-Y_2)}{L}\right]}\right]} & nL+Y_2>t_0>nL-Y_2\\
\f{c\pi L}{12\epsilon} & nL+Y_1>t_0>nL+Y_2\\
\f{c}{3}\log{\left[\f{L}{\pi}\sin{\left[\f{\pi(Y_1-Y_2)}{L}\right]}\right]} & (n+1)L-Y_1>t_0>nL+Y_1\\
\f{c\pi L}{12\epsilon} & (n+1)L-Y_2>t_0>(n+1)L-Y_1 \\
\end{cases},\\
&\text{For~(c),}~~S_B\approx\begin{cases}
\f{c}{3}\log{\left[\f{L}{\pi}\sin{\left[\f{\pi(Y_1-Y_2)}{L}\right]}\right]} & nL+Y_2>t_0>nL-Y_2\\
\f{c\pi L}{12\epsilon} & (n+1)L-Y_1>t_0>nL+Y_2\\
\f{c\pi L}{6\epsilon}& nL+Y_1>t_0>(n+1)L-Y_1\\
\f{c\pi L}{12\epsilon} & (n+1)L-Y_2>t_0>nL+Y_1 \\
\end{cases},\\
\end{split}
\end{equation}
where $n$ is an integer. 

\begin{figure}[tbp]
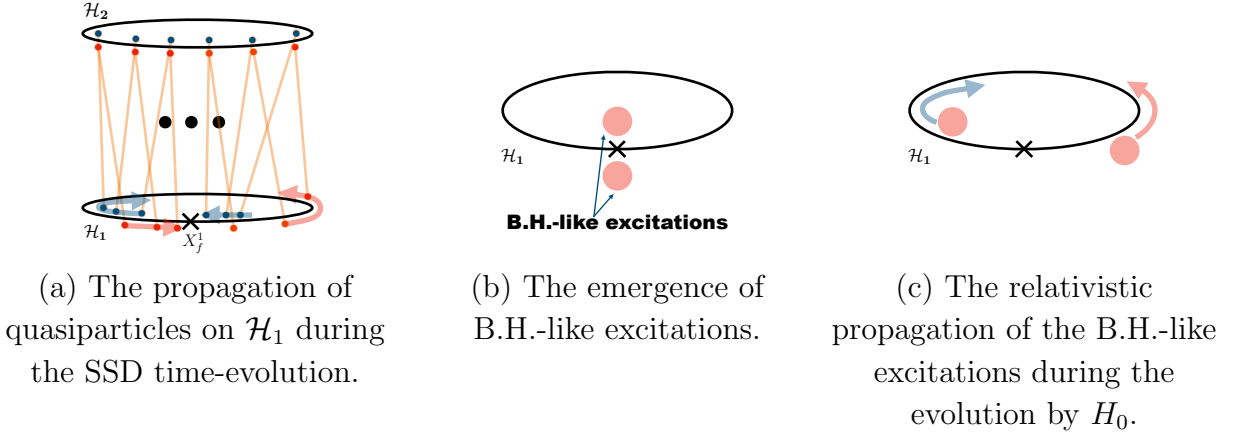

  \begin{tabular}{ccc}
    \begin{minipage}[t]{0.31\hsize}
      \centering
      \includegraphics[keepaspectratio, scale=0.025]{Figure/Section5/Figure_propagation_of_Bells.pdf}
      \\ 
      (a) The propagation of quasiparticles on $\mathcal{H}_1$ during the SSD time-evolution.
    \end{minipage} & 
                     
                     \begin{minipage}[t]{0.31\hsize}
                       \centering
                       \includegraphics[keepaspectratio, scale=0.025]{Figure/Section5/Figure_Emergence_of_BH-like-excitation.pdf}
                       \\ 
                       (b) The emergence of B.H.-like excitations.
                     \end{minipage} 
                     
                     \begin{minipage}[t]{0.32\hsize}
                       \centering
                       \includegraphics[keepaspectratio, scale=0.025]{Figure/Section5/Figure_Propagation_of_BH-like-excitation.pdf}
                       \\
                       (c) The relativistic propagation of the B.H.-like excitations during the evolution by $H_0$.
                     \end{minipage}
  \end{tabular}
  \caption{The emergence and the time-evolution of the two B.H.-like excitations. }
  \label{Fig:emergence-and-propagation-of-B.H.-like-excitations}
\end{figure}

The periodic behavior (\ref{eq:Asym-beh-SB3}) can be understood
from the relativistic propagation of the two local objects that have a huge amount of information,
i.e., 
B.H.-like excitations. 
Here, we introduce an effective model that describes the time evolution of $S_B$
induced by $H_0$ for the large $t_1$-regime.
This model describes the leading behavior of $S_B$ in the coarse-grained regime.
At $t_1=t_0=0$, in the coarse-grained regime, the leading behavior of the TFD
state (\ref{eq:def_of_TFD})
may be approximated by the state consisting of the product of Bell pairs,
\be \label{eq:TFD-Bell-pairs}
\ket{\text{TFD}} \approx \prod_{\tilde{x}}\ket{\text{Bell};\tilde{x}}_L \ket{\text{Bell};\tilde{x}}_R, 
\ee
where $\tilde{x}$ is defined as $\tilde{x}\equiv \f{x}{\epsilon}$, and
$\ket{\text{Bell};\tilde{x}}_{L,R}$
denote the Bell pairs consisting of the two quasiparticles at $\tilde{x}$ in $\mathcal{H}_1$ and $\mathcal{H}_2$ respectively as in 
Fig.\ \ref{Fig:emergence-and-propagation-of-B.H.-like-excitations}.
During the unitary time-evolution by $H^1_0$ or $H^1_{\text{SSD}}$,
the quasiparticles on $\mathcal{H}_1$ of $\ket{\text{Bell};\tilde{x}}_L$ and $\ket{\text{Bell};\tilde{x}}_R$
correspond to the left- and right-moving particles respectively. 
These particles move to the left and right at the speed determined by $H^1$ or $H^1_{\text{SSD}}$.
During the evolution by $H^1_{\text{SSD}}$, all particles on $\mathcal{H}_1$ move to $x=X^1_f$ and accumulate around $x=X^1_f$ \cite{Goto:2021sqx}.
For large $t_1$, two B.H.-like excitations emerge around $x= X^1_f$.
After these excitations emerge, the entanglement entropy for the subsystem including $x= X^1_f$ is approximated by $\f{c \pi L}{6\epsilon}$.
Subsequently, we evolve this system with $H_{0}$, and then one of the B.H.-like excitation moves to the left
and the other moves to the right at the speed of light.
In the time regime where one of these excitations is in $B$, $S_B$ is
approximated by $\f{c \pi L}{12\epsilon}$,
while in the time-regime where both are in $B$, $S_B$ is approximated by $\f{c \pi L}{6\epsilon}$.
In the time regime where no excitations are in $B$, $S_B$ is approximated by
$\f{c}{3}\log{\left[\f{L}{\pi}\sin{\left[\f{\pi(Y_1-Y_2)}{L}\right]}\right]}$.


\subsection{Bipartite and tripartite mutual information, and genuine tripartite entanglement \label{Sec:BandTOMI}}

In the previous sections, we have developed an effective picture in terms of B.H.-like excitations
to describe the time evolution of entanglement entropy for a single interval.
We note that the above behavior is universal for any CFT.
We now generalize it to the time evolution of BMI and TMI.
Here, the distinction between integrable (e.g., the free fermion theory) and chaotic theories (holographic theories) becomes important. In the free fermion theory, the B.H.-like excitations are just clusters of quasiparticles. On the other hand, this is not the case in holographic CFTs and the interior of B.H.-like excitations should have a strong scrambling effect.

\begin{figure}[tbp]
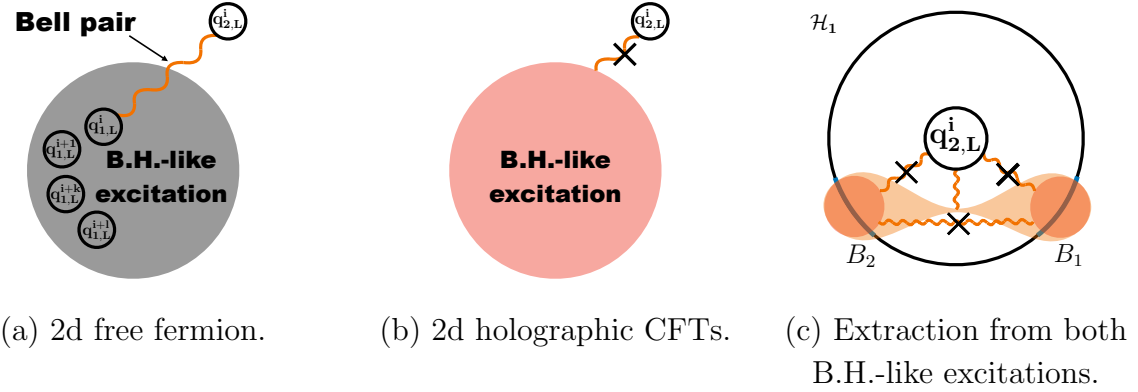

    \begin{tabular}{ccc}
      \begin{minipage}[t]{0.31\hsize}
        \centering
        \includegraphics[keepaspectratio, scale=0.03]{Figure/Section5/Figure_extraction_single_free.pdf}
        \\
   (a) $2$d free fermion.
      \end{minipage} & 
     
     \begin{minipage}[t]{0.31\hsize}
        \centering
        \includegraphics[keepaspectratio, scale=0.03]{Figure/Section5/Figure_extraction_single_hole.pdf}
        \\
    (b) $2$d holographic CFTs.
      \end{minipage} 
      
      \begin{minipage}[t]{0.31\hsize}
        \centering
        \includegraphics[keepaspectratio, scale=0.03]{Figure/Section5/Figure_extraction_double.pdf}
        \\
    (c) Extraction from both B.H.-like excitations.
      \end{minipage}
    \end{tabular}
    \caption{
    Retrieval of a single Bell pair from B.H.-like excitations
    in the 2d free fermion theory (a)
    and 2d holographic CFTs (b).
    The panel (c) illustrates 
    the retrieval of a single Bell pair from both B.H.-like excitations in $2$d holographic CFTs. 
    The subsystems, $B_1$ and $B_2$, are the symmetric intervals in (\ref{B1B2}).
    The red-shadowed region is the region where 
    quantum information is
    scrambled. The orange-shadowed region illustrates $q^i_{1,L}$ is shared by two B.H.-like excitations.}
    \label{Fig:emergence-and-propagation-of-B.H.-like-excitations2}
  \end{figure}
 
\subsubsection{Bipartite mutual information\label{sec:physical-interpretation}}

Let us first consider BMI $I_{A,B}$ for the time-evolved state \eqref{eq:stata-in-system3}.
Here $A$ and $B$ are the sub-regions (single intervals) of ${\cal H}_1$ and ${\cal H}_2$.
BMI $I_{A,B}$ can be thought of as measuring the number of Bell pairs shared by $A$ and $B$. 
Let us consider extracting a Bell pair from the B.H.-like excitations. 
At $t_0 = t_1 = 0$, the system is in the TFD state approximated by (\ref{eq:TFD-Bell-pairs}).
Consider a single Bell pair shared by quasiparticles, $q^i_{1,D}$ and $q^i_{2,D}$.
Here, $q^i_{j=1,2,D=L,R}$ 
denote a quasiparticle on $i$-th site of $\mathcal{H}_j$, and $D=L/R$ refers left/right-moving quasiparticles.
When evolved with $H^1_{\text{SSD}}$, 
the B.H.-like excitations emerge around $x=X^1_f$.
Then, we attempt to extract from 
the B.H.-like excitations the Bell pair that $q^i_{1,D}$ and $q^i_{2,D}$ initially share.
In the free fermion theory, if a single B.H.-like excitation includes $q^i_{1,L}$, then we can extract the Bell pair shared by $q^i_{1,L}$ and $q^i_{2,L}$ from this excitation [Fig.\ \ref{Fig:emergence-and-propagation-of-B.H.-like-excitations2} (a)]. 
The situation is crucially different for $2$d holographic CFTs;
we cannot extract this Bell pair from only the single B.H.-like excitation
[Fig.\ \ref{Fig:emergence-and-propagation-of-B.H.-like-excitations} (b)].
This is because quasiparticles in $\mathcal{H}_1$ are locally hidden in the two B.H.-like excitations by the scrambling effect.
In fact, as we will show momentarily,
when only a single B.H.-like excitation is in $B$, $I_{A,B}$ is zero
On the other hand, 
when both of the B.H.-like excitations are in $B$,
$A$ and $B$ share the Bell pairs initially shared by $A$ and $\mathcal{H}_1$.

\begin{figure}[tbp]
    \begin{tabular}{ccc}
      \begin{minipage}[t]{0.33\hsize}
        \centering
        \includegraphics[keepaspectratio, scale=0.4]{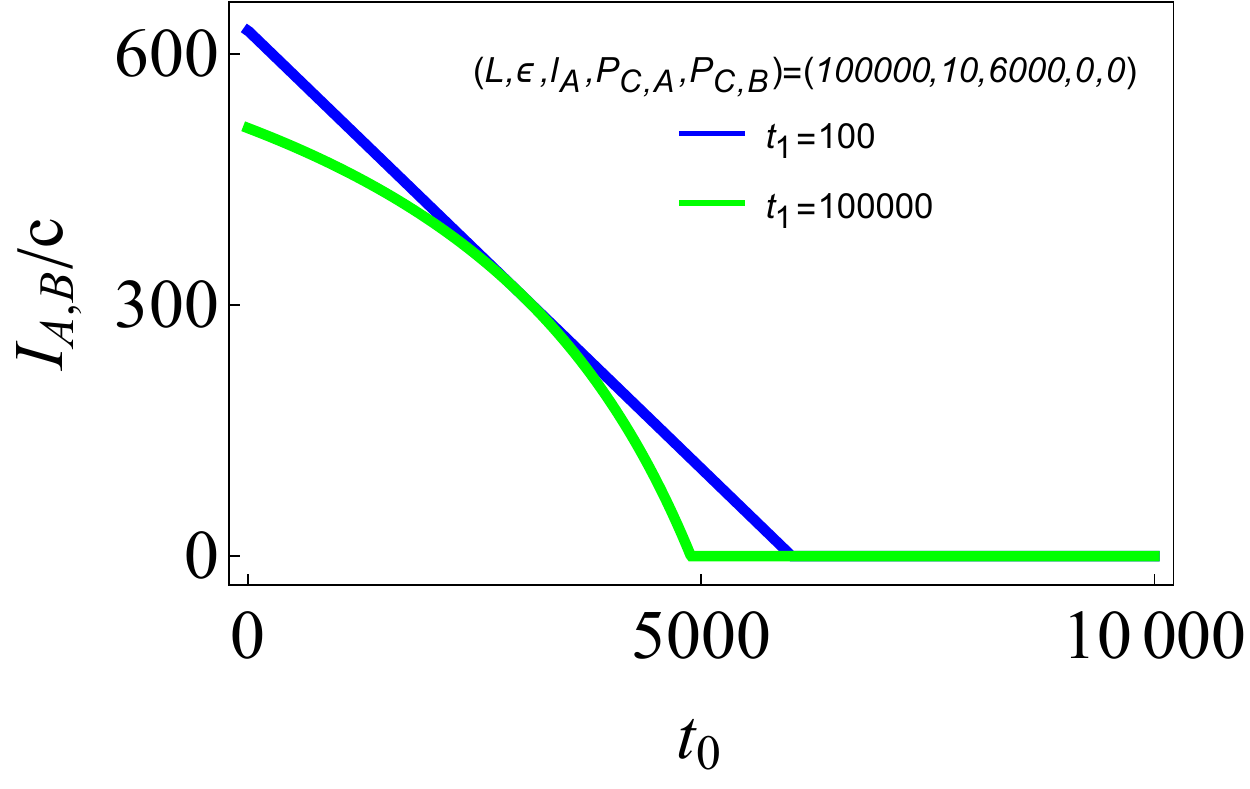}
   (1) $I_{A,B}$ with small $t_1$ in (a).
      \end{minipage} & 
     
     \begin{minipage}[t]{0.33\hsize}
        \centering
        \includegraphics[keepaspectratio, scale=0.4]{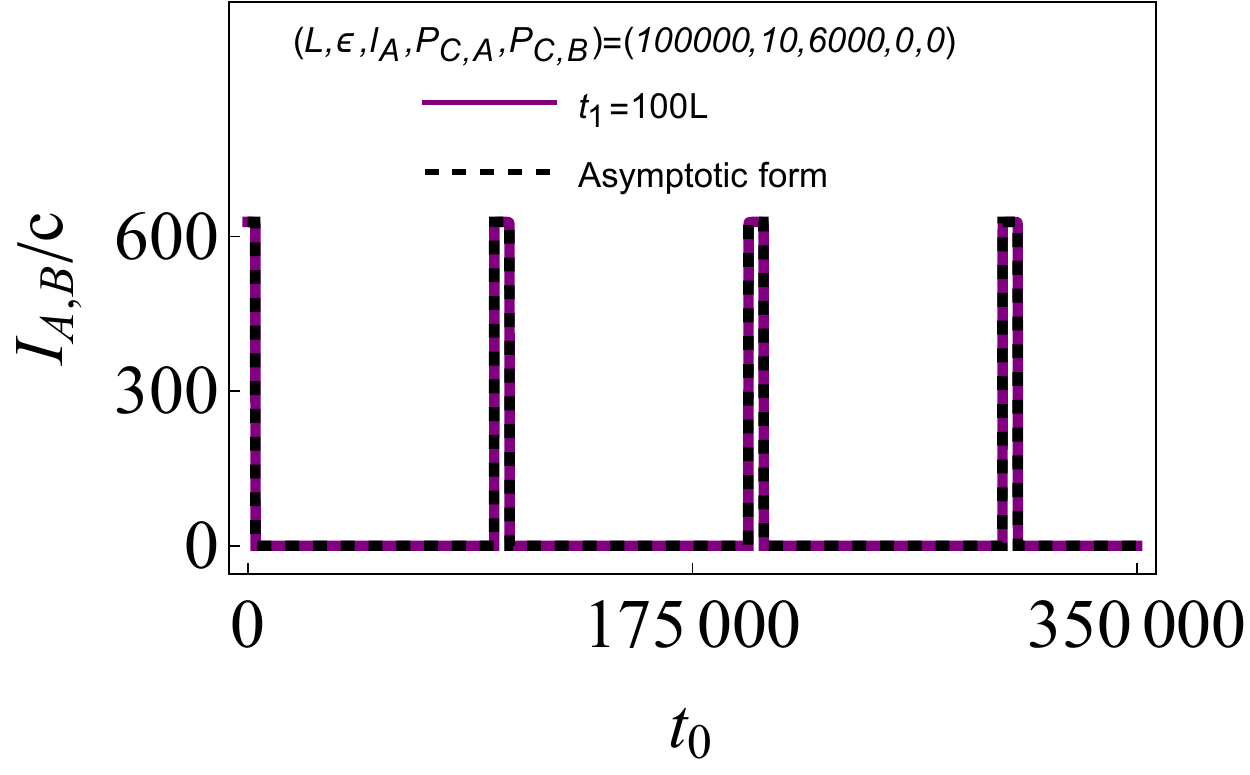}
    (2) $I_{A,B}$ with large $t_1$ in (a).
      \end{minipage} 
      
      \begin{minipage}[t]{0.33\hsize}
        \centering
        \includegraphics[keepaspectratio, scale=0.4]{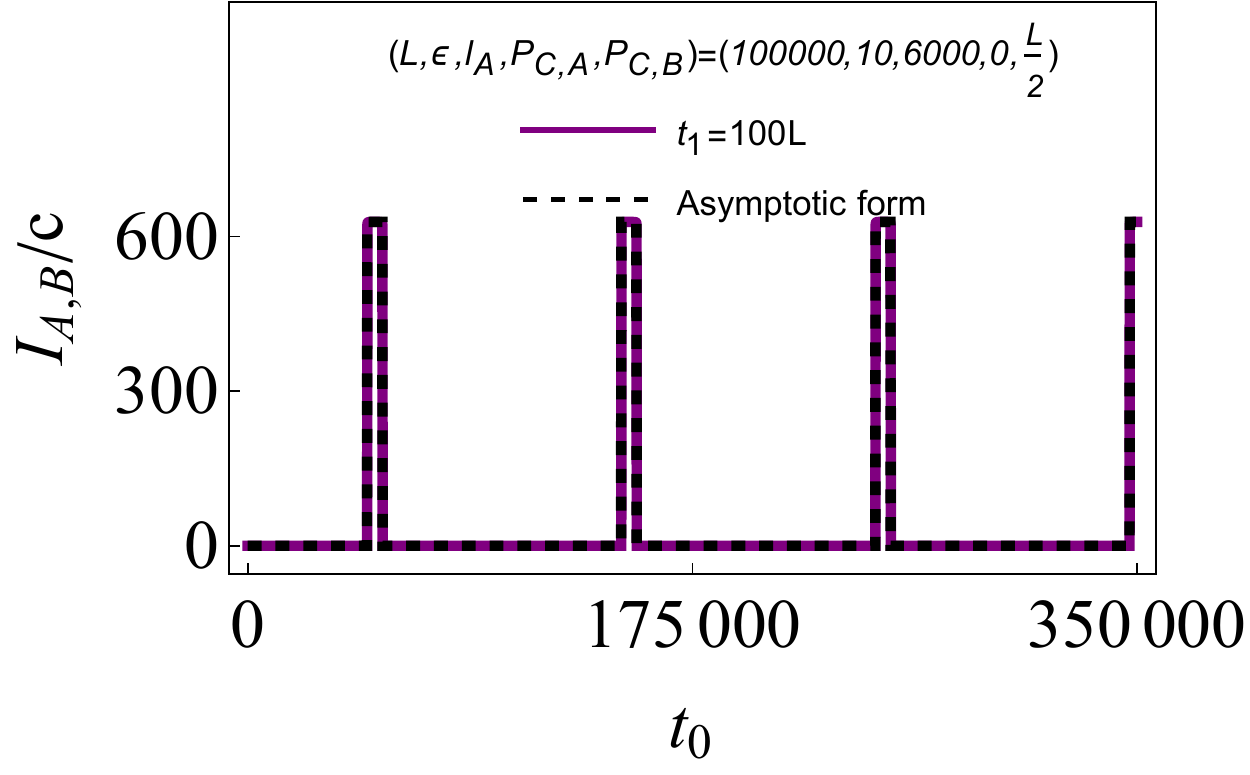}
    (3) $I_{A,B}$ with large $t_1$ in (c).
      \end{minipage}
    \end{tabular}
    \caption{
    The BMI $I_{A,B}$ of  \eqref{eq:stata-in-system3} as a function of $t_0$ for various choices of $t_1$.  For simplicity, we take $l_A$ to be the same as $l_B$.
In (2) and (3), the dashed lines are 
the asymptotic behavior in (\ref{eq:aym-IAB-single-in3}).}
    \label{Fig:time-dependence-of-IAB-single-in3}
  \end{figure}

In Fig.\ \ref{Fig:time-dependence-of-IAB-single-in3}, we plot $I_{A,B}$
as a function of $t_0$ for
various choices of $t_1$, $A$ and $B$.
Here, for the configurations of $A$ and $B$,  
we consider the following three cases:
(a) $x=X^1_f \in B$; (b) $\f{L}{2}>Y_1>Y_2>0$; (c) $x=X^{2}_f \in B$.
For (b), we assume that $A$ and $B$ are the disjoint intervals for simplicity.
Then, $I_{A,B}$ is approximately zero.
For (a) and (c), for large $t_1$, 
$I_{A,B}$ is approximated by the following periodic function of $t_0$ 
with period $L$:
\be \label{eq:aym-IAB-single-in3}
\begin{split}
    \text{For~(a)}~~&~~ I_{A,B}\approx\begin{cases}
\f{c\pi l_A}{3\epsilon} & \text{for}~~(n+1)L-Y_1>t_0>nL-Y_2\\
0 &\text{for}~~(n+1)L-Y_2>t_0>(n+1)L-Y_1,\\
\end{cases},\\
\text{For~(c)}~~&~~ I_{A,B}\approx\begin{cases}
0 & \text{for}~~\left(n+1\right)L-Y_1>t_0>nL-Y_2\\
\f{c\pi l_A}{3\epsilon} &\text{for}~~nL+Y_1>t_0>\left(n+1\right)L-Y_1\\
0 & \text{for}~~(n+1)L-Y_2>t_0>nL+Y_1\\
\end{cases}.
\end{split}
\ee
The dashed lines in Fig.\ \ref{Fig:time-dependence-of-IAB-single-in3} illustrate these asymptotic behaviors.
In these cases, there are the $t_0$-regimes where both B.H.-like excitations are
in $B$, and in these regimes,
$I_{A,B}$ is approximated by $\f{c\pi l_A}{3\epsilon}$, while in (b), there are no such $t_0$-regimes.

\subsubsection{Tripartite mutual information}

\begin{figure}[tbp]
  \begin{tabular}{cc}
    \begin{minipage}[t]{1\hsize}
      \centering
      \includegraphics[keepaspectratio, scale=0.6]{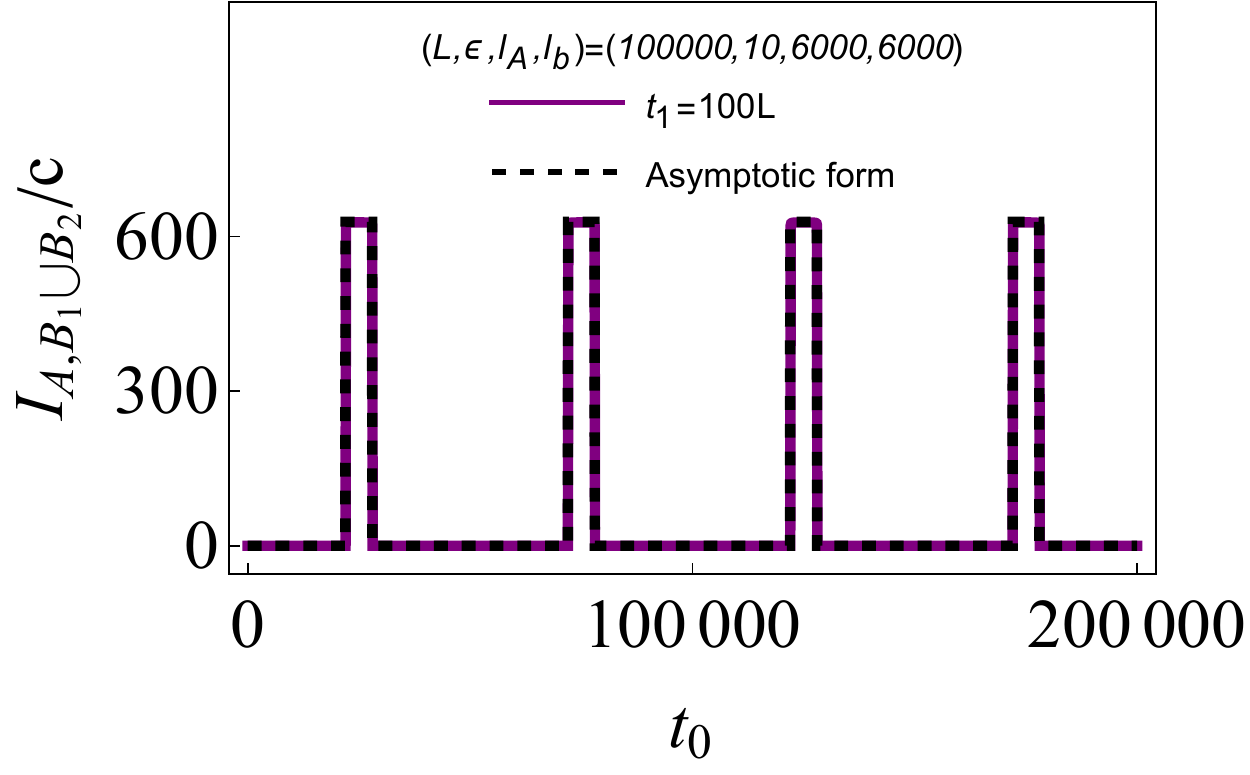}
      
    \end{minipage} &
                     
  \end{tabular}
  \caption{The BMI $I_{A,B_1 \cup B_2}$ in the large $t_1$ regime as a function of $t_0$. The solid line illustrates the $t_1$-dependence of $I_{A,B_1\cup B_2}$ for  $t_1=10^2L$. 
    In this figure, $l_{B_1}=l_{B_2}=l_{b}$, $P_{C,B_1}=\f{3L}{4}$, and $P_{C,B_2}=\f{L}{4}$. 
    The center of $A$ is $x=X^f_1$.
    The dashed line illustrates the $t_0$-dependence of $I_{A,B_1\cup B_2}$ in (\ref{IAB_u2_double}). \label{Fig:BOMI_system3_double_intervals}}
\end{figure}

In summary, the two B.H.-like excitations in $2$d holographic CFTs should be
(local) excitations,
each of which has half the entropy 
of a black hole,
and information inside this excitation should be scrambled. 
We cannot extract Bell pairs from a single B.H.-like excitation,
while we can extract them from both excitations
[Fig.\ \ref{Fig:emergence-and-propagation-of-B.H.-like-excitations2} (c)].
To further discuss this,
let us now consider the case where $B$ consists of two intervals, $B_1$ and $B_2$,
$B=B_1 \cup B_2$.
Specifically, we consider 
the case where 
$B$ is given by a union of symmetric double intervals,
\be \label{B1B2}
B_1=\left\{x\bigg{|}L>L-Y_2>x>L-Y_1>\f{L}{2}\right\},~~B_2=\left\{x\bigg{|}\f{L}{2}>Y_1>x>Y_2>0\right\},
\ee
where $\f{L}{2}>Y_1>Y_2>0$. 
In Fig.\ \ref{Fig:BOMI_system3_double_intervals}, we take $Y_1>Y_2>X_1>X_2>0$.
In this case, for small $t_1$, $I_{A,B=B_1\cup B_2}$ is practically zero,
while for large $t_1$ the $t_0$-dependence of $I_{A,B}$ is approximated by 
the following periodic function of $t_0$ with $L$:
\be \label{IAB_u2_double}
I_{A,B=B_1\cup B_2}\approx
\begin{cases}
0 & nL+Y_2>t_0>nL-Y_2\\
\f{c\pi l_A}{3\epsilon} & nL+Y_1>t_0>nL+Y_2\\
0 & (n+1)L-Y_1>t_0>nL+Y_1\\
\f{c\pi l_A}{3\epsilon}  & (n+1)L-Y_2>t_0>(n+1)L-Y_1 \\
\end{cases}.
\ee
In this case, there are $t_0$ regimes where both the B.H.-like excitations are in $B=B_1\cup B_2$.
In these $t_0$ regimes, the $I_{A,B_1\cup B_2}$ is approximated by $\f{c\pi l_A}{3\epsilon}$.
This suggests that we may be able to reconstruct $I_{A,B}$ from all quasiparticles in $A$ and $\mathcal {H}_1$ even under the $2$d holographic time evolution.
Also, we can see from the $t_0$-dependence of $I_{A,B}$ for large $t_1$ that the time evolution of $I_{A,B}$ may follow the relativistic propagation of the local excitations as in \cite{Nozaki:2014hna,Nozaki:2014uaa,He:2014mwa}.

By combining BMI for single intervals $A,B$,
and for a single interval $A$ and
double interval $B=B_1 \cup B_2$,
we can discuss TMI.
By considering local and global TMI defined below, 
let us show that the amount of information scrambled by the dynamics depends on the observers.
Define local TMI for (\ref{B1B2}) as 
\be \label{localTMI-system3}
I_{A,B_1,B_2}=I_{A,B_1}+I_{A,B_2}-I_{A,B_1\cup B_2},
\ee
where $A$ denotes the subsystem of $\mathcal{H}_2$.
(Here, let us the $t_0$ dependence of TMI in the large $t_1$ regime.)
In the large $t_1$ regime, $I_{A,B_{i=1,2}}$ is approximated by zero, while the $t_0$-dependence of $I_{A,B_1\cup B_2}$ is given by (\ref{IAB_u2_double}).
Thus, the local TMI, $I_{A,B_1,B_2}$ in the large $t_1$ regime is zero 
for $nL+Y_2>t_0>nL-Y_2$ or $(n+1)L-Y_1>t_0>nL+Y_1$, while it is approximated by $-\f{c\pi l_A}{3\epsilon}$ 
for $nL+Y_1>t_0>nL+Y_2$ or $(n+1)L-Y_2>t_0>(n+1)L-Y_1$.
The negativity of $I_{A, B_1, B_2}$ is the signature of scrambling.
On the other hand, define global TMI as 
\be \label{globalTMI-system3}
I_{A,B,\overline{B}}=I_{A,B}+I_{A,\overline{B}}-I_{A,B\cup \overline{B}},
\ee
where $\overline{B}$ is the complement to $B$ in $\mathcal{H}_1$.
This global TMI is a stationary constant value and zero.
One possible interpretation for the $t_0$-dependence of local and global TMI is that when two B.H.-like excitations are in $B$, the quasiparticles on $\mathcal{H}_1$ of the Bell pairs initially shared by $A$ and $\mathcal{H}_1$ may be locally-hidden in $B$, while there may be no quasiparticles locally-hidden in $\mathcal{H}_1$.

By contrast, the local and global TMI for free fermions is zero for the double interval setup in Fig. \ref{Fig:BOMI_system3_double_intervals} in agreement with the quasiparticle picture.


\subsubsection{Genuine tripartite entanglement\label{sec:tripartite-entanglement}}

Let us also note
the following behavior  
of BMI in the large $t_1$ regime.
For the symmetric double intervals in (\ref{B1B2}),
$I_{A,B_{i=1,2}}$ and $I_{B_1,B_2}$ are approximately zero.
On the other hand, the $t_0$-dependence of $I_{A,B_1\cup B_2}$ is given by (\ref{IAB_u2_double}).
One possible interpretation for these BMI is that the system in its steady state may have only tripartite entanglement, which we call genuine tripartite entanglement.
This genuine tripartite entanglement may be a characteristic property of the system in the steady state during the $2$d SSD holographic time evolution.
In contrast, in the $2$d free fermion theory,
there are the $t_0$-regimes where $I_{A,B_{i=1,2}}$ and $I_{B_1,B_2}$ becomes positive 
(see Appendix \ref{App:sum-of-results}).
The reduced density of the system in this steady state may be given by 
\be
\rho_{A,B_1,B_2} \approx \rho_{A}\otimes \rho_{B_1} \otimes \rho_{B_2}+\sigma_{A,B_1,B_2},
\ee
where 
$\sigma_{A,B_1,B_2}$ is Hermitian
and traceless, 
$\Tr_{A}\sigma_{A,B_1,B_2}=\Tr_{B_{i=1,2}}\sigma_{A,B_1,B_2}=0$.
The periodic behavior of $I_{A, B_1\cup B_2}$ in $t_0$
may come from the periodicity of $\sigma_{A,B_1,B_2}$,
$
\sigma_{A,B_1,B_2}(t+L)=\sigma_{A,B_1,B_2}(t).
$

\subsubsection{An atypical state}
At the end of this section, let us consider the entanglement structure of the steady state under the evolution by $H^1$.
In Table \ref{summary}, we summarize the entanglement property of the system with various $t_1$ and large $t_0$.
From Table \ref{summary}, we can see that if the system is highly inhomogeneous, then we can not evolve it with even $2$d holographic Hamiltonian to the typical state. 
This atypical state may have a quantum nature because the $t_0$-dependence of $S_B$ and $I_{A,B}$ is periodic (quantum revival).
\begin{table}[t]
  \begin{center}
    \caption{Summary of the properties in Setup 3}
      \begin{tabular}{|c|c|c|c|} \hline
        $t_1$ & $\rho_{A \cup B}$ &  $S_B$ & $I_{A,B}$ \\ \hline
        small & $\rho_{A} \otimes \rho_B$ & Approximately stationary & Chaotic \\
        large & not factorize & Quantum revival & Quantum revival \\ \hline
      \end{tabular}
    \label{summary}
  \end{center}
\end{table}

\section{Line tension picture \label{Sec:Line-tension-picture}}
\begin{figure}[h!!]
\begin{center}
\includegraphics[scale=0.25]{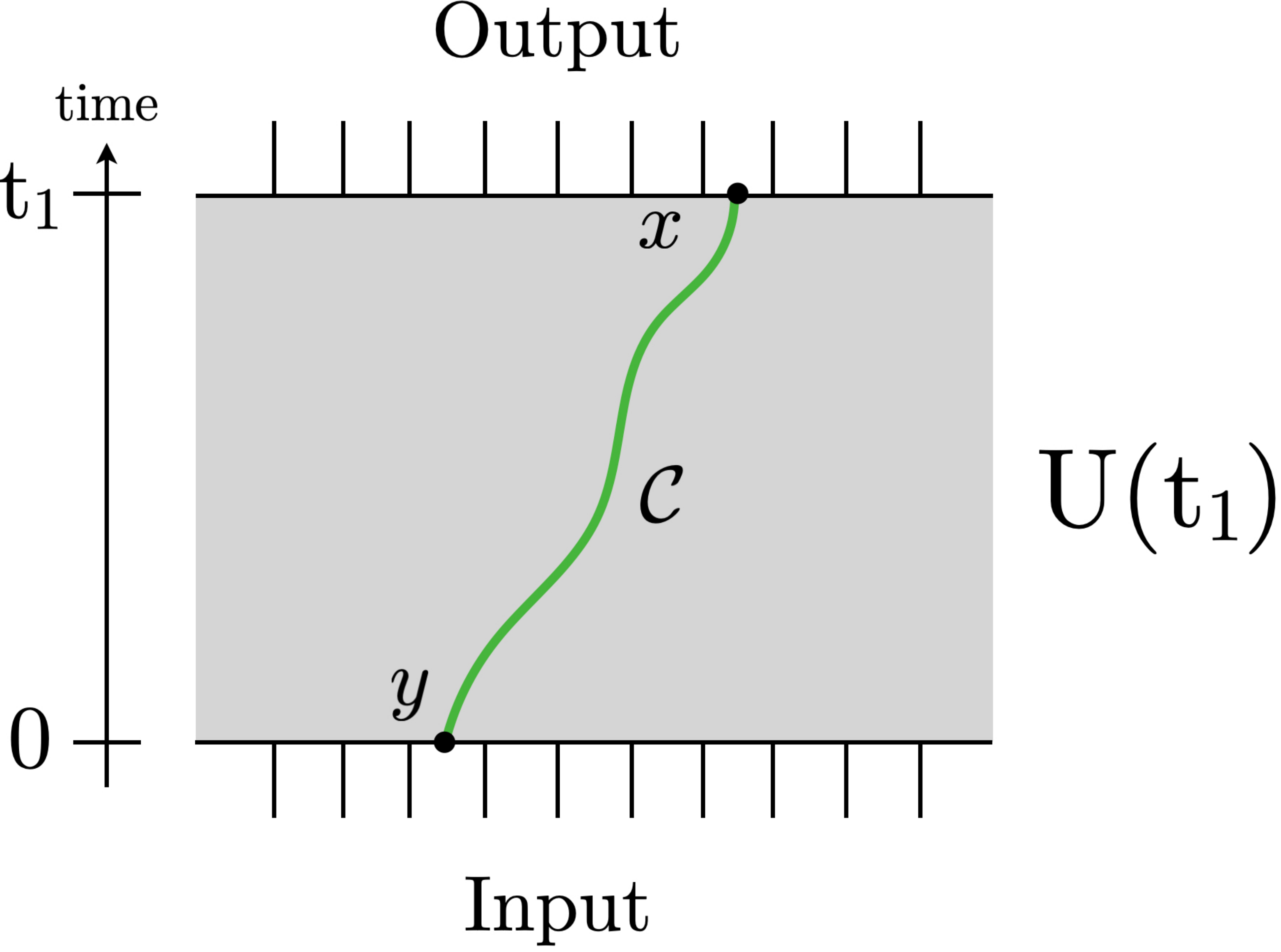}
\caption{A curve ${\cal C}$ that divides the unitary circuit into two parts. In the line-tension picture, the entanglement entropy $S_U(x,y,t_1)$ is given by the integral of the line tension ${\cal T}(v)$ along the curve ${\cal C}$. }
\label{LT1}
\end{center}
\end{figure}

In Sec.\ \ref{Section:thesystem1},
we studied the time-dependence of BMI
after quantum quench
with
the inhomogeneous Hamiltonian
as the post quench Hamiltonian.
We observed that the BMI
is
not fully described by the quasiparticle picture.
In this section, we propose a generalization of the so-called line-tension picture to a random unitary circuit with the SSD time-evolution.
In a chaotic system, the entanglement production is effectively described by the line-tension picture introduced in \cite{
2018arXiv180300089J,2017PhRvX...7c1016N,Mezei:2018jco,PhysRevX.8.031058,PhysRevX.8.021013}.  
To explain the basic idea of the line-tension picture, here we assume that the spatial direction is homogeneous and infinite.  For simplicity, we assume that the system is time-evolved by the unitary operator $U(t_1)$ from $t=0$ to $t=t_1$. We divide the infinite line where the system lives into two pieces at position $x$ at $t=t_1$. We also divide the line at position $y$ at $t=0$.  Now the entanglement entropy $S_U(x,y,t_1)$ of the unitary operator is computed as
\ba
S_U(x,y,t_1)={\rm min}_{\cal C}\int_{\cal C}dt\,  {\cal T}(v)\, ,
\ea
 where the minimization is taken over all the possible curves ${\cal C}$ that connects the point $(x,t_1)$ and $(y,0)$. 
The symbol ${\cal T}(v)$ is the line-tension associated to a curve ${\cal C}$ that connects the point $(x,t_1)$ and $(y,0)$ as in Fig. \ref{LT1}. 
The curve ${\cal C}$ in spacetime has a velocity $v=dx/dt$ and a line-tension ${\cal T}(v)$ that depends on $v$. In our case, when the spacetime is uniform, the minimal curve is given by a straight line with a constant velocity $v=(x-y)/t_1$.

The details of the function ${\cal T}(v)$ depend on the system and are estimated in a chaotic system using random unitary circuits which illustrate the phenomenon of quantum information scrambling. In the scaling limit and in the limit of large bond dimension $q$, the line-tension is simply given by counting the number of bonds cut which is
\begin{align}
\mathcal{T}(v)= \begin{cases}\log q & v<1 \\ v \log q & v>1\, .\end{cases} \label{linetension}
\end{align}
To compute the entanglement of the unitary operator in a holographic CFTs using the line-tension picture, we need to identify the bond dimension (the local Hilbert space dimension) $q$ in the random unitary circuit. This can be accomplished by comparing the rates at which the information gets scrambled.
While the entanglement entropy grows at a rate of $\log q$ in random unitary circuits, it is known that in holographic CFTs the entanglement of the unitary operator (computed as the entanglement between two CFTs in the time-evolved thermofield double state) grows at a rate of $\f{c\pi}{6\epsilon}$. Here, $\epsilon$ is dimensionless as it has been written in units of the lattice spacing.
Therefore, we make the identification
\ba
q\sim e^{\f{c\pi}{6\epsilon}}\, .
\ea
Notice that $\log q$ simply equal to the entropy density given by the Cardy formula $S_{\rm Cardy}/(2\pi R)=\f{c\pi}{6\epsilon}$. Using this, one can correctly reproduce the growth of the entanglement in holographic CFTs which is
\ba
S_U(x,y,t_1)\sim \f{c\pi}{6\epsilon}t_1\, .
\ea
\subsection{Line tension picture with inhomogenity \label{Subsection:Line_tension_inhomo }}
\begin{figure}[t]
\begin{center}
\includegraphics[scale=0.3]{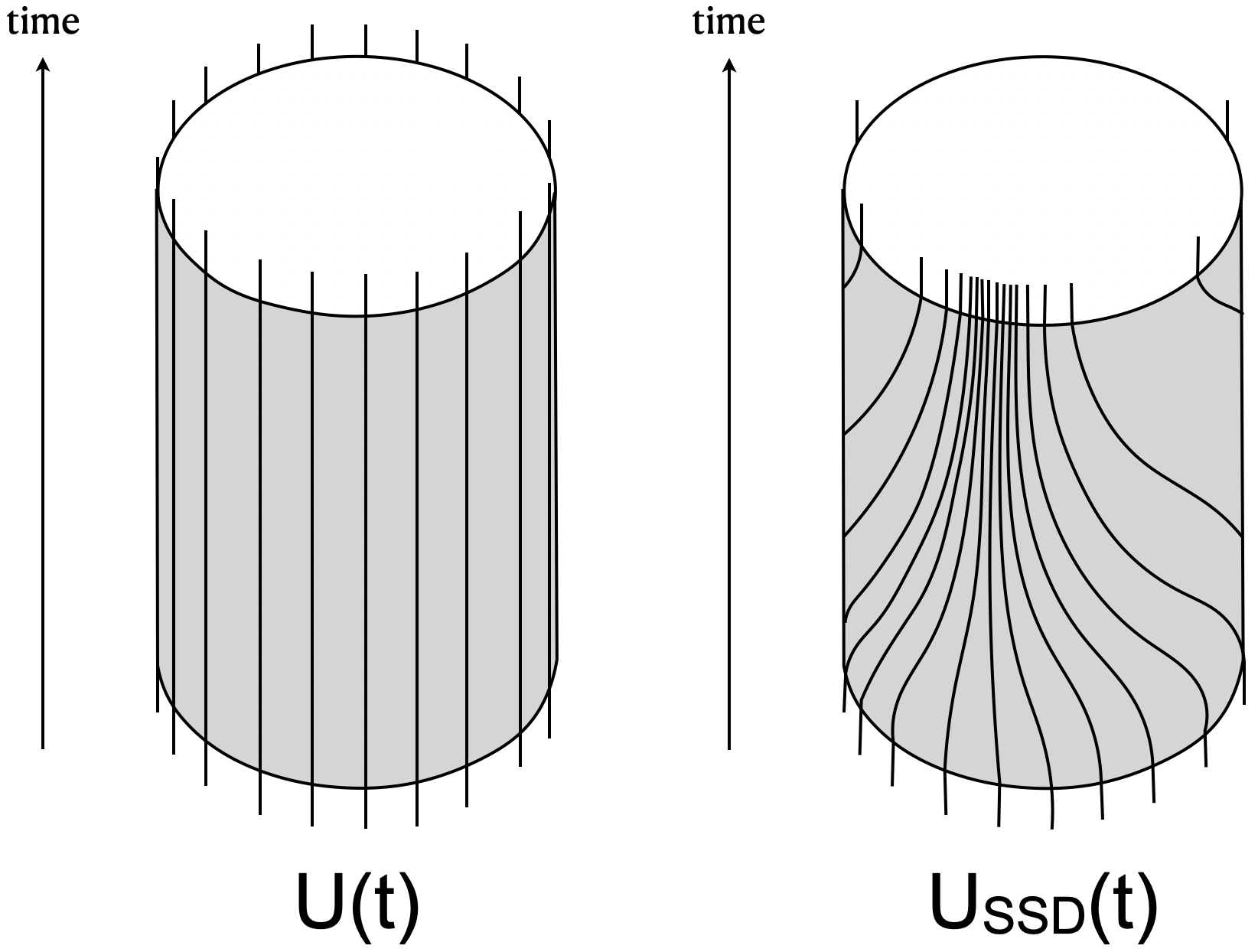}
\caption{SSD time-evolution deforms the spacetime in the line-tension picture  non-uniformly}
\label{UnitarySSD}
\end{center}
\end{figure}

In the above we assumed the homogeneity and infiniteness of the space direction.
Here, we describe how to generalize the line-tension picture to the situation where the spatial direction is inhomogeneous and compact, which fits the SSD time evolution in a compact space discussed in our paper. One can make similar arguments in the cases of other inhomogeneous Hamiltonians, and we will briefly comment on these in the last part of this section.  The main idea is as follows. The aforementioned line-tension picture was based on a geometric representation of a random unitary circuit consisting of quantum gates uniformly arranged in the spatial direction. In the Schrodinger picture, the spatial direction is deformed non-uniformly by the SSD time evolution, and to give a line-tension picture that captures the dynamics of entanglement by the SSD time evolution, we should consider line-tension picture in a deformed inhomogeneous spacetime, see Fig.\ \ref{UnitarySSD}.

 We look for an appropriate spacetime generated by the SSD time evolution. We are especially interested in the coordinate whose metric is conformally flat. As in \cite{Goto:2021sqx}, we introduce new coordinates in which action of the SSD Hamiltonian is simple.
The evolution under the SSD Hamiltonian is simplified by introducing the Poincar\'e coordinate $(z_P,\bar{z}_P)$. The boundary global coordinate $(w,\bar{w})$ and the boundary Poincar\'e coordinate $(z_P,\bar{z}_P)$ are related as
\begin{align}
    z_P=L\cot\left(\frac{i\pi w}{L}\right)\, ,\quad  \bar{z}_P=-L\cot\left(\frac{i\pi \bar{w}}{L}\right)\, ,\label{poincare}
\end{align}
where $z_P=x_P-i\tau_P$ and $\bar{z}_P=x_P+i\tau_P$ are the complex coordinates in the Poincar\'e coordinate. The symbol $\tau_P$ is the Euclidean time coordinate, and $x_P$ is the spatial  coordinate ($-\infty<x_P<\infty$) in the plane where the Poincar\'e coordinate is defined.
Notice that in this Poincar\'e coordinate, the two fixed points of the SSD Hamiltonian are located at the origin and the spatial infinity. 

Now let us see how the Poincar\'e coordinate simplifies the translation under the SSD Hamiltonian.
The flow of the Poincar\'e time is generated by the following Hamiltonian
\begin{align}
    H_P=\int^\infty_{-\infty}dx_P T_{\tau_P\tau_P}(x_P)=-\int dz_P T(z_P)-\int d\bar{z}_P \overline{T}(\bar{z}_P)\, .
\end{align}
We use the usual transformation rule for the energy-momentum tensor
\begin{align}
  \left(\frac{d z_{P}}{d w}\right)^{2} T\left(z_{P}\right)
  &=
    T(w)-\frac{c}{24 \pi}
    \mathrm{Sch}\left(z_{P}, w \right)
    \nonumber\\ &=T(w)+\frac{\pi c}{12L^2}\, ,
\end{align}
with $\frac{d z_{P}}{d w}=-\frac{i\pi}{\sin^2\left(\frac{i\pi w}{L}\right) },\frac{d \bar{z}_{P}}{d \bar{w}}=\frac{i\pi}{\sin^2\left(\frac{i\pi \bar{w}}{L}\right) }$ and move to the original global coordinate $(w,\bar{w})$ as
\begin{align}
H_{P} &=-\oint d w\left(\frac{d w}{d z_{P}}\right)\left(T(w)+\frac{\pi c}{12L^2}\right)-\oint d \bar{w} \left(\frac{d \bar{w}}{d \bar{z}_{P}}\right)\left(\overline{T}(\bar{w})+\frac{\pi c}{12L^2}\right) \nonumber\\
&=\int \frac{d x}{2\pi }2\sin^2\left(\frac{\pi x}{L}\right)T_{\tau\tau}(x)+\frac{c}{12L}
\nonumber\\
&=\frac{1}{2\pi}\left(H_{\rm SSD}+\frac{c\pi }{6L}\right)\, .\label{PoiSSD}
\end{align}
Therefore, the SSD Hamiltonian generates the time-flow in the Poincar\'e
coordinate defined as (\ref{poincare}). This indicates that the line-tension picture in the Poincar\'e
coordinate appropriately captures the entanglement dynamics under the SSD time-evolution.
  The metric is given by
\begin{align}
ds^2=dwd\bar{w}=\f{dz_P d\bar{z}_P}{\pi^2|1+z^2_P/L^2|^2}\, .\label{Poincare}
\end{align}
We propose the entanglement entropy computed in the line-tension picture in a curved spacetime with metric $g_{z\bar{z}}$ is given by the following line integral
\begin{align}
S_A=\min _{\partial \gamma_A=\partial A } \int_{\gamma_A} ds  \mathcal{T}(v)\, .
\end{align}
where $\gamma_A$ is the curve anchored at the edges of the subregion $A$ and homologous to $A$.

Specifically, using a pair of coordinates $(z(s),\bar{z}(s))$ on the two-dimensional spacetime, we obtain 
\begin{align}
S_A=\min _{\partial \gamma_A=\partial A } \int_{\gamma_A} \frac{dz}{z'} \mathcal{T} + \int_{\gamma_A} \frac{d\bar{z}}{\bar{z}'} \mathcal{T}\, , \label{lineent}
\end{align}
where $z'=dz/ds$ and $\bar{z}'=d\bar{z}/ds$.

\begin{figure}[h]
\begin{center}
\includegraphics[scale=0.3]{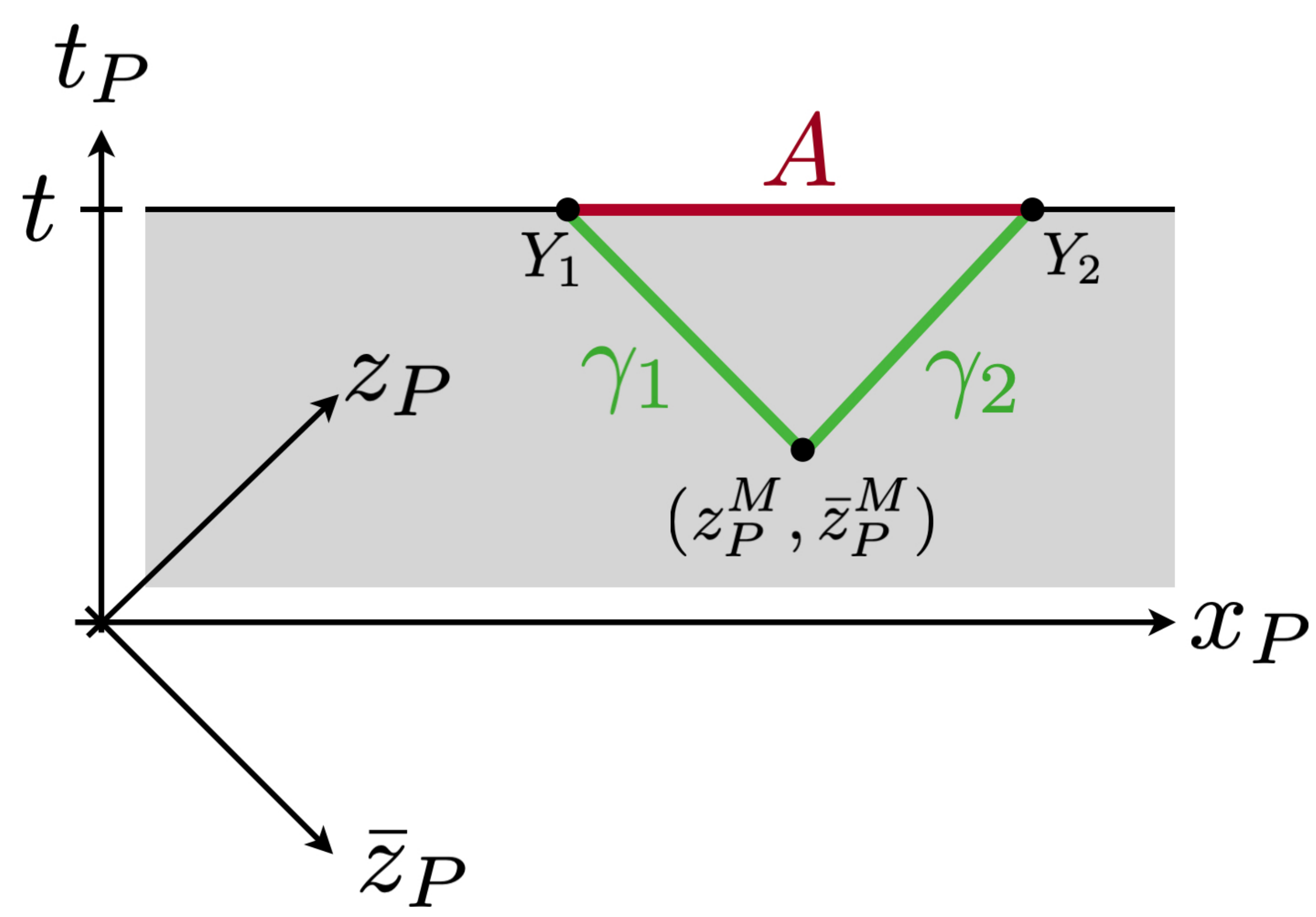}
\caption{Configuration of the minimal curve in the line-tension picture}
\label{Line_example}
\end{center}
\end{figure}

In our case, we have a curved metric (\ref{Poincare}) with line tension (\ref{linetension}).  Let us compute the entanglement entropy by simply taking a subregion as $A=[Y_1,Y_2]$ at time $t$ after the SSD time evolution. 
The entanglement: $S^1_{A}$ is given by the space-like (or light-like) curve with  $\mathcal{T}(v)= \log q=\frac{c}{6\epsilon}$ as Fig.\ \ref{Line_example}. Segments of the curve $\gamma_1$ and $\gamma_2$ intersects at the point $(z_P^M,\bar{z}_P^M)$. 
The entanglement entropy for the subregion $A$ is computed as 
\begin{align}
   S_{A}&=\int_{\gamma_2} \frac{dz}{z'} \mathcal{T} + \int_{\gamma_1} \frac{d\bar{z}}{\bar{z}'} \mathcal{T}\nonumber\\
   &=\frac{c}{6\epsilon}\left[\int_{z_P^M}^{iL\cot\frac{\pi Y_2}{L}-2\pi t} \f{dz_P}{1+(z_P/L)^2} + \int^{\bar{z}_P^M}_{-iL\cot\frac{\pi Y_1}{L}+2\pi t} \f{d\bar{z}_P}{1+(\bar{z}_P/L)^2}\right]\, , \label{wnew1}
\end{align}
This can be simplified to the new coordinate system $(w^{\rm New},\bar{w}^{\rm New})$ 
with flat metric defined 
by pulling the curved coordinate $(z_P,\bar{z}_P)$ back to $w$ coordinate after the SSD ($=$Poincar\'e) time evolution
\footnote{We add an extra $2\pi$ factor in front of $t$.
This comes from the $2\pi$ difference between $H_{\rm SSD}$ 
and the Poincar\'e Hamiltonian $H_{P}$ (\ref{PoiSSD}).}
\begin{align}
    z_P+2\pi t=L\cot\left(\frac{i\pi w^{\rm New}}{L}\right)\, ,\quad  \bar{z}_P-2\pi t=L\cot\left(\frac{i\pi \bar{w}^{\rm New}}{L}\right)\, .
\end{align}
I.e., $w^{\rm New}$ and $\bar{w}^{\rm New}$ are related by the original $w$ and $\bar{w}$ as 
\begin{align}
    w^{\rm New}=\frac{L}{i\pi}\cot ^{-1}\left[\cot\frac{i\pi w}{L}-\frac{2\pi}{L}t\right]\, ,\ \bar{w}^{\rm New}=\frac{L}{i\pi}\cot ^{-1}\left[\cot\frac{i\pi \bar{w}}{L}+\frac{2\pi}{L}t\right]\, .
\end{align}
 $w^{\rm New}$ and $\bar{w}^{\rm New}$ are nothing but $w^{{\rm New},\alpha}_{x,\epsilon}$ and $\bar{w}^{{\rm New},\alpha}_{x,\epsilon}$ with $\alpha=1,\epsilon=0,\tau_0=0$ and $\tau_1=it$. This can be explicitly checked by the formula $\cot ^{-1}(z)=\frac{i}{2}\log \left[\left(z-i\right)/\left(z+i\right)\right]$ with $z=\cot\frac{i\pi w}{L}-\frac{2\pi}{L}t$.

 Since we can treat $t$ just as a parameter in the integral, we have $dw^{\rm New}=dw$, thus we can compute the integral as
 \begin{align}
   S_{A}&=\frac{c\pi}{6\epsilon}\left[-i\int_{iX_{M}^{\rm New} }^{iX_{Y_2}^{\rm New}} dw^{\rm New} +i \int_{-iX^{\rm New}_{Y_1}}^{-iX_{M}^{\rm New}} d\bar{w}^{\rm New}\right]\nonumber\\
   &=\frac{c\pi}{6\epsilon}\left[X_{Y_2}^{\rm New}-X^{\rm New}_{Y_1}\right]\, , 
\end{align}
where $X_{M}^{\rm New}$ is the intersection point $(z_P^M,\bar{z}_P^M)$ in the $X^{\rm New}$ coordinate. This correctly reproduces the result obtained by the holographic computations (\ref{EE_on_torus}) in the leading order of the coarse-grained limit. 

\begin{figure}[t]
\begin{center}
\includegraphics[scale=0.3]{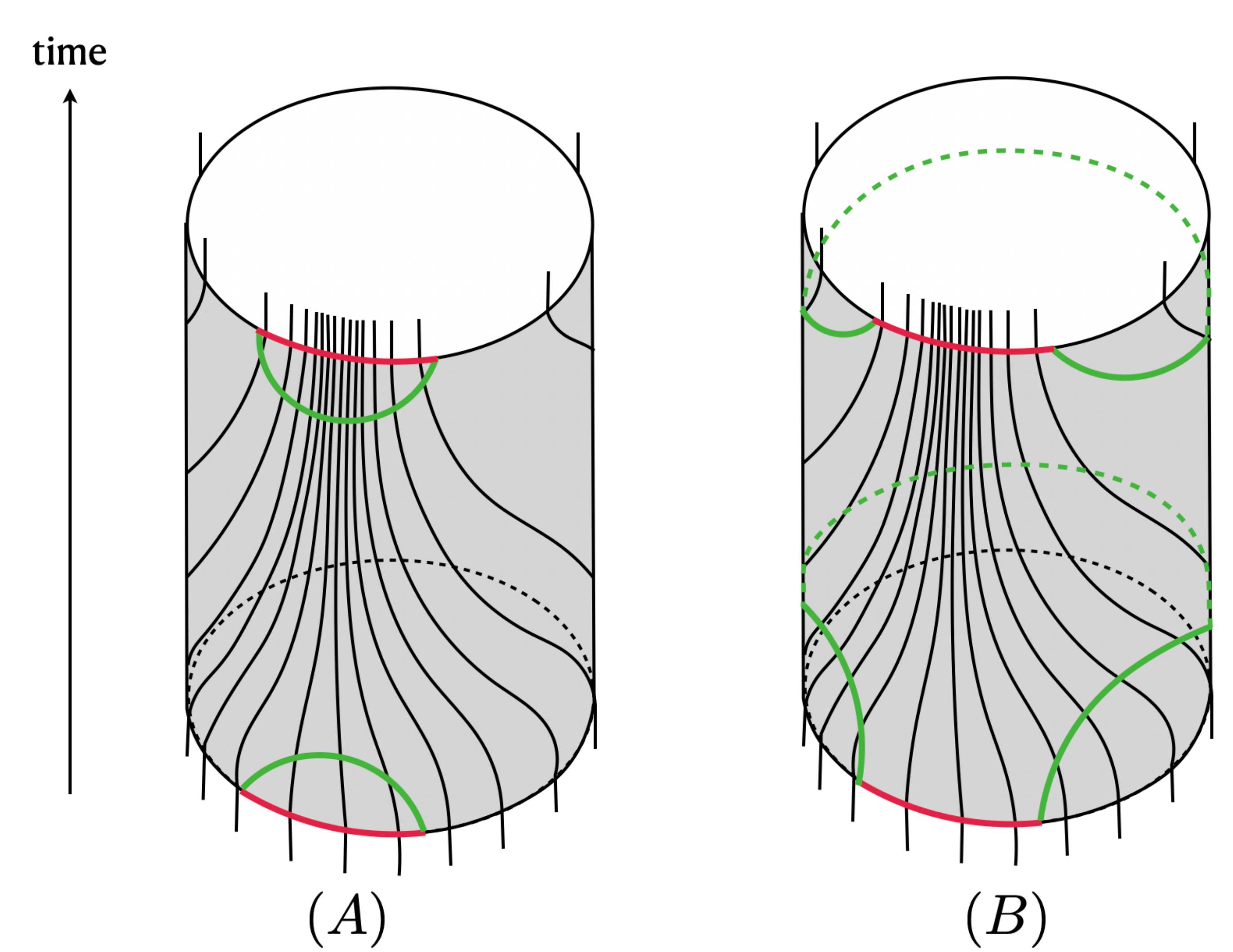}
\caption{Two candidates for the minimal curve that computes $S_{A\cup B}$.}
\label{SSDconf}
\end{center}
\end{figure}

We have more interesting configurations for the entanglement entropy for double intervals $A$ and $B$ placed at $t=0$ and time $t$ respectively. Let us consider a sufficiently late time when the disconnected configuration dominates over the connected ones as Fig.\ \ref{SSDconf}. Two candidates for the curve that would give the entanglement entropy are drawn in Fig.\ \ref{SSDconf}. In the case of the uniform Hamiltonian (see the left panel in Fig.\ \ref{UnitarySSD}, if you take small enough regions, the left configuration in Fig.\ \ref{SSDconf} always dominates, and we have the trivial mutual information, i.e., $S_{A\cup B}=S_A+S_B$. This is not the case for the SSD Hamiltonian. As you can see in Fig.\ \ref{SSDconf}, if the subregion contains the fixed point $X^1_f$ of the SSD Hamiltonian, the vertical lines representing ``gates'' originally aligned uniformly are condensed around $X^1_f$. The amount of the entanglement counts the number of the lines cut by the minimal curve. Therefore no matter how small a subregion is taken, at sufficiently late times, it is more efficient to take a curve like the right one in Fig.\ \ref{SSDconf} that is not homologous to subregions A and B respectively (while the union of each curve is homologous to $A\cup B$) than to take the right one. Such curves give nontrivial mutual information. This is the characteristic entanglement behavior of systems driven by Hamiltonians with fixed points, such as the SSD Hamiltonian. It correctly reproduces the holographic calculations.

This prescription of the line-tension picture proposed in this section can be generalized to other inhomogeneous time evolutions. In the case of the cosine-square deformation, an appropriate coordinate system that simplifies the action of its Hamiltonian is given by the coordinate transformation
\begin{align}
    \tilde{z}_P=L\tan\left(\frac{i\pi w}{L}\right)\, ,\quad  \bar{\tilde{z}}_P=L\tan\left(\frac{i\pi \bar{w}}{L}\right)\, .
\end{align}
and the $w_{\theta}^{\rm New}$ and $\bar{w}_\theta^{\rm New}$ are defined as
\begin{align}
    \tilde{z}_P+2\pi t=L\tan\left(\frac{i\pi w^{\rm New}}{L}\right)\, ,\quad  \bar{\tilde{z}}_P-2\pi t=L\tan\left(\frac{i\pi \bar{w}^{\rm New}}{L}\right)\, .
\end{align}
As we pointed out in \cite{Goto:2021sqx}, in the case of the general  M\"obius Hamiltonian, we can find an appropriate coordinate system by the coordinate transformation
\begin{align}
    \tan \frac{z_\theta}{2 L \cosh 2 \theta}=e^{-2 \theta}\cot \left(\frac{i \pi w}{L}\right), \quad \tan \frac{\bar{z}_\theta}{2 L \cosh 2 \theta}=-e^{-2 \theta} \cot \left(\frac{i \pi \bar{w}}{L}\right)
\end{align}
instead of (\ref{poincare}) in the case of the SSD Hamiltonian. The  M\"obius Hamiltonian generates the simple time translation in the $(z_\theta,\bar{z}_\theta)$ coordinate.
The $(w_{\theta}^{\rm New},\bar{w}_\theta^{\rm New})$ coordinates analogous to (\ref{wnew1}) is defined as
\begin{align}
    \tan \frac{z_\theta+2\pi t}{2 L \cosh 2 \theta}=e^{-2 \theta}\cot \left(\frac{i \pi w_\theta^{\rm New}}{L}\right), \quad \tan \frac{\bar{z}_\theta-2\pi t}{2 L \cosh 2 \theta}=-e^{-2 \theta} \cot \left(\frac{i \pi \bar{w}_\theta^{\rm New}}{L}\right)\, ,
\end{align}
which simplifies the integral that computes the entanglement.
One can check that
$w_{\theta}^{\rm New}$ and $\bar{w}_\theta^{\rm New}$ are nothing but $w^{{\rm New},\alpha}_{x,\epsilon}$ and $\bar{w}^{{\rm New},\alpha}_{x,\epsilon}$ with $\alpha=0,\epsilon=0$ and $\tau_1=it$.

\section{Gravitational description\label{Section:gravitational-description}}

\begin{figure}[tbp]
    \begin{tabular}{ccc}
      \begin{minipage}[t]{0.5\hsize}
        \centering
        \includegraphics[keepaspectratio, scale=0.6]{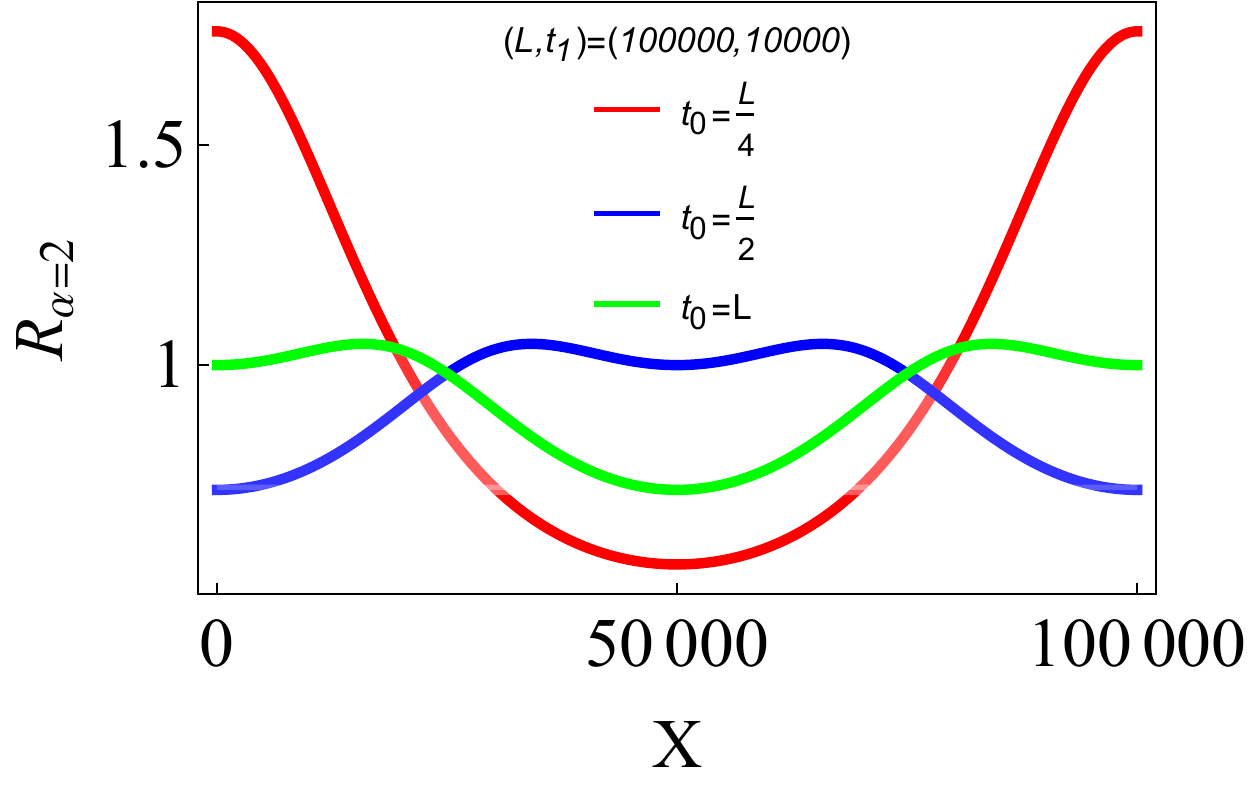}
        
   (a) Small $t_1$-regime.
      \end{minipage} & 
     
     \begin{minipage}[t]{0.5\hsize}
        \centering
        \includegraphics[keepaspectratio, scale=0.6]{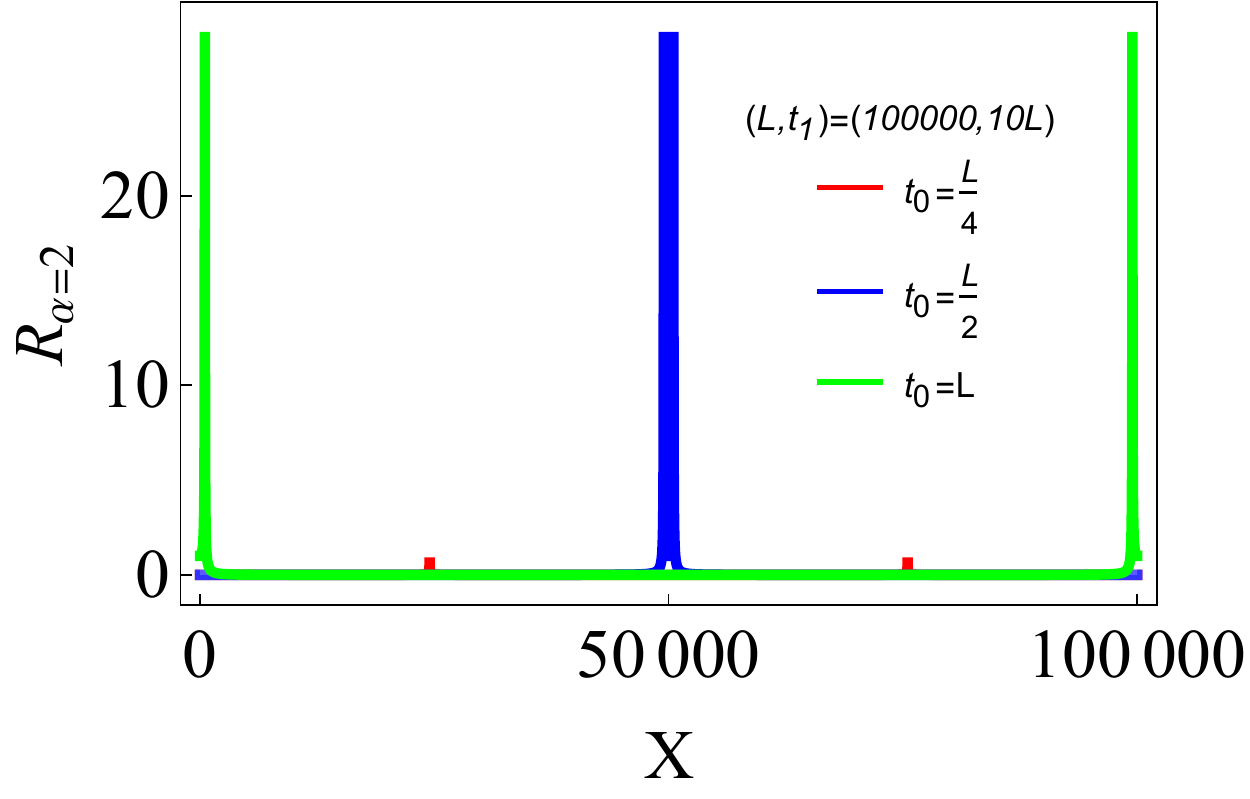}
        
    (b) Large $t_1$-regime.
      \end{minipage} 
     
    \end{tabular}
    \caption{The spatial dependence of the black hole horizon for various $t_1$ and $t_0$ as a function of $X$. Here, $R_{\alpha=2}$ is defined by $r_{\alpha=2,\text{Horizon}}/r_0$. For the large $t_1$, the spatial locations where the peaks of $R_{\alpha=2}$ emerge are approximately equal to the locations of the B.H.-like excitations. }
    \label{Fig:BH_for_system3}
  \end{figure}

Let us now turn to the gravitational dual descriptions of Setups 1, 2, and 3. 
As in \cite{1999AIPC..484..147B,2012JHEP...12..027R}, these dual geometries are constructed from the expectation value of energy density under the evolution by the Hamiltonians considered.
Equivalently, these geometries are given by a map from the BTZ-black hole 
in $w^{\text{New}}_{x,\epsilon}$ 
and $\overline{w}^{\text{New}}_{x,\epsilon}$ 
to the time-dependent one in terms of $t_{i=0,1,2}$.
The dual geometry of the reduced density matrix associated with $\mathcal{H}_2$ is a stationary BTZ-black hole. 
Since $\rho_{\mathcal{H}_1}$ is a mixed state, 
its gravity dual should be a black hole geometry.
The details of the complicated metric associated with $\rho_{\mathcal{H}_1}$ are reported in Appendix \ref{sec:dual-geometries}. 
Here, we describe the spacetime-profile of the black hole horizon in these dual geometries.
Let us introduce the radial coordinate, $r'$, that guarantees the asymptotic geometry near the AdS boundary is given by the pure AdS$_3$ 
or the modified geometry, 
the metric of which is given
by replacing the time-component 
of the pure AdS$_3$ 
with $g_{tt}=-4L^2r'^2\sin^4{\left(\pi X/L\right)}$.
Then, the spatial and temporal dependence of 
the black hole horizon in the dual geometries 
for Setup 1 and 2
is almost the same as that in \cite{Goto:2021sqx}.
In Fig.\ \ref{Fig:BH_for_system3}, 
we plot the black hole horizon corresponding to 
Setup 3 for various $t_0$ and $t_1$ as a function of $x$.
The spacetime-dependence of the black hole horizon for 
Setup 4 
is reported in Appendix \ref{sec:gravity-dutal-app}.

The extremes in the spatial direction of the black hole horizon for $\alpha=2$ are given by
\be
r'_{\alpha=2,\text{Horizon}} =\f{r_0\sqrt{L^2+\pi^2t_1^2}}{\left|L \cos{\left(\f{2\pi t_0}{L}\right)}-\pi t_1 \sin{\left(\f{2\pi t_0}{L}\right)}\right|}.
\ee
The details of the analysis on the black hole horizon are reported in Appendix \ref{sec:Asym-behavior-of-ralpha2}.
In the large $t_1$-regime, $r'_{\alpha=2,\text{Horizon}}$ is extremized along the trajectories of the B.H.-like excitations.
These extremes along these trajectories are approximated by
\be
\begin{split}
    r'_{\alpha=2,\text{Horizon}} \approx \begin{cases}
    \f{r_0}{\left|\sin{\left(\f{2\pi t_0}{L}\right)}\right|} ~\text{for}~t_0\neq \f{nL}{2}\\
    \f{\pi r_0 t_1}{L}~\text{for}~t_0= \f{nL}{2}
     \end{cases}.
\end{split}
\ee
Thus, if the B.H.-like excitations are at $x \neq X^{i=1,2}_f$, then $r'_{\alpha=2,\text{Horizon}}$ depends on only $t_0$, while if these excitations are at $x = X^{i=1,2}_f$, then $r'_{\alpha=2,\text{Horizon}}$ depends on only $t_1$, and it linearly increases with $t_1$.
\section{Wormhole growth}

In addition to the horizon,
another geometrical object 
of our interest is 
a wormhole
connecting the two
Hilbert spaces.
Here, as a measure of wormhole growth, we use the ``free" energy 
defined from
a two-point function as
\be 
F(X_1,Y_1)=-\log{\left[\left\langle \mathcal{O}_1(Y_1)\mathcal{O}_2(X_1) \right \rangle\right]},
\ee
where $\mathcal{O}_{i=1,2}$ are local primary operators 
in $\mathcal{H}_{i=1,2}$
with 
conformal dimension
$(h_{\mathcal{O}},h_{\mathcal{O}})$.
In the Heisenberg picture, 
this free energy islgiven by the universal and non-universal pieces as
\be \label{eq:free-energy-correlator}
F(X_1,Y_1)=-h_{\mathcal{O}}\log{\left[\f{dw^{\text{New}}_{Y_1,\epsilon}}{dw_{Y_1}}\f{d\overline{w}^{\text{New}}_{Y_1,\epsilon}}{d\overline{w}_{Y_1}}\right]}+2h_{\mathcal{O}} G(X_1,Y_1).
\ee
Here, we consider light operators 
with $c \gg h_{\mathcal{O}} \gg 1$.
In this regime, 
the non-universal piece $G$ is determined by the length of geodesics in the stationary BTZ black hole.

\subsection{Setup 1}
Let us begin by analyzing the free energy in (\ref{eq:free-energy-correlator}) for Setup 1.
We consider the $t_1$-dependence of (\ref{eq:free-energy-correlator}) with general $Y_1$.
We assume that $\f{L}{2}>X_1>Y_1>0$.
Under the evolution by $H^1_{\text{M\"obius}}$ with $\theta\neq\infty$, 
the imaginary parts of 
$w^{\text{New}}_{Y_1,\epsilon}$ and $\overline{w}^{\text{New}}_{Y_1,\epsilon}$ of  (\ref{eq:free-energy-correlator}) monotonically increase with $t_1$.
In the large $t_1$-regime, $F(X_1,Y_1)$ is approximately given by a monotonically-increasing function of $t_1$,
\be 
\label{eq:free-energy-system1-late}
F(X_1,Y_1) \approx 
\f{h_{\mathcal{O}}\pi}{\epsilon}
\left(
\text{Im}\left[w^{\text{New}}_{Y_1,\epsilon}\right]
+\text{Im}\left[\bar{w}^{\text{New}}_{Y_1,\epsilon}\right]\right)
+4h_{\mathcal{O}}\log{\left(\f{2\epsilon}{\pi}\right)}.
\ee
In small $t_1$-regime where 
$0>\text{Im}\,[w^{\text{New}}_{Y_1,\epsilon}]-X_1>-X_1$, $L>\text{Im}\,[\overline{w}^{\text{New}}_{Y_1,\epsilon}]>0$, 
and $\f{L}{2}>X_1-X^{\text{New}}_{Y_1,\epsilon}$, $F(X_1,Y_1)$ is approximately given by a function following the trajectory of the local operator,
\be \label{eq:free-energy-system1-early}
F(X_1,Y_1) \approx \f{h_{\mathcal{O}}\pi\left(X_1-X^{\text{New}}_{Y_1,\epsilon}\right)}{\epsilon}+4h_{\mathcal{O}}\log{\left(\f{2\epsilon}{\pi}\right)}.
\ee
In this $t_1$-regime, $F(X_1,Y_1)$ may decrease with $t_1$.

In the SSD limit $\theta\to \infty$, 
if $Y_1=0$, $F(Y_1=0,X_1)$ is a stationary constant, 
and approximated as 
$F(Y_1=0,X_1)\approx \f{h_{\mathcal{O}} \pi X_1}{\epsilon}$.
Unless $Y_1=0$, for large $t_1$, the imaginary parts of 
$w^{\text{New}}_{Y_1,\epsilon}$
and 
$\bar{w}^{\text{New}}_{Y_1,\epsilon}$
reduce to 
$\text{Im}\,[w^{\text{New}}_{Y_1,\epsilon}] \approx L$ 
and $\text{Im}\,[\overline{w}^{\text{New}}_{Y_1,\epsilon}] \approx 0$. 
Consequently, the $t_1$-dependence of $F(X_1,Y_1)$ in this limit is approximated by 
\be \label{eq:free-energy-system1-asym}
F(X_1,Y_1)\approx 
4h_{\mathcal{O}}
\log
\left[
\frac{2\pi t_1}{L}
\sin \left(\f{\pi Y_1}{L}\right)
\right]
+\f{\pi h_{\mathcal{O}}L}{\epsilon}+4h_{\mathcal{O}}\log{\left(\f{2\epsilon}{\pi}\right)}.
\ee
Thus, $F(X_1,Y_1)$ 
is approximately stationary except for the logarithmic growth with $t_1$. 

\begin{figure}[tbp]
     \begin{center}
        \includegraphics[width=7cm]{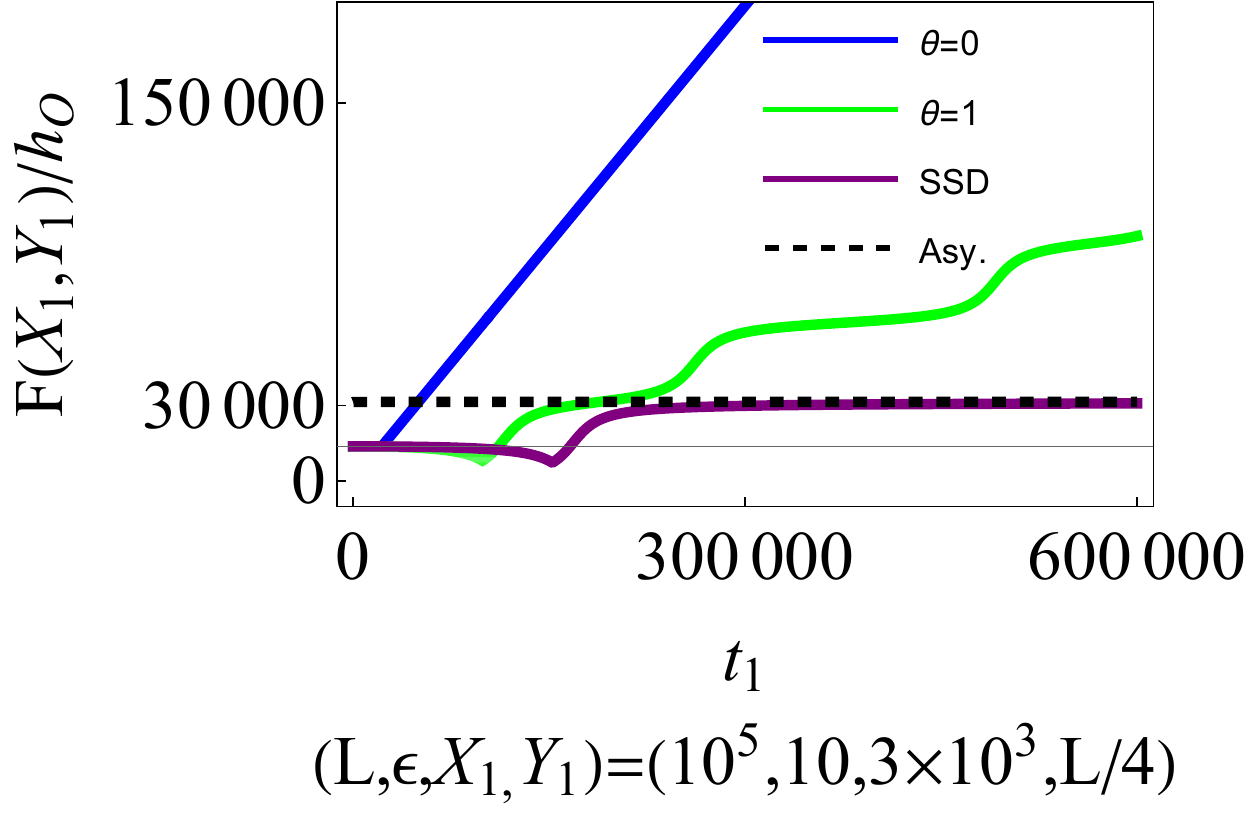}
        \hspace{0.5cm}
          \includegraphics[width=7cm]{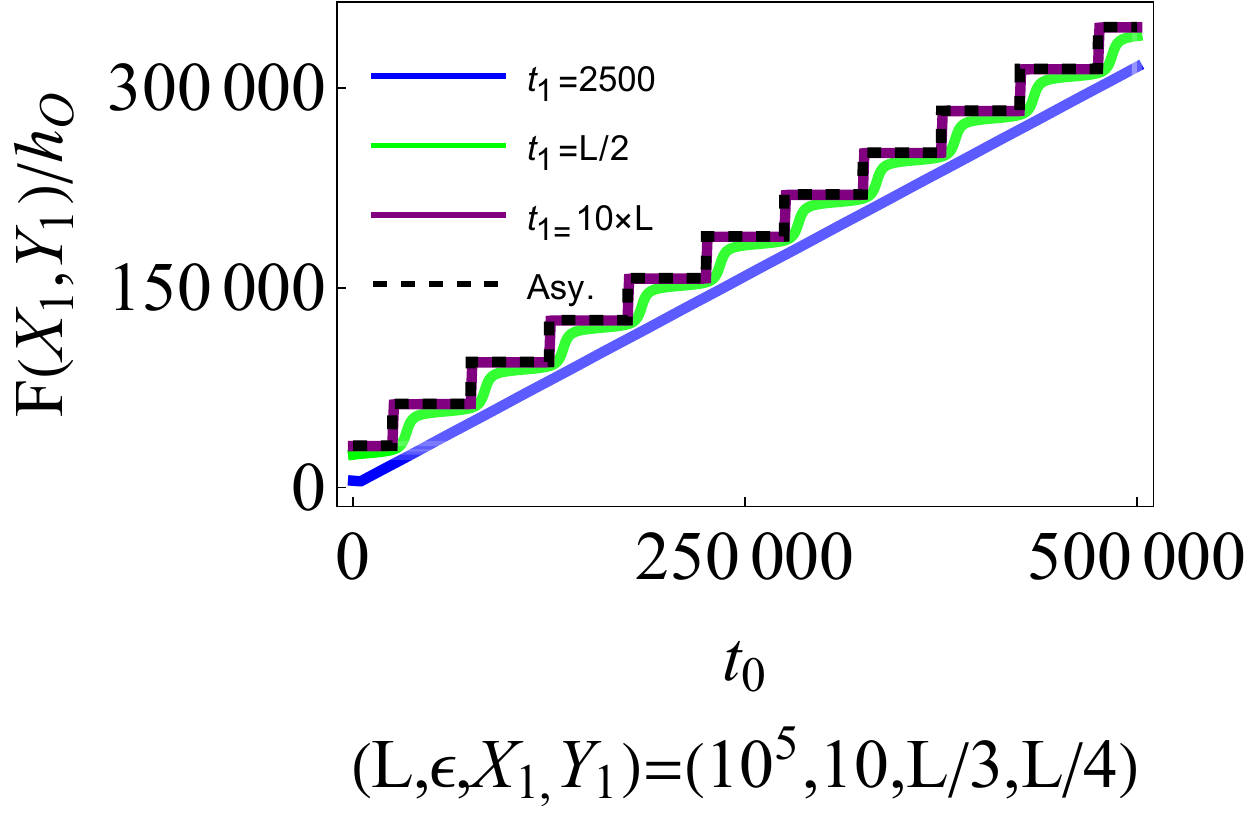}
          \end{center}

         \caption{(Left) 
         The $t_1$-dependence of $F(X_1,Y_1)$ for various $\theta$ in Setup 1. The dashed line illustrates the $t_1$-dependence of $F(X_1,Y_1)/h_{\mathcal{O}}$ in (\ref{eq:free-energy-system1-asym}).
         (Right)
          The $t_1$-dependence of $F(X_1,Y_1)$ for various $t_1$ in Setup 3. The dashed line illustrates the asymptotic behavior of $F(X_1,Y_1)/h_{\mathcal{O}}$ in (\ref{eq:free-energy-system3-asym}).
         \label{Free-energy-system1}}
      \end{figure}

From these analyses,
we can see
that M\"obius/SS deformation may prevent the wormhole from growing with $t_1$.
In Fig.\ \ref{Free-energy-system1}, we plot $F(X_1,Y_1)$ 
in Setup 1 
as a function of $t_1$.
We can see 
for larger $\theta$, 
the growth 
of $F(X_1,Y_1)$ 
is slower.

%
%

\subsection{Setup 2}

In Setup 2, $F(X_1,Y_1)$ grows lineary 
with $t_0$ under the evolution 
by $H^{1}$, 
and then grows with $t_1$ under the evolution by $H^{1}_{\text{M\"obius}}$ as in the previous section. 

\subsection{Setup 3 \label{sec:worm-hole-system3}}
Let turn to the analysis on $F(X_1,Y_1)$ in Setup 3.
We, as before, assume $\f{L}{2}>X_1>Y_1>0$.
As in Setup 1, for various $t_1$, 
the imaginary parts of 
$w^{\text{New}}_{Y_1,\epsilon}$ and $\overline{w}^{\text{New}}_{Y_1,\epsilon}$ of 
\eqref{eq:free-energy-correlator} 
monotonically increase with $t_0$. 
Therefore, the early-time behavior of $F(X_1,Y_1)$ may be approximated by (\ref{eq:free-energy-system1-early}), while the late-time $t_0$-dependence is approximated by (\ref{eq:free-energy-system1-late}).
For large $t_1$, the $t_0$-dependence of $F(X_1,Y_1)$ is given by the asymptotic form,
\be 
\label{eq:free-energy-system3-asym}
\begin{split}
    F(X_1,Y_1)
    &\approx 
    4h_{\mathcal{O}} \log{\left(\f{2\epsilon}{\pi}\right)}
    +h_{\mathcal{O}}
    \log 
    \left[
    \frac{16 \pi^4 t_1^4}{L^4} 
    \sin^2\left(\frac{\pi  (t_0-Y_1)}{L}\right) 
    \sin^2\left(\frac{\pi  (t_0+Y_1)}{L}\right)
    \right] \\
    &
    \quad
    +\begin{cases}
    \f{\pi h_{\mathcal{O}} (2m+1)L}{\epsilon}, 
    & mL+Y_1>t_0>mL-Y_1 
    \\
    \f{2\pi h_{\mathcal{O}} (m+1)L}{\epsilon}, 
    & (m+1)L-Y_1>t_0>mL+Y_1 \\
    \end{cases},
\end{split}
\ee
where $m$ is an integer.
In Fig.\ \ref{Free-energy-system1},
we 
plot $F(X_1,Y_1)$ of Setup 3 for various $t_1$ as a function of $t_0$.
We can see that 
for larger $t_1$,
$F(X_1, Y_1)$ is not given by the simple linear growth,
but  
approximated by a sequence of step-functions.
%
%
%

\begin{figure}[tbp]
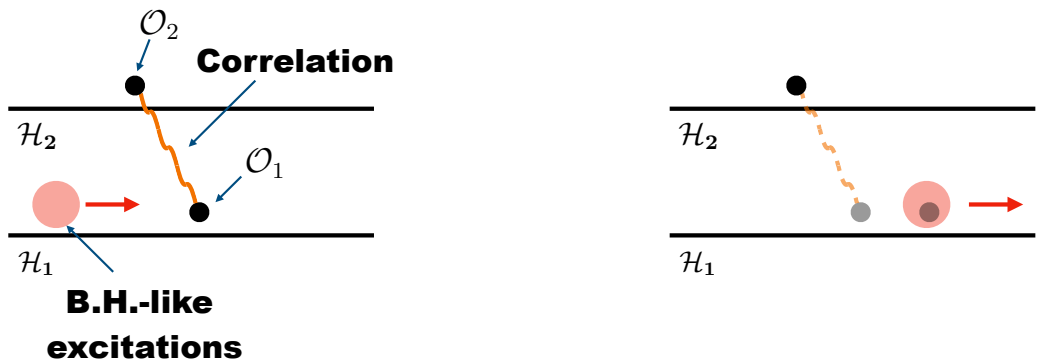

    \begin{tabular}{ccc}
      \begin{minipage}[t]{0.5\hsize}
        \centering
        \includegraphics[keepaspectratio, scale=0.04]{Figure/Section6/Before-scrambling.pdf}
        
   (a) Before hitting the local operator.
      \end{minipage} & 
     
     \begin{minipage}[t]{0.5\hsize}
        \centering
        \includegraphics[keepaspectratio, scale=0.04]{Figure/Section6/After-scrambling.pdf}
        
    (b) After hitting the local operator.
      \end{minipage} 
     
    \end{tabular}
    \caption{The destruction of the non-local correlation between $\mathcal{O}_1$ and $\mathcal{O}_2$ by the B.H.-like excitations.}
    \label{Fig:scrambling-on-local-operator}
\end{figure}

The asymptotic behavior \eqref{eq:free-energy-system3-asym} 
can be interpreted by using the description in Section \ref{sec:physical-interpretation}.
For large $t_1$, at $t_0=0$, two B.H.-like excitations emerge near $x=X^1_f$ and move towards the left and right
at the speed of light under the evolution by $H^1$ 
(Fig.\ \ref{Fig:scrambling-on-local-operator}). 
Here, we assume that the size of these excitations is $\mathcal{O}(\epsilon)$. 
Then, in the coarse-grained region, these excitations are approximated as the local excitations.
Recall that we have operators $\mathcal{O}_{i=1,2}$
on $\mathcal{H}_{i=1,2}$
that are inserted as at $Y_1$ and $X_1$, respectively. 
At $t_0\approx mL\pm Y_1$ where $m$ is an integer, the left- and right-moving B.H.-like excitations hit $\mathcal{O}_1$, and simultaneously the information about $\mathcal{O}_1$ is scrambled in the interior of the B.H.-like excitations. 
As a consequence, the correlation between $\mathcal{O}_1$ and $\mathcal{O}_2$ is weakened 
each time these B.H.-like excitations pass $\mathcal{O}_1$. 
The information about $\mathcal{O}_1$ may be delocalized and encoded in these B.H.-like excitations. 
        
\section{Discussions and future directions \label{Section:Discussions-and-future-directions }}

In this paper, we studied three quantum quench processes
with the inhomogeneously-deformed Hamiltonians in 2d CFT.
Of particular interest for us is interested in the interplay between
inhomogeneous deformation and quantum information scrambling. 
With these setups, we discussed the operator entanglement,
the recovery of quantum information, and the dynamics of B.H.-like excitations.
As mentioned in Ref.\ \cite{Goto:2021sqx},
these inhomogeneously-deformed Hamiltonians may be engineered both in digital and
analog quantum simulators, such as cold atoms and Rydberg atoms.
Simulating our quench processes in these systems opens up the possibility of
studying quantum aspects of black holes in the lab. 
In particular, from the findings in our paper,
among others, 
it would be interesting to look into the following aspects:
\begin{itemize}

\item {\it Quantum black hole:}
  The $t_0$-dependence of the correlation function may be described by the propagation of the B.H.-like excitation (see Section \ref{sec:worm-hole-system3}). 
  In the frame where one of the B.H.-like excitations is stationary, a local operator falls into and is radiated from this excitation.
  As in \cite{Hayden_2007}, this excitation has the almost same amount of entropy as the black hole and its interior may have a strong scrambling effect. 
  Therefore, if we can create these excitations in the experimental systems, then we may simulate the dynamics of black holes in the laboratories.

\item {\it Genuine tripartite entanglement:} Let us consider the application of the genuine tripartite entanglement obtained in this paper. 
  In $2$d holographic CFTs, for the large $t_1$-regime, the local BMI is approximately zero, while the global BMI can be $\mathcal{O}(\f{1}{\epsilon})$ in the certain $t_1$-intervals (see \ref{sec:tripartite-entanglement}).
  One possible interpretation for this entanglement property of the steady state
  is that in the $t_1$-regime where only $I_{A,B_1\cup B_2}$ is
  $\mathcal{O}(\f{1}{\epsilon})$,
  three persons belonging to $A$, $B_1$, and $B_2$, respectively, may be able to share the quantum information, while only two of them may not. 
  In the other words, without the cooperation of these three people, they may never get the quantum information correctly. 
  This entanglement property may be applied to secure quantum communications.
\end{itemize}


Finally, we conclude by listings some of the future directions:
\begin{itemize}
\item {\it Multipartite entanglement}: It would be interesting to create a system where the local MI is effectively zero, while the global BMI that is shared by $n(>3)$-parties is $\mathcal{O}(\f{1}{\epsilon})$. 
  If the number of fixed points increases \cite{Han_2020}, then the number of parties sharing the global BMI might increase.
  
\item{\it Quantum scars}: In this paper, we discovered the systems which are not evolved to the typical state with a 2$d$ homogeneous holographic Hamiltonian. These states may be interpreted as quantum scar states. It would be interesting to establish the relationship between these states considered in this paper and the quantum scar states \cite{2018PhRvB..98o5134T,2018arXiv180609624M,2019PhRvL.122q3401L,2018arXiv180701815H,2018NatPh..14..745T,2021arXiv210803460P,2022JHEP...12..163D,2022arXiv221103630C,2022arXiv221205962L}.
\end{itemize}

\section*{Acknowledgements}

K.G.~is supported by JSPS KAKENHI Grant-in-Aid for Early-Career Scientists (21K13930) and Research Fellowships of Japan Society for the Promotion of Science for Young Scientists (22J00663).
M.N.~is supported by funds from the University of Chinese Academy of Sciences (UCAS), funds from the Kavli Institute for Theoretical Sciences (KITS).
S.R.~is supported by the National Science Foundation under 
Award No.\ DMR-2001181, and by a Simons Investigator Grant from
the Simons Foundation (Award No.~566116).
This work is supported by
the Gordon and Betty Moore Foundation through Grant
GBMF8685 toward the Princeton theory program. 
K.T.~is supported by JSPS KAKENHI Grant No.~21K13920 and MEXT KAKENHI Grant No.~22H05265. This material is based upon work supported by the National Science Foundation under Grant No. NSF-DMR 2018358 and by an
appointment to the YST Program at the APCTP through the Science and Technology Promotion Fund and Lottery Fund of the Korean Government, as well as the Korean Local Governments -
Gyeongsangbuk-do Province and Pohang City (MT).

\appendix
\section{Evolution of operators induced by $U^1_{E,\alpha}e^{-\epsilon H}$ \label{App:Evo_o_Ope}}

The Euclidean time evolution operators 
considered in the main text and Appendices 
are
(see 
\eqref{Euclidean time evolutions} and 
\eqref{eq:evolvedstatein3-CSD}) 
\be
\begin{split}
    U^1_{E}=\begin{cases}
    e^{-H^1_{\text{M\"obius}}\tau_1} 
    & \alpha=0
    \\
    e^{-H^1_{\text{SSD}}\tau_1}e^{-H^1_0\tau_0} & 
    \alpha =1
    \\ 
    e^{-H^1_0\tau_0}e^{-H^1_{\text{SSD}}\tau_1} 
    & 
    \alpha =2
    \\
    e^{-H^1_{\text{CSD}}\tau_2}e^{-H^1_{\text{SSD}}\tau_1} & 
    \alpha=3
    \end{cases},
    \qquad
     \tilde{U}^1_{E}=\begin{cases}
    e^{H^1_{\text{M\"obius}}\tau_1} & 
    \alpha=0
    \\
    e^{H^1_0\tau_0}e^{H^1_{\text{SSD}}\tau_1} & 
    \alpha =1
    \\ 
    e^{H^1_{\text{SSD}}\tau_1}e^{H^1_0\tau_0} & 
    \alpha =2
    \\
    e^{H^1_{\text{SSD}}\tau_1}e^{H^1_{\text{CSD}}\tau_2} & 
    \alpha=3
    \end{cases}.
\end{split}
\ee
In these appendices, we use the index $\alpha=0,1,2,3$ to distinguish 
these cases.
The new complex variables $(w^{\text{New},\alpha}_{x, \epsilon},\overline{w}^{\text{New},\alpha}_{x, \epsilon})$ in (\ref{transformation_of_operator}) are given by 
\be \label{new_coordinate_in_euclidean}
\begin{split}
&w^{\text{New},0}_{x, \epsilon} =   \epsilon+\f{L}{2\pi} \log{
\left[ \f{[(1-\lambda_1)\cosh{(2\theta)-(\lambda_1+1)}]z_{x}+(\lambda_1-1)\sinh{(2\theta)}}{(1-\lambda_1)\sinh{(2\theta)}z_{x}+[(\lambda_1-1)\cosh{(2\theta)}-(\lambda_1+1)]}\right] },
\\
&\bar{w}^{\text{New},0}_{x, \epsilon} =  \epsilon+ \f{L}{2\pi} \log{\left[ \f{[(1-\lambda_1)\cosh{(2\theta)-(\lambda_1+1)}]\overline{z}_{x}+(\lambda_1-1)\sinh{(2\theta)}}{(1-\lambda_1)\sinh{(2\theta)}\overline{z}_{x}+[(\lambda_1-1)\cosh{(2\theta)}-(\lambda_1+1)]}\right] },
\\
&w^{\text{New},1}_{x, \epsilon}=\epsilon+\tau_0+\f{L}{2\pi}\log{\left[\f{\pi \tau_1(1-z_x)-Lz_x}{\pi \tau_1(1-z_x)-L}\right]},
\\
&
\overline{w}^{\text{New},1}_{x, \epsilon}=\epsilon+\tau_0+\f{L}{2\pi}\log{\left[\f{\pi \tau_1(1-\overline{z}_x)-L\overline{z}_x}{\pi \tau_1(1-\overline{z}_x)-L}\right]},
\\
&w^{\text{New},2}_{x, \epsilon}=\epsilon+\f{L}{2\pi}\log{\left[\f{\pi \tau_1(1-e^{\f{2\pi \tau_0}{L}}z_x)-e^{\f{2\pi \tau_0}{L}}Lz_x}{\pi \tau_1(1-e^{\f{2\pi \tau_0}{L}}z_x)-L}\right]},
\\
&
\overline{w}^{\text{New},2}_{x, \epsilon}=\epsilon+\f{L}{2\pi}\log{\left[\f{\pi \tau_1(1-e^{\f{2\pi \tau_0}{L}}\overline{z}_x)-e^{\f{2\pi \tau_0}{L}}L\overline{z}_x}{\pi \tau_1(1-e^{\f{2\pi \tau_0}{L}}\overline{z}_x)-L}\right]},\\
&w^{\text{New},3}_{x, \epsilon} =\epsilon +\f{L}{2\pi}\log{\left[\frac{z_x \left(L^2+\pi  L (\tau_1+\tau_2)+2 \pi ^2 \tau_1 \tau_2\right)+\pi  (L (\tau_2-\tau_1)+2 \pi  \tau_1 \tau_2)}{L^2-\pi  L \tau_1-\pi  L \tau_2+\pi  z_x (L (\tau_1-\tau_2)+2 \pi  \tau_1 \tau_2)+2 \pi ^2 \tau_1 \tau_2}\right]},\\
&\overline{w}^{\text{New},3}_{x, \epsilon}=\epsilon +\f{L}{2\pi}\log{\left[\frac{\overline{z}_x \left(L^2+\pi  L (\tau_1+\tau_2)+2 \pi ^2 \tau_1 \tau_2\right)+\pi  (L (\tau_2-\tau_1)+2 \pi  \tau_1 \tau_2)}{L^2-\pi  L \tau_1-\pi  L \tau_2+\pi  \overline{z}_x (L (\tau_1-\tau_2)+2 \pi  \tau_1 \tau_2)+2 \pi ^2 \tau_1 \tau_2}\right]},
\end{split}
\ee
where the variables and parameters, $z$, $\overline{z}$, and $\lambda_{1}$, are defined by 
\be
\begin{split}
&z_x=e^{\f{2\pi w_x}{L}},
\quad \overline{z}_x=e^{\f{2\pi \overline{w}_x}{L}},  
\quad 
\lambda_{1} = \exp{\left(\f{2\pi \tau_1}{L \cosh{(2\theta)}}\right)}.
\end{split}
\ee

\subsection{Real time evolution of operators \label{App:Evo_o_Ope_ana}}

After the analytic continuation,
$
\tau_{i=0,1,2}=it_{i=0,1,2},
$
only the imaginary parts of $w^{\text{New},\alpha}_{x, \epsilon}$ and $\overline{w}^{\text{New},\alpha}_{x, \epsilon}$ depend on $t_{i=0,1,2}$.
The dependence of $w^{\text{New},\alpha}_{x,\epsilon}$ and $\overline{w}^{\text{New},\alpha}_{x,\epsilon}$ on $t_{i=0,1,2}$ is given by
\be \label{transformation_of_cc}
\begin{split}
&w^{\text{New},0}_{x, \epsilon} =\epsilon +i\f{L \varphi_{x,0}}{\pi}, 
\quad
\overline{w}^{\text{New},0}_{x, \epsilon}=\epsilon +i\f{L \overline{\varphi}_{x,0}}{\pi},\\
&w^{\text{New},1}_{x,\epsilon}=\epsilon+it_0+i \f{L\varphi_{x,1}}{\pi},
\quad
\overline{w}^{\text{New},1}_{x,\epsilon}=\epsilon+it_0+\f{L\overline{\varphi}_{x,1}}{\pi},\\
&w^{\text{New},2}_{x,\epsilon}=\epsilon+i \f{L\varphi_{x,2}}{\pi},
\quad 
\overline{w}^{\text{New},2}_{x,\epsilon}=\epsilon+i\f{L\overline{\varphi}_{x,2}}{\pi}, 
\\
&
w^{\text{New},3}_{x,\epsilon}=\epsilon +i \f{L \varphi_{x,3}}{\pi}, \quad 
\overline{w}^{\text{New},3}_{x,\epsilon}=\epsilon +i \f{L \overline{\varphi}_{x,3}}{\pi},
\end{split}
\ee
where the variables, $\varphi_{x,\alpha}$, $\overline{\varphi}_{x,\alpha}$, $r_{x,\alpha}$, and $\overline{r}_{x,\alpha}$, are defined by
\be
\begin{split}
&r_{x,0}=\Bigg{[}\left(-\cos{\left(\f{\pi t_1}{L\cosh{2\theta}}\right)}\cos{\left(\f{\pi x}{L}\right)}+\sin{\left(\f{\pi t_1}{L\cosh{2\theta}}\right)}\sin{\left(\f{\pi x}{L}\right)}e^{2\theta}\right)^2\\
&~~~~+\left(\cos{\left(\f{\pi t_1}{L\cosh{2\theta}}\right)}\sin{\left(\f{\pi x}{L}\right)}+\sin{\left(\f{\pi t_1}{L\cosh{2\theta}}\right)}\cos{\left(\f{\pi x}{L}\right)}e^{-2\theta}\right)^2\Bigg{]}^{\f{1}{2}},\\
&\cos{\varphi_{x,0}}=\f{\cos{\left(\f{\pi t_1}{L\cosh{2\theta}}\right)}\cos{\left(\f{\pi x}{L}\right)}-\sin{\left(\f{\pi t_1}{L\cosh{2\theta}}\right)}\sin{\left(\f{\pi x}{L}\right)}e^{2\theta}}{r_{x,0}},\\
&\sin{\varphi_{x,0}}=\f{\cos{\left(\f{\pi t_1}{L\cosh{2\theta}}\right)}\sin{\left(\f{\pi x}{L}\right)}+\sin{\left(\f{\pi t_1}{L\cosh{2\theta}}\right)}\cos{\left(\f{\pi x}{L}\right)}e^{-2\theta}}{r_{x,0}},\\
&\overline{r}_{x,0}=\Bigg{[}\left(\cos{\left(\f{\pi t_1}{L\cosh{2\theta}}\right)}\cos{\left(\f{\pi x}{L}\right)}+\sin{\left(\f{\pi t_1}{L\cosh{2\theta}}\right)}\sin{\left(\f{\pi x}{L}\right)}e^{2\theta}\right)^2\\
&~~~~+\left(\cos{\left(\f{\pi t_1}{L\cosh{2\theta}}\right)}\sin{\left(\f{\pi x}{L}\right)}-\sin{\left(\f{\pi t_1}{L\cosh{2\theta}}\right)}\cos{\left(\f{\pi x}{L}\right)}e^{-2\theta}\right)^2\Bigg{]}^{\f{1}{2}},\\
&\cos{\overline{\varphi}_{x,0}}=\f{\cos{\left(\f{\pi t_1}{L\cosh{2\theta}}\right)}\cos{\left(\f{\pi x}{L}\right)}+\sin{\left(\f{\pi t_1}{L\cosh{2\theta}}\right)}\sin{\left(\f{\pi x}{L}\right)}e^{2\theta}}{\overline{r}_{x,0}},\\
&\sin{\overline{\varphi}_{x,0}}=\f{-\cos{\left(\f{\pi t_1}{L\cosh{2\theta}}\right)}\sin{\left(\f{\pi x}{L}\right)}+\sin{\left(\f{\pi t_1}{L\cosh{2\theta}}\right)}\cos{\left(\f{\pi x}{L}\right)}e^{-2\theta}}{\overline{r}_{x,0}},\\
\end{split}
\ee
\be
\begin{split}
&r_{x,2}=\sqrt{4\pi^2t_1^2\sin^2{\left(\f{\pi(t_0+x)}{L}\right)}-4\pi Lt_1\sin{\left(\f{\pi(t_0+x)}{L}\right)}\cos{\left(\f{\pi(t_0+x)}{L}\right)}+L^2},\\
&\cos{\varphi_{x,2}}= \f{-2\pi t_1 \sin{\left(\f{\pi(t_0+x)}{L}\right)}+L\cos{\left(\f{\pi(t_0+x)}{L}\right)}}{r_{x,2}},\sin{\varphi_{x,2}}=\f{L\sin{\left(\f{\pi(t_0+x)}{L}\right)}}{r_{x,2}}.\\
&\overline{r}_{x,2}=\sqrt{4\pi^2t_1^2\sin^2{\left(\f{\pi(t_0-x)}{L}\right)}-4\pi Lt_1\sin{\left(\f{\pi(t_0-x)}{L}\right)}\cos{\left(\f{\pi(t_0-x)}{L}\right)}+L^2},\\
&\cos{\overline{\varphi}_{x,2}}=- \f{2\pi t_1 \sin{\left(\f{\pi(t_0-x)}{L}\right)}-L\cos{\left(\f{\pi(t_0-x)}{L}\right)}}{\overline{r}^1_x},\sin{\overline{\varphi}_{x,2}}=\f{L\sin{\left(\f{\pi(t_0-x)}{L}\right)}}{\overline{r}_{x,2}},\\
&\varphi_{x,1}=\varphi_{x,2}|_{t_0=0},~~\overline{\varphi}_{x,1}=\overline{\varphi}_{x,2}|_{t_0=0},\\
\end{split}
\ee
\be
\begin{split}
&r_{x,3}=\bigg{[}\left(\left(L^2-4 \pi ^2 t_1 t_2\right) \cos \left(\frac{\pi  x}{L}\right)-2 \pi  L t_1 \sin \left(\frac{\pi  x}{L}\right)\right)^2+\left(L^2 \sin \left(\frac{\pi  x}{L}\right)+2 \pi  L t_2 \cos \left(\frac{\pi  x}{L}\right)\right)^2\bigg{]}^{\f{1}{2}},\\
&\overline{r}_{x,3}=\bigg{[}\left(\left(L^2-4 \pi ^2 t_1 t_2\right) \cos \left(\frac{\pi  x}{L}\right)+2 \pi  L t_1 \sin \left(\frac{\pi  x}{L}\right)\right)^2+\left(-L^2 \sin \left(\frac{\pi  x}{L}\right)+2 \pi  L t_2 \cos \left(\frac{\pi  x}{L}\right)\right)^2
\bigg{]}^{\f{1}{2}},\\
&\cos{\varphi_{x,3}}=\f{1}{r_{x,3}}\left[(L^2-4\pi^2 t_1 t_2)\cos{\left(\f{\pi x}{L}\right)}-2\pi L t_1 \sin{\left(\f{\pi x}{L}\right)}\right],\\
&\sin{\varphi_{x,3}}=\f{2\pi L\cos{\left(\f{\pi x}{L}\right)}}{r_{x,3}}\left[t_2+\f{L}{2\pi}\tan{\left(\f{\pi x}{L}\right)}\right],\\
&\cos{\overline{\varphi}_{x,3}}=\f{1}{\overline{r}_{x,3}}\left[(L^2-4\pi^2 t_1 t_2)\cos{\left(\f{\pi x}{L}\right)}+2\pi L t_1 \sin{\left(\f{\pi x}{L}\right)}\right],\\
&\sin{\overline{\varphi}_{x,3}}=\f{2\pi L\cos{\left(\f{\pi x}{L}\right)}}{\overline{r}_{x,3}}\left[t_2-\f{L}{2\pi}\tan{\left(\f{\pi x}{L}\right)}\right].
\end{split}
\ee

\section{The details of calculations and results in $2$d holographic CFTs \label{App:hol_CFT}}
Let us present the details of the calculations and results in $2$d holographic CFTs.

\subsection{Non-universal piece of OEE in $2$d holographic CFTs \label{App:NUP_hol_CFT}}
We now present the details of the non-universal pieces, $S_{\text{dis}}$ and $S_{\text{con}}$, of the entanglement entropy.
Let us concentrate on $S_{\text{dis}}$.
This non-universal piece, $S_{\text{dis}}$, is given by
\be
\begin{split}
    &S_{\text{dis}}=\text{Min}\Bigg{[}
    S^{1}_{\text{dis}}, S^{2,\pm}_{\text{dis}}, S^{3,\pm}_{\text{dis}}, S^{4,\pm}_{\text{dis}}
    \Bigg{]},\\
\end{split}
\ee
where $\tilde{S}^i_{\text{dis}}$ are defined by
\be
\begin{split}
    \tilde{S}^{1}_{\text{dis}}&=\f{c}{6}\log{\left[\left|\sin{\left[\f{\pi}{2\epsilon}\left(w^{\text{New}, \alpha}_{Y_1, \epsilon}-w^{\text{New}, \alpha}_{Y_2, \epsilon}\right)\right]}\right|^2\left|\sin{\left[\f{\pi}{2\epsilon}\left(w_{X_1}-w_{X_1}\right)\right]}\right|^2\right]},\\
    \tilde{S}^{2,\pm}_{\text{dis}}&=\f{c}{6}\log{\left[\left|\sin{\left[\f{\pi}{2\epsilon}\left(\pm iL-\left(w^{\text{New}, \alpha}_{Y_1, \epsilon}-w^{\text{New}, \alpha}_{Y_2, \epsilon}\right)\right)\right]}\right|^2\left|\sin{\left[\f{\pi}{2\epsilon}\left( \pm iL-\left(w_{X_1}-w_{X_1}\right)\right)\right]}\right|^2\right]},\\
    \tilde{S}^{3,\pm}_{\text{dis}}&=\f{c}{6}\log{\left[\left|\sin{\left[\f{\pi}{2\epsilon}\left(w^{\text{New}, \alpha}_{Y_1, \epsilon}-w^{\text{New}, \alpha}_{Y_2, \epsilon}\right)\right]}\right|^2\left|\sin{\left[\f{\pi}{2\epsilon}\left( \pm iL-\left(w_{X_1}-w_{X_1}\right)\right)\right]}\right|^2\right]},\\
    \tilde{S}^{4,\pm}_{\text{dis}}&=\f{c}{6}\log{\left[\left|\sin{\left[\f{\pi}{2\epsilon}\left(\pm iL-\left(w^{\text{New}, \alpha}_{Y_1, \epsilon}-w^{\text{New}, \alpha}_{Y_2, \epsilon}\right)\right)\right]}\right|^2\left|\sin{\left[\f{\pi}{2\epsilon} \left(w_{X_1}-w_{X_1}\right)\right]}\right|^2\right]}
\end{split}
\ee

\if[0]
\be
\begin{split}
    &\tilde{S}^1_{\text{dis}}=\text{Min}\Bigg{[}\f{c}{6}\log{\left[\left|\sin{\left[\f{\pi}{2\epsilon}\left(iL-\left(w^{\text{New}, \alpha}_{Y_1, \epsilon}-w^{\text{New}, \alpha}_{Y_2, \epsilon}\right)\right)\right]}\right|^2\left|\sin{\left[\f{\pi}{2\epsilon}\left(iL-\left(w_{X_1}-w_{X_1}\right)\right)\right]}\right|^2\right]},\\
    &\f{c}{6}\log{\left[\left|\sin{\left[\f{\pi}{2\epsilon}\left(w^{\text{New}, \alpha}_{Y_1, \epsilon}-w^{\text{New}, \alpha}_{Y_2, \epsilon}\right)\right]}\right|^2\left|\sin{\left[\f{\pi}{2\epsilon}\left(w_{X_1}-w_{X_1}\right)\right]}\right|^2\right]}\Bigg{]} \\
    &\tilde{S}^2_{\text{dis}}=\text{Min}\Bigg{[}\f{c}{6}\log{\left[\left|\sin{\left[\f{\pi}{2\epsilon}\left(iL-\left(w^{\text{New}, \alpha}_{Y_1, \epsilon}-w^{\text{New}, \alpha}_{Y_2, \epsilon}\right)\right)\right]}\right|^2\left|\sin{\left[\f{\pi}{2\epsilon}\left(w_{X_1}-w_{X_1}\right)\right]}\right|^2\right]}.\\
\end{split}
\ee
\fi
Then, let us turn to $S_{\text{con}}$. This contribution from the geodesics connecting the endpoints of the subsystems on the different Euclidean time slices is given by
\be
\begin{split}   
    &S_{\text{con}}=\text{Min}\left[\tilde{S}^{1}_{\text{con}},\tilde{S}^{2,\pm}_{\text{con}},\tilde{S}^{3,\pm}_{\text{con}},\tilde{S}^{4,\pm}_{\text{con}}\right]
\end{split}
\ee
where $\tilde{S}^{i}_{\text{con}}$ are defined by
\be
\begin{split}
    &\tilde{S}^{1}_{\text{con}}=\f{c}{6}\log{\left[\left|\sin{\left[\f{\pi}{2\epsilon}\left(w^{\text{New}, \alpha}_{Y_1, \epsilon}-w_{X_1}\right)\right]}\right|^2\left|\sin{\left[\f{\pi}{2\epsilon}\left(w^{\text{New}, \alpha}_{Y_2, \epsilon}-w_{X_2}\right)\right]}\right|^2\right]},\\
    &\tilde{S}^{2,\pm}_{\text{con}}=\f{c}{6}\log{\left[\left|\sin{\left[\f{\pi}{2\epsilon}\left(\pm iL-\left(w^{\text{New}, \alpha}_{Y_1, \epsilon}-w_{X_1}\right)\right)\right]}\right|^2\left|\sin{\left[\f{\pi}{2\epsilon}\left(\pm iL-\left(w^{\text{New}, \alpha}_{Y_2, \epsilon}-w_{X_2}\right)\right)\right]}\right|^2\right]},\\
     &\tilde{S}^{3,\pm}_{\text{con}}= \f{c}{6}\log{\left[\left|\sin{\left[\f{\pi}{2\epsilon}\left(w^{\text{New}, \alpha}_{Y_1, \epsilon}-w_{X_1}\right)\right]}\right|^2\left|\sin{\left[\f{\pi}{2\epsilon}\left(\pm iL-\left(w^{\text{New}, \alpha}_{Y_2, \epsilon}-w_{X_2}\right)\right)\right]}\right|^2\right]},\\
     &\tilde{S}^{4,\pm}_{\text{con}}=\f{c}{6}\log{\left[\left|\sin{\left[\f{\pi}{2\epsilon}\left(\pm iL-\left(w^{\text{New}, \alpha}_{Y_1, \epsilon}-w_{X_1}\right)\right)\right]}\right|^2\left|\sin{\left[\f{\pi}{2\epsilon}\left(w^{\text{New}, \alpha}_{Y_2, \epsilon}-w_{X_2}\right)\right]}\right|^2\right]}.
\end{split}
\ee 




\subsection{The definition of $\theta_C$ \label{Section:theta-critical}}
Here, we describe the definition of $\theta_C$ that is introduced in Section \ref{Section:theta-position-dependence}.
Let $B$ be a subsystem including $X^{1}_f$ of $\mathcal{H}_1$, and also let $A$ be a subsystem including the origin of $\mathcal{H}_2$.
Furthermore, let us assume that $S_{\text{dis}}$ for the small $t_1$ is given by
\be \label{eq:sdis-for-critical}
S_{\text{dis}}= \f{c \pi}{6\epsilon}\left[2L-\left(X^{\text{New},\alpha=1}_{Y_1,\epsilon}-X^{\text{New},\alpha=1}_{Y_2,\epsilon}+(X_1-X_2)\right)\right]
    \ee
The time for (\ref{eq:sdis-for-critical}) to be maximized is determined by $\partial_{t_1}\left[-\left(X^{\text{New},\alpha=1}_{Y_1,\epsilon}-X^{\text{New},\alpha=1}_{Y_2,\epsilon}\right)\right]=0$. 
Let $t_{1,\text{Max}}$ denote this time, and this time depends on $\theta$, $Y_1$, $Y_2$, and $L$.
Let us define $\theta_C$ as the $\theta$ satisfying $\left(X^{\text{New},\alpha=1}_{Y_1,\epsilon}-X^{\text{New},\alpha=1}_{Y_2,\epsilon}\right)=L-(X_1-X_2)=l_A$ at $t=t_{1,\text{Max}}$.

\section{The entanglement dynamics for (\ref{eq:evolvedstatein3-CSD}) \label{App:system4}}

\subsection{The $t_2$-dependence of entanglement entropy \label{App:EE-system4}}
\begin{figure}[htbp]
    \begin{tabular}{ccc}
      \begin{minipage}[t]{0.3\hsize}
        \centering
        \includegraphics[keepaspectratio, scale=0.25]
        {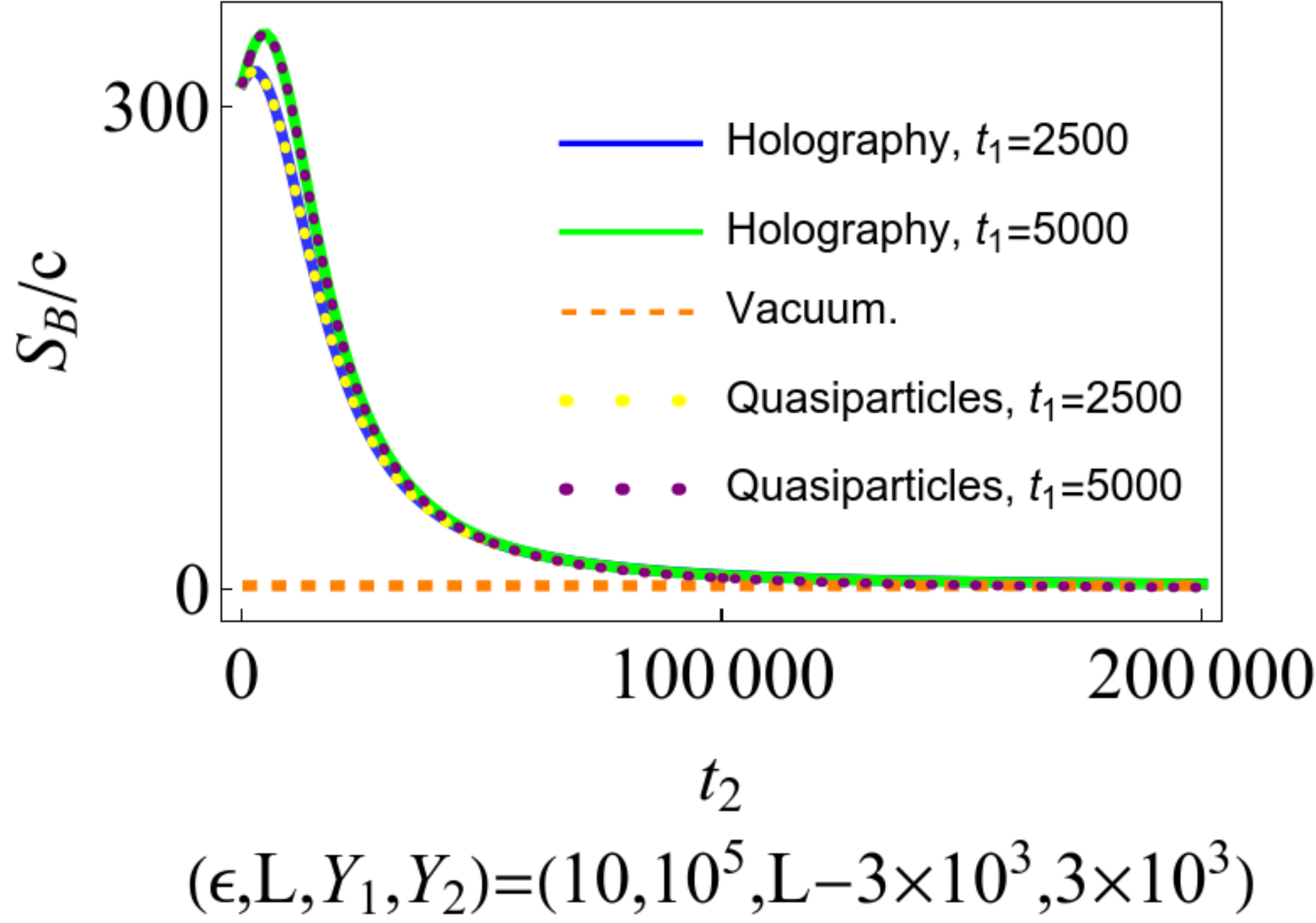}
    
    $(a_1)$ For the small $t_1$-regime. 
      \end{minipage} & 
     
     \begin{minipage}[t]{0.33\hsize}
        \centering
        \includegraphics[keepaspectratio,scale=0.28]
        {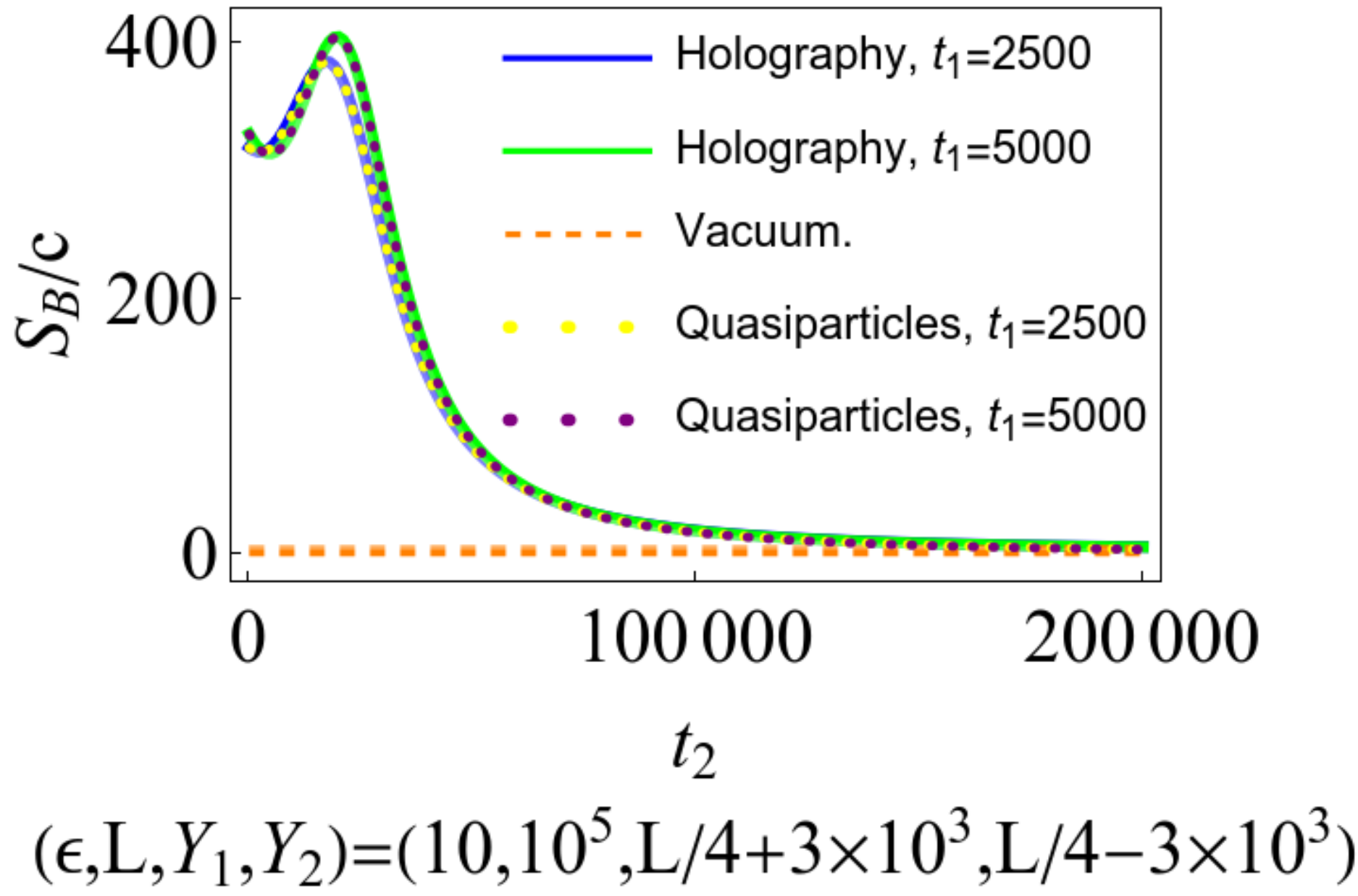}
     
     $(b_1)$ For the small $t_1$-regime. 
      \end{minipage} 
      
      \begin{minipage}[t]{0.33\hsize}
        \centering
        \includegraphics[keepaspectratio,scale=0.29]
        {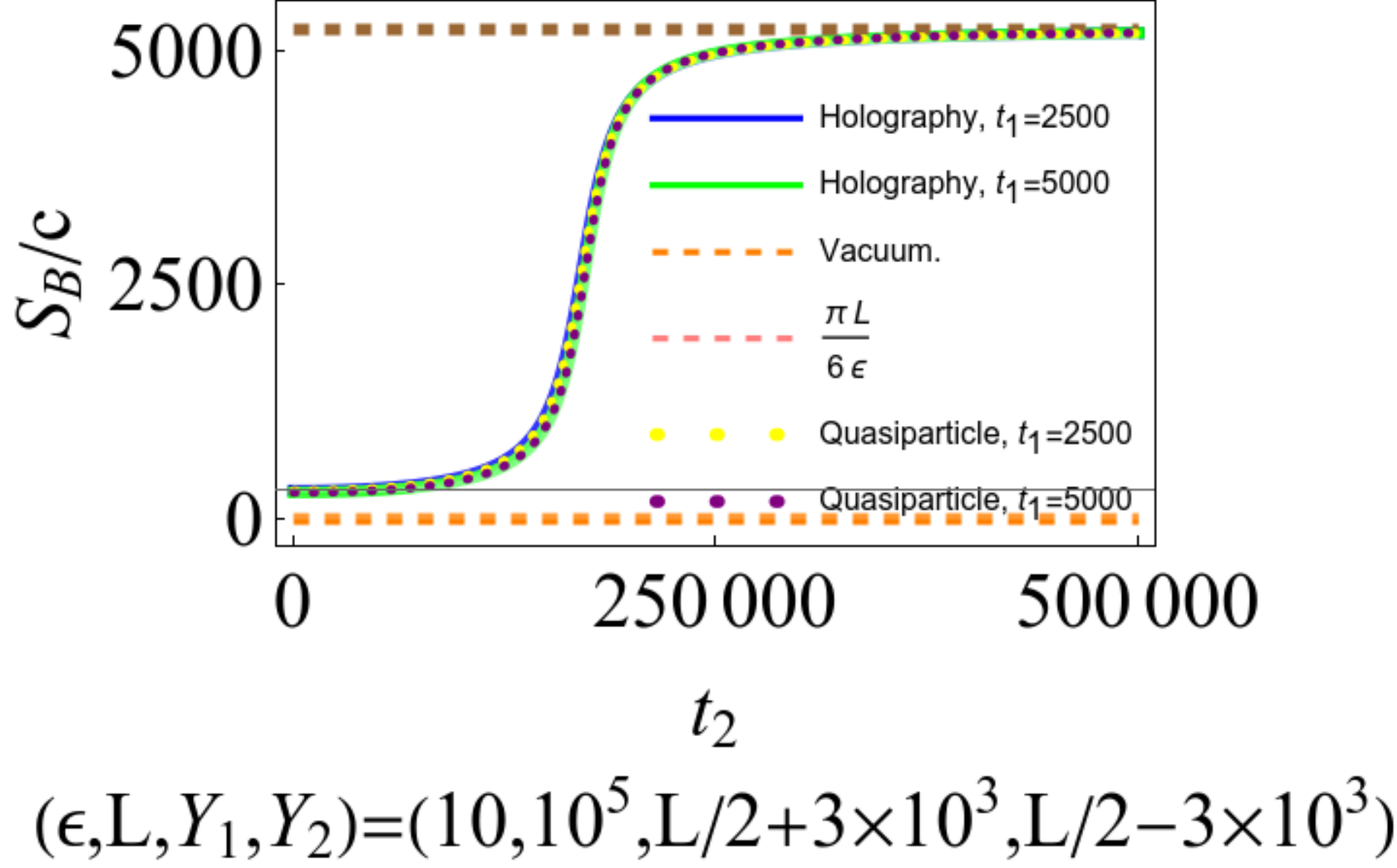}
     
      $(c_1)$ For the small $t_1$-regime. 
      \end{minipage}\\
      \begin{minipage}[t]{0.33\hsize}
        \centering
        \includegraphics[keepaspectratio, scale=0.24]{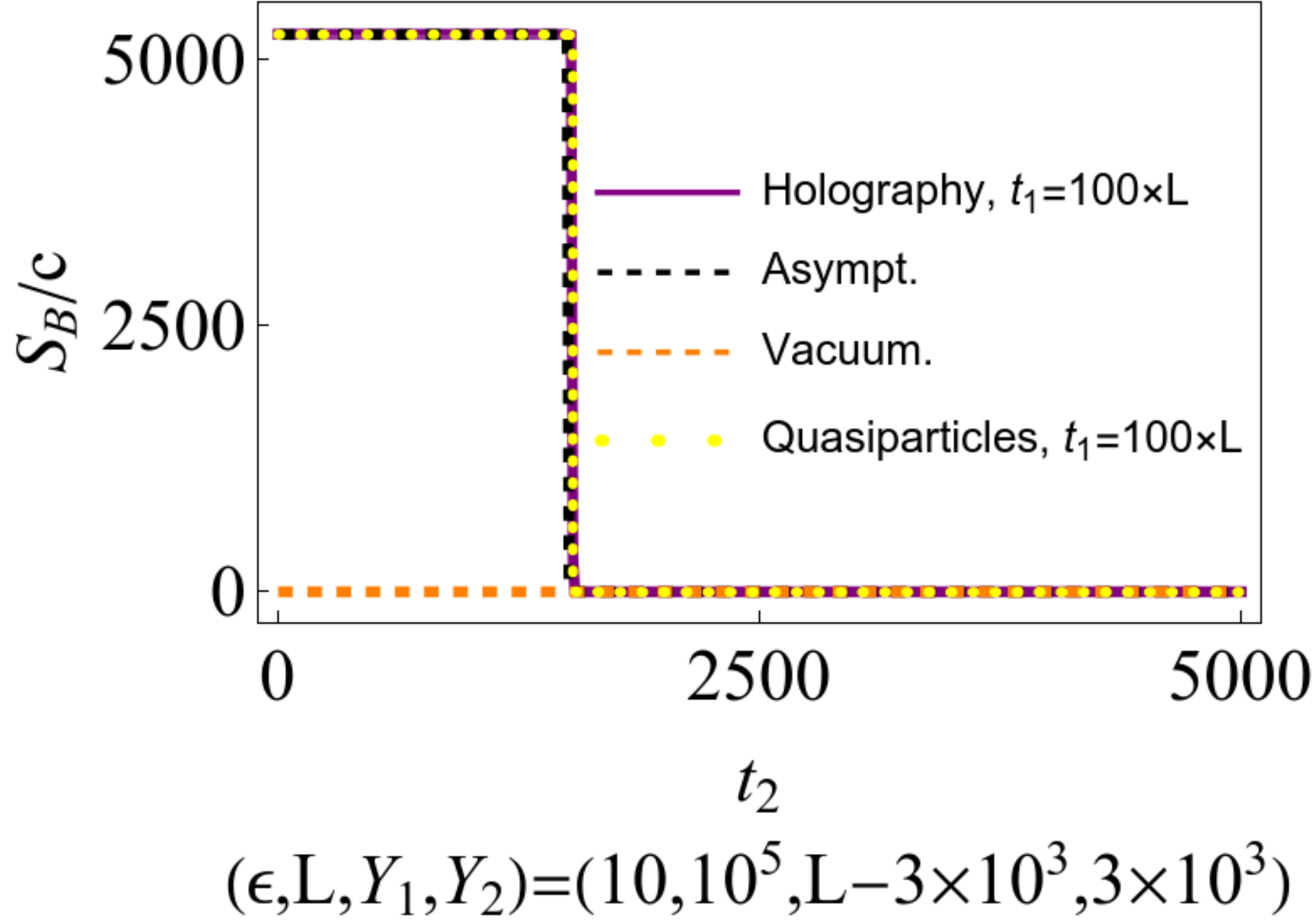}
   
    $(a_2)$ For the large $t_1$-regime. 
      \end{minipage} & 
     
     \begin{minipage}[t]{0.33\hsize}
        \centering
        \includegraphics[keepaspectratio, scale=0.30]{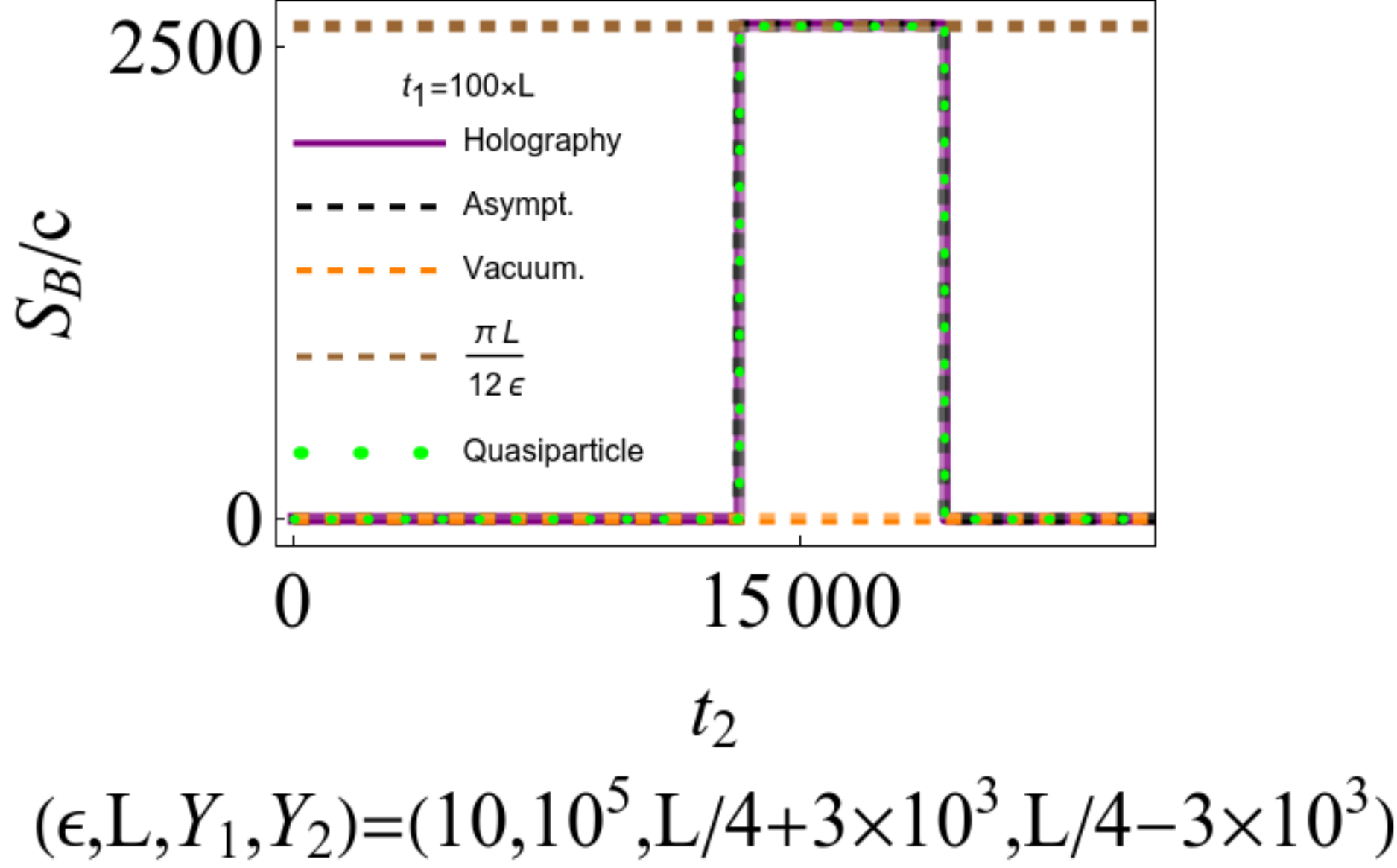}
     
     $(b_2)$ For the large $t_1$-regime. 
      \end{minipage} 
      
      \begin{minipage}[t]{0.33\hsize}
        \centering
        \includegraphics[keepaspectratio, scale=0.30]{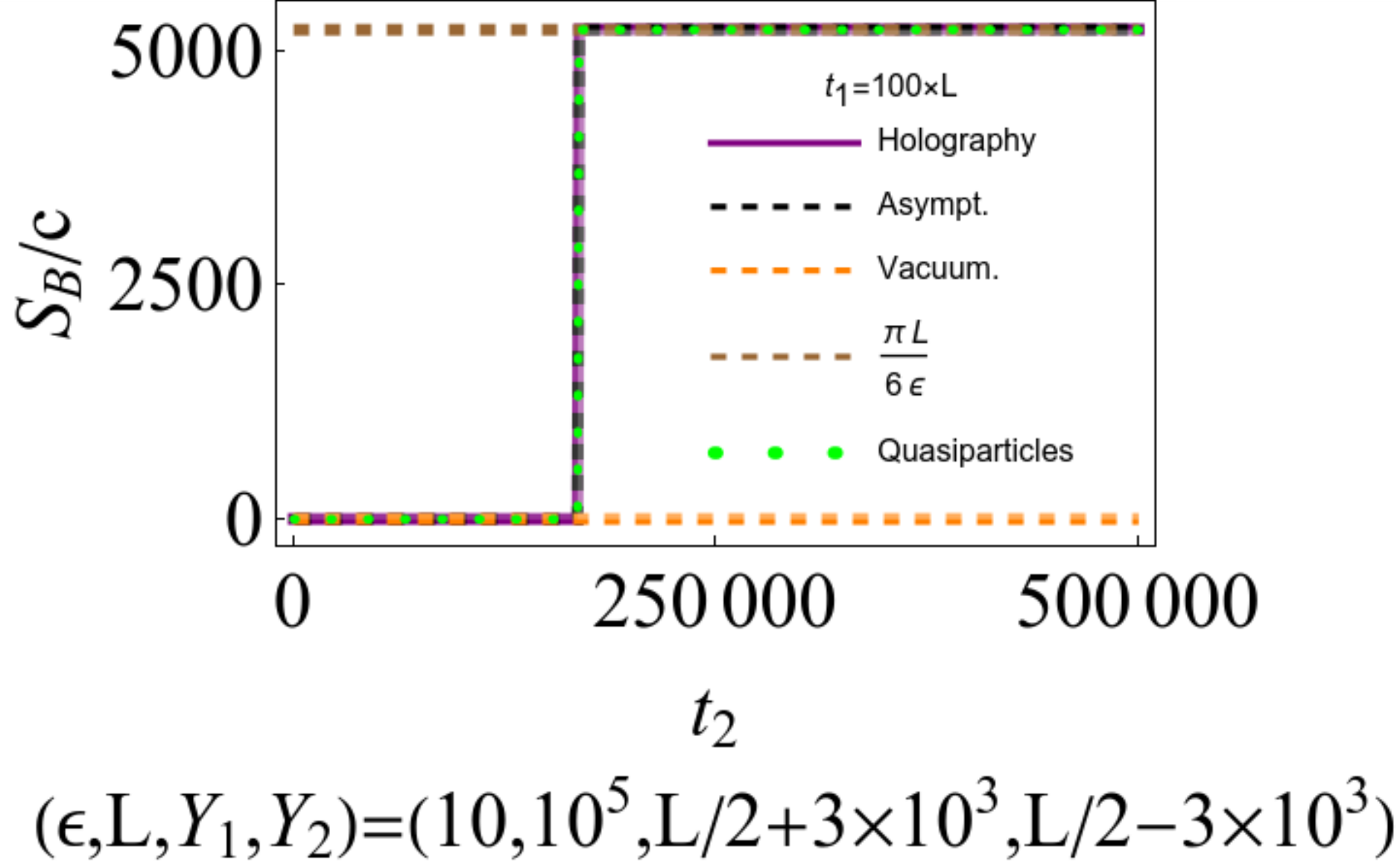}
     
      $(c_2)$ For the large $t_1$-regime. 
      \end{minipage}\\
    \end{tabular}
    \caption{The entanglement entropy, $S_B$, for various $t_1$ as the function of $t_2$. The panels, ($a_{i=1,2}$) show the $t_2$-dependence of $S_B$ for case (a), ($b_i$) show the one for (b), and ($c_i$) show the one for (c). In the top panels, we show the $t_2$-dependence of $S_B$ for the small $t_1$-regime, while in the bottom panels, we show the $t_2$-dependence of $S_B$ for the large $t_1$-regime. The black dashed line illustrates the asymptotic behavior of $S_B$ in (\ref{eq:asybh-single-CSD}) in the large $t_1$ limit. The orange, brown, and dashed lines illustrate the entanglement entropy for the vacuum state, thermal entropy, and half of it, respectively.}
    \label{Fig:t2-dependence-SB-in-3}
  \end{figure}

Let us consider the state (\ref{eq:evolvedstatein3-CSD}).
We report the $t_2$-dependence of entanglement entropy of (\ref{eq:evolvedstatein3-CSD}). for the subsystems considered in Section \ref{Section:EE}.
In Fig. \ref{Fig:t2-dependence-SB-in-3}, we depict $S_B$ for various $t_1$ as a function of $t_2$.
In the $t_1$-limit, the $t_2$-dependence of $S_B$ is approximated by 
\be \label{eq:asybh-single-CSD}
\begin{split}
     \text{If}~ x=X^1_f \in B, & S_B\approx\begin{cases}
    \f{c\pi L}{6\epsilon} & \text{for}~ \f{L}{2\pi}\left|\tan{\left(\f{\pi Y_2}{L}\right)}\right|>t_2>0 \\
    \f{c\pi L}{12\epsilon} & \text{for}~ \f{L}{2\pi}\left|\tan{\left(\f{\pi Y_1}{L}\right)}\right|>t_2>\f{L}{2\pi}\left|\tan{\left(\f{\pi Y_2}{L}\right)}\right|\\
    \f{c}{3}\log{\left[\sin{\left[\f{\pi(Y_1-Y_2)}{L}\right]}\right]}  & \text{for}~  t_2>\left|\f{L}{2\pi}\tan{\left(\f{\pi Y_1}{L}\right)}\right|
    \end{cases},\\
     \text{If}~ \f{L}{2}>Y_1>x>Y_2>0, & S_B\approx\begin{cases}
    \f{c}{3}\log{\left[\sin{\left[\f{\pi(Y_1-Y_2)}{L}\right]}\right]}  & \text{for}~ \f{L}{2\pi}\left|\tan{\left(\f{\pi Y_2}{L}\right)}\right|>t_2>0 \\
    \f{c\pi L}{12\epsilon} & \text{for}~ \f{L}{2\pi}\left|\tan{\left(\f{\pi Y_1}{L}\right)}\right|>t_2>\f{L}{2\pi}\left|\tan{\left(\f{\pi Y_2}{L}\right)}\right|\\
    \f{c}{3}\log{\left[\sin{\left[\f{\pi(Y_1-Y_2)}{L}\right]}\right]}  & \text{for}~  t_2>\f{L}{2\pi}\left|\tan{\left(\f{\pi Y_1}{L}\right)}\right|
    \end{cases}, \\
     \text{If}~ x=X^2_f \in B, & S_B\approx\begin{cases}
    \f{c}{3}\log{\left[\sin{\left[\f{\pi(Y_1-Y_2)}{L}\right]}\right]}  & \text{for}~ \f{L}{2\pi}\left|\tan{\left(\f{\pi Y_2}{L}\right)}\right|>t_2>0 \\
    \f{c\pi L}{12\epsilon} & \text{for}~ \f{L}{2\pi}\left|\tan{\left(\f{\pi Y_1}{L}\right)}\right|>t_2>\f{L}{2\pi}\left|\tan{\left(\f{\pi Y_2}{L}\right)}\right|\\
    \f{c\pi L}{6\epsilon} & \text{for}~  t_2>\f{L}{2\pi}\left|\tan{\left(\f{\pi Y_1}{L}\right)}\right|
    \end{cases}, 
\end{split}
\ee
where  $\f{L}{2}>L-Y_1>Y_2>0$. 
We can see from the $t_2$-dependence of $S_B$ that except for the vacuum entropy, it may be described by the propagation of quasiparticles at the velocities, $v_{L,R}(x)=\pm 2\cos^2{\left(\f{\pi x}{L}\right)}$.
Here $v_{L,R}(x)$ denote the speeds of left- and right-moving quasiparticles, respectively.

\subsection{The $t_2$-dependence of BMI \label{App:MI-system4}}
Now, we report the $t_2$-dependence of BMI for the subsystems discussed in Section \ref{Sec:BandTOMI}.
\subsubsection{The single interval \label{App:MI-system4-single}}
\begin{figure}[htbp]
    \begin{tabular}{ccc}
      \begin{minipage}[t]{0.5\hsize}
        \centering
        \includegraphics[keepaspectratio, scale=0.6]{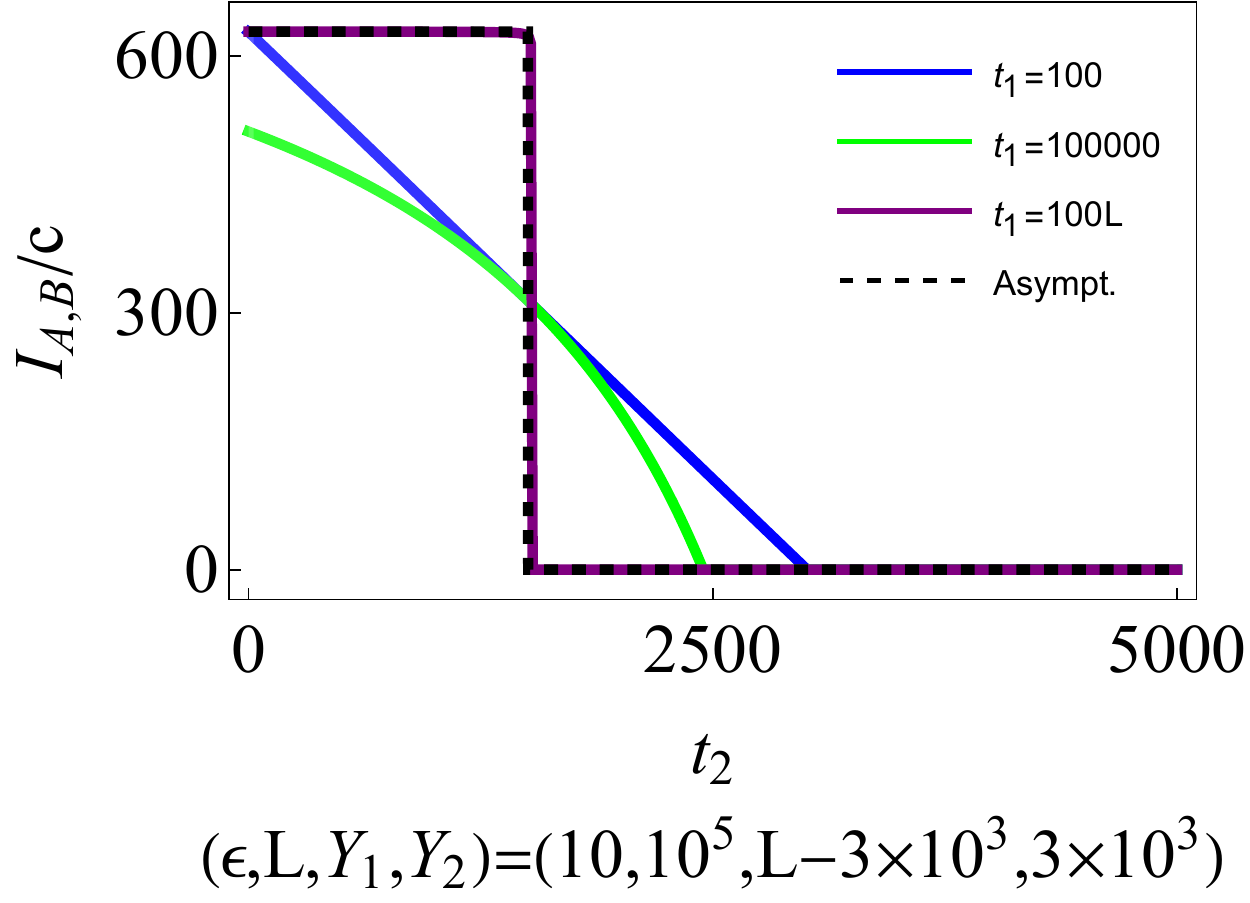}
  
   (a) $t_2$-dependence of $I_{A,B}$ in (A).
      \end{minipage} & 
     
     \begin{minipage}[t]{0.5\hsize}
        \centering
        \includegraphics[keepaspectratio, scale=0.66]{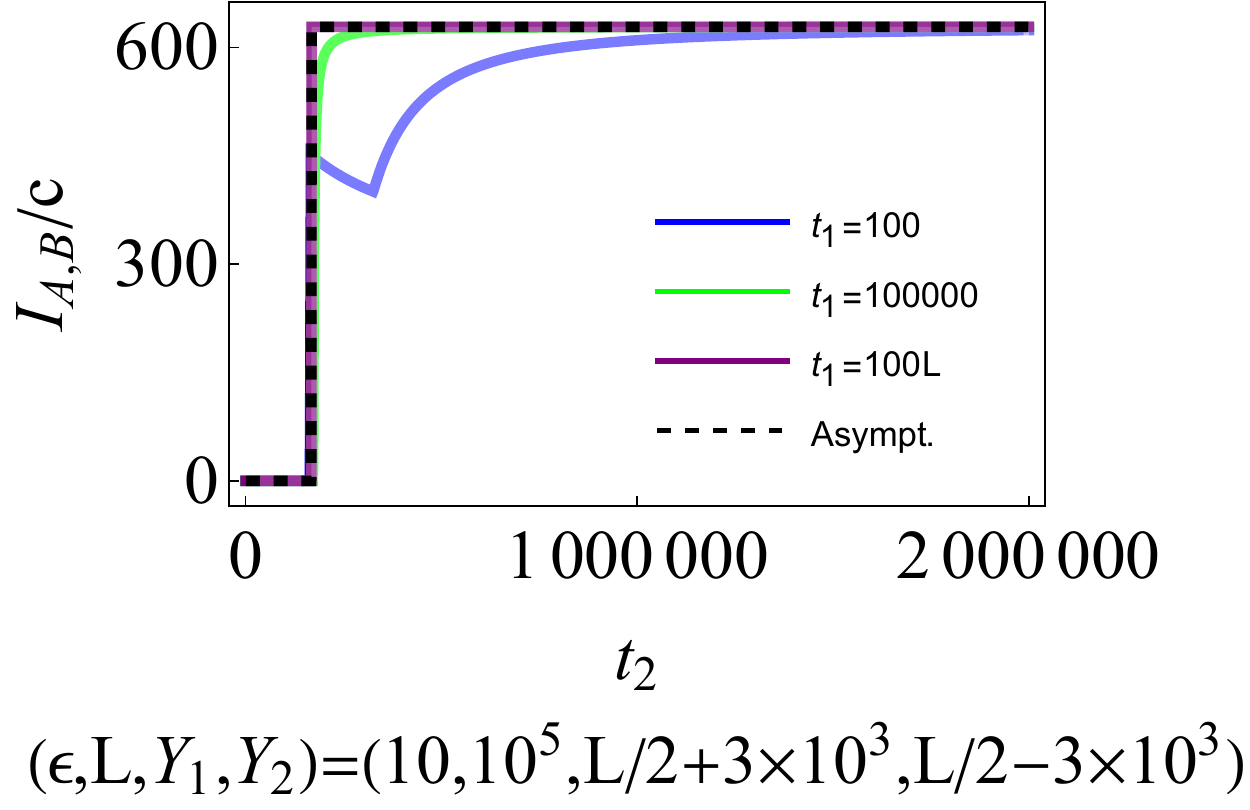}
   
    (b) $t_2$-dependence of $I_{A,B}$ in (C).
      \end{minipage} 
      
    \end{tabular}
    \caption{BMI, $I_{A,B}$, of (\ref{eq:evolvedstatein3-CSD}) for various $t_1$ as the function of $t_2$. For simplicity, $l_A=l_B$.
    In (a), the solid lines illustrate the $t_2$-dependence of $I_{A,B}$ with $t_1=10,1000,10^6$ for (A). 
    In (b), the solid lines illustrate the $t_2$-dependence of $I_{A,B}$ with $t_1=10,1000,10^6$ for (C). 
     The dashed line illustrates the asymptotic behavior in (\ref{eq:aym-IAB-single-in3-CSD}).}
    \label{Fig:time-dependence-of-IAB-single-in4}
  \end{figure}

For the single intervals considered in Section \ref{sec:physical-interpretation}, we depict the  $I_{A,B}$ for various $t_1$ as a function of $t_2$ in Fig.\ \ref{Fig:time-dependence-of-IAB-single-in4}. 
For (C), $I_{A,B}$ is approximately zero.
In the large $t_1$ limit, the asymptotic behavior of $I_{A,B}$ for the single interval is given by
\be \label{eq:aym-IAB-single-in3-CSD}
\begin{split}
\text{If}~ x=X^1_f \in B, ~& I_{A,B} \underset{t_1\gg 1}{\approx}\begin{cases}
\f{c\pi l_A}{3\epsilon} &\f{L}{2\pi}\left|\tan{\left(\f{\pi Y_2}{L}\right)}\right|>t_2>0\\
0 &t_2>\f{L}{2\pi}\left|\tan{\left(\f{\pi Y_2}{L}\right)}\right|\\
     \end{cases},\\
\text{If}~ x=X^2_f \in B, ~& I_{A,B} \underset{t_1\gg 1}{\approx}\begin{cases}
0&\f{L}{2\pi}\left|\tan{\left(\f{\pi Y_1}{L}\right)}\right|>t_2>0\\
\f{c\pi l_A}{3\epsilon} &t_2>\f{L}{2\pi}\left|\tan{\left(\f{\pi Y_1}{L}\right)}\right|\\
     \end{cases},
\end{split}
\ee
where $L-Y_1>Y_2>0$.
For (A) and (C), there are $t_2$-regimes where both B.H.-like excitations introduced in Section \ref{sec:physical-interpretation} are in $B$, while for (B), there are none.
\subsubsection{The double intervals \label{App:MI-system4-double}}
\begin{figure}[htbp]
    \begin{tabular}{cc}
      \begin{minipage}[t]{1\hsize}
        \centering
        \includegraphics[keepaspectratio, scale=0.5]{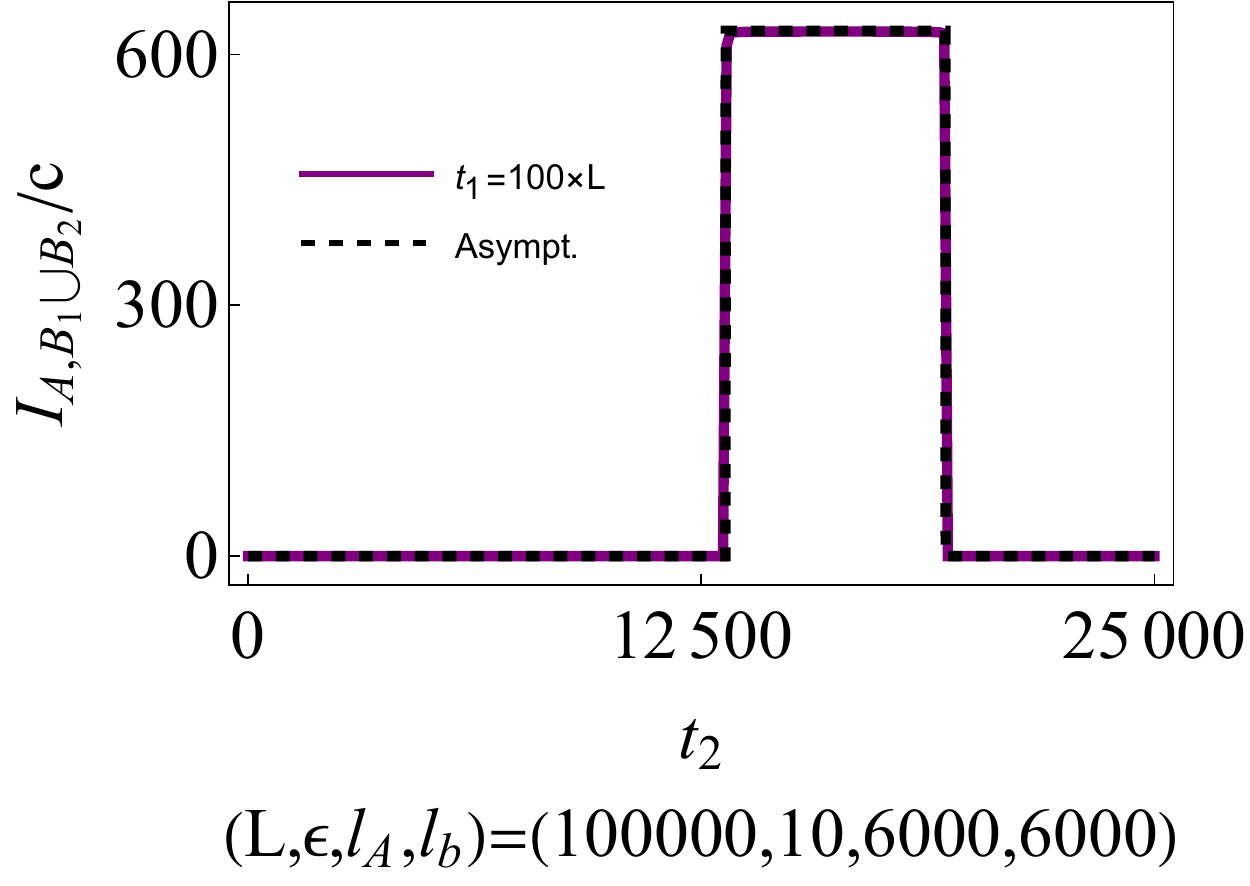}
    
      \end{minipage} &
     
    \end{tabular}
    \caption{The BMI, $I_{A,B_1 \cup B_2}$, in the large $t_1$-regime as the function of $t_2$. The solid line illustrates the $t_2$-dependence of $I_{A,B_1 \cup B_2}$ for  $t_1=10^7$. 
    In this figure, $l_{B_1}=l_{B_2}=l_b$, $P_{C,B_1}=\frac{3L}{4}$ and $P_{C,B_2}=\f{L}{4}$, and $P_{C,A}=X^1_f$. 
    The dashed line illustrates the $t_2$-dependence of $I_{A,B_1\cup B_2}$ in (\ref{eq:asym_bomi_double-CSD}). \label{Fig:BOMI_system4_double_intervals}}
  \end{figure}

Let us now turn to the $t_2$-dependence of $I_{A,B_1\cup B_2}$ for the subsystems in (\ref{B1B2}).
In Fig.\ \ref{Fig:BOMI_system4_double_intervals}, we depict  $I_{A,B_1\cup B_2}$ for large $t_1$ as a function of $t_2$.
The asymptotic behavior of $I_{A,B=B_1\cup B_2}$ for the $t_1$-regime is given by
\be \label{eq:asym_bomi_double-CSD}
\begin{split}
    I_{A,B=B_1\cup B_2} \approx \begin{cases}
    0 & \f{L}{2\pi} \left|\tan{\left(\f{\pi Y_2}{L}\right)}\right|>t_2>0 \\
    \f{c\pi l_A}{3\epsilon} & \f{L}{2\pi} \left|\tan{\left(\f{\pi Y_1}{L}\right)}\right|\ge t_2\ge\f{L}{2\pi} \left|\tan{\left(\f{\pi Y_2}{L}\right)}\right| \\
    0 & t_2>\f{L}{2\pi} \left|\tan{\left(\f{\pi Y_1}{L}\right)}\right|
    \end{cases},
\end{split}
\ee
where $\f{L}{2}>Y_1>Y_2>0$.
In this case, there is a time regime where both of the black-hole-like excitations can be in $B=B_1 \cup B_2$.
However, there is no time regime where both of the black-hole-like excitations are in only $B_1$ or $B_2$.
\subsection{The $t_2$-dependence of TMI \label{App:TMI-system4}}
We present the asymptotic behavior of TMI in the large $t_1$ limit.
The TMI which we consider are $I_{A,B,\overline{B}}$ and $I_{A,B_1,B_2}$.
They are defined by (\ref{localTMI-system3}) and (\ref{globalTMI-system3}), respectively.
The value of the global TMI for the large $t_1$ is zero.
In the early $t_2$-regime, $\f{L}{2\pi} \left|\tan{\left(\f{\pi Y_2}{L}\right)}\right|>t_2>0$, the local MI is zero, in the intermediate $t_2$-interval, $\f{L}{2\pi} \left|\tan{\left(\f{\pi Y_1}{L}\right)}\right|>t_2>\f{L}{2\pi} \left|\tan{\left(\f{\pi Y_2}{L}\right)}\right|$, it is approximated by $-2 S^{\text{Reg.}}_A$, and then, in the late $t_2$-regime, $t_2>\f{L}{2\pi} \left|\tan{\left(\f{\pi Y_1}{L}\right)}\right|$, it is zero. We can see from the $t_2$-dependence of the global TMI that as in the case of (\ref{eq:stata-in-system3}), there is no non-locally-hidden correlation between $A$, $B$ and $\overline{B}$. 
Furthermore, we can see the time $t_2$-dependence of the local TMI that there may exist the non-locally-hidden correlation shared by $A$, $B_1$, and $B_2$.

By contrast, both the local and global TMI for the setup in Fig. \ref{Fig:BOMI_system4_double_intervals} vanishes for both physical spin structures $\nu=3,4$ in the free fermion CFT as expected. This is because the entanglement is carried by bell pairs in the free theory and hence there is no tripartite entanglement.
\subsection{Growth of wormhole for (\ref{eq:evolvedstatein3-CSD}) \label{App:grwoth-wh-system4}}
\begin{figure}[tbp]
        \begin{center}
          \includegraphics[width=8cm]{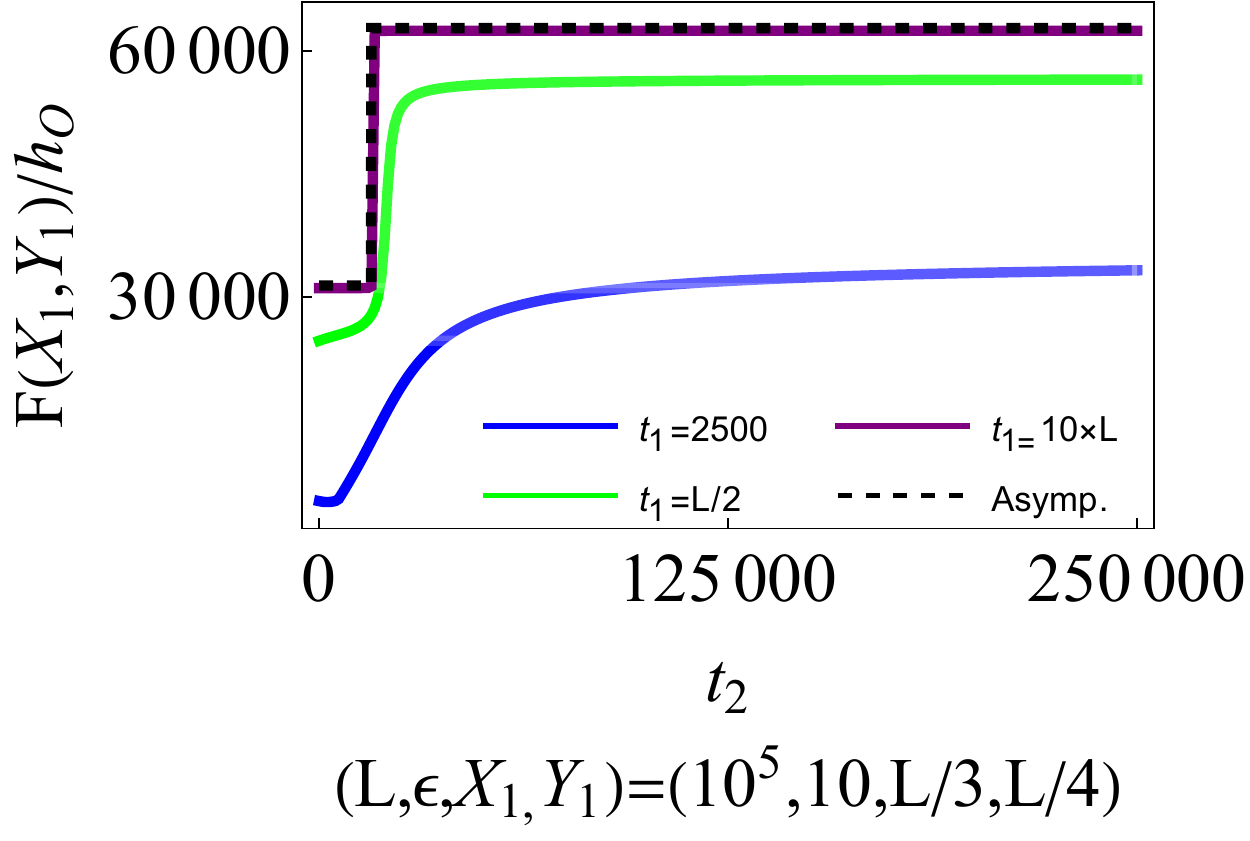}
          
          
          \caption{The $t_2$-dependence of $F(X_1,Y_1)$ for various $t_1$ for (\ref{eq:evolvedstatein3-CSD}). The dashed line illustrates the $t_2$-dependence of $F(X_1,Y_1)/h_{\mathcal{O}}$ in (\ref{eq:evolvedstatein3-CSD}).\label{Free-energy-system4}}
        \end{center}
        \end{figure}

Let us present the $t_2$-dependence of $F(X_1,Y_1)$ for the various $t_1$ for (\ref{eq:evolvedstatein3-CSD}).
In Fig.\ \ref{Free-energy-system4}, we depict $F(X_1,Y_1)$ for the various $t_1$ as a function of $t_2$.
For the large $t_1$-regime, the $t_2$-dependence of $F(X_1,Y_1)$ is given by 
\be \label{eq:wg_asymptotic_system4}
\begin{split}
F(X_1,Y_1)\approx4h_{\mathcal{O}} \log{\left(\f{2\epsilon}{\pi}\right)}&+h_{\mathcal{O}}\log \left(\frac{16 \pi ^4 t_1^4 \left(L^2 \sin ^2\left(\frac{\pi  Y_1}{L}\right)-4 \pi ^2 t_2^2 \cos ^2\left(\frac{\pi  Y_1}{L}\right)\right)^2}{L^8}\right)\\
&+\begin{cases}
\f{h_{\mathcal{O}L} \pi }{ \epsilon} &~\text{for}~ \f{L}{2\pi} \tan{\left(\f{\pi Y_1}{L}\right)}>t_2>0\\
\f{2\pi h_{\mathcal{O}}L}{\epsilon} &~\text{for}~t_2>\f{L}{2\pi} \tan{\left(\f{\pi Y_1}{L}\right)}
\end{cases}.
\end{split}
\ee

\section{Non-chaotic theories \label{App:free_fermion}}
Let us present the details of the calculations and results in $2$d free fermion CFT.
\subsection{The entanglement entropy in $2$d free fermion CFT \label{App:NUP_in_free}}

\begin{table}[]
    \centering
    \begin{tabular}{|c|c|c|}
    \hline
         $\nu$& sector & ($\nu_1$,$\nu_2$) \\
         \hline
         1&(R,R)  &(0,0) \\
    2&(R,NS)  &(0,$\frac{1}{2}$) \\
    3&(NS,NS)  &($\frac{1}{2}$,$\frac{1}{2}$) \\
    4&(NS,R)  &($\frac{1}{2}$,0) \\
    \hline
    \end{tabular}
    \caption{Spin structures of the fermion on a torus.}
    \label{FermionSpinStructure}
\end{table}

In this section, we outline the technique for calculating entanglement entropy for free Dirac fermions using bosonization as explained in \cite{Herzog2013}. There are two possible boundary conditions one can impose on the fermions along each cycle of the torus, namely, the periodic (R) or anti-periodic (NS) boundary conditions,
\begin{equation}
    \psi\left(\frac{w}{2\epsilon}+1\right) = e^{2i\pi\nu_1}\psi\left(\frac{w}{2\epsilon}\right), \qquad \psi\left(\frac{w}{2\epsilon}+\tau\right) = e^{2i\pi\nu_2}\psi\left(\frac{w}{2\epsilon}\right).
\end{equation}
The four possibilities are summarized in table \ref{FermionSpinStructure}. In this coordinate system, the cycle along $\tau=iL/2\epsilon$ corresponds to the spatial direction.

Let $A$ and $B$ denote the subsystems of $\mathcal{H}_2$ and $\mathcal{H}_1$, respectively.
The edges of $A$ are denoted by $X_1$ and $X_2$. while those of $B$ are denoted by $Y_1$ and $Y_2$.
Here, we assume that $X_1>X_2>0$ and $Y_1 >Y_2 >0$.
The R\'{e}nyi entanglement entropy is given by a two-point function of the twist operators on the 2-torus. This is equivalent to the partition function on the orbifolded theory with a branch cut running along the entanglement cut.  Such a partition function can be computed using bosonization \cite{Herzog2013}. The resulting operator entanglement entropy can be divided into one piece that depends on the spin-structure and another that does not. For a subsystem $\mathcal{V}$, the former shall be referred to as the non-universal piece $S^{(n)}_{\mathcal{V},\nu,\text{non-univ.}}$ while the latter will be referred to as the universal piece $S^{(n)}_{\mathcal{V},\text{univ.}}$ so that $S^{(n)}_{V,\nu}=S^{(n)}_{\mathcal{V},\text{univ.}}+S^{(n)}_{\mathcal{V},\nu,\text{non-univ.}}$. For an interval, $B$, in the first Hilbert space and $A$ in the second Hilbert space as well as their union, the universal and non-universal pieces are given by
\begin{align}\label{SAuniv}
    S_{A,\text{univ.}}^{(n)} 
    =&\frac{n+1}{12n}\log 
    \left|\frac{2\epsilon
    \theta_1\left(\frac{w_{X_1}-w_{X_2}}{2\epsilon}\bigg|\tau\right)
    }{\partial_z \theta_1(0|\tau)}\right|^2,
    \\ \nonumber
    S_{A,\nu,\text{non-univ.}}^{(n)} 
    =&\frac{1}{1-n}\sum_{k=-\frac{n-1}{2}}^{\frac{n-1}{2}} \log\left|
    \frac{
    \theta_\nu\left(\frac{k}{N}\frac{w_{X_1}-w_{X_2}}{2\epsilon}\bigg|\tau\right)
    }{\theta_\nu(0|\tau)}\right|^2,
    \\ \nonumber
    S_{B,\text{univ.}}^{(n)}  
    =&-\frac{c(n+1)}{24 n} \log\left[\prod_{i=1,2}\left|\frac{dw_{Y_i}^{\text{New},\alpha}}{dw_{Y_i}}\frac{d\bar{w}_{Y_i}^{\text{New},\alpha}}{d\bar{w}_{Y_i}}\right|\right]
    +\frac{n+1}{12n}\log\left|\frac{2\epsilon 
    \theta_1\left(\frac{w_{Y_1}^{\text{New},\alpha}-w_{Y_2}^{\text{New},\alpha}}{2\epsilon}\bigg|\tau\right)
    }{\partial_z \theta_1(0|\tau)}\right|^2, \\ \nonumber
    S_{B,\nu,\text{non-univ.}}^{(n)}  
    =&\frac{1}{1-n}\sum_{k=-\frac{n-1}{2}}^{\frac{n-1}{2}} \log\left|
    \frac{
    \theta_\nu\left(\frac{k}{N}\frac{w_{Y_1}^{\text{New},\alpha}-w_{Y_2}^{\text{New},\alpha}}{2\epsilon}\bigg|\tau\right)
    }{\theta_\nu(0|\tau)}\right|^2, 
    \end{align}
    \begin{align}
    S_{A\cup B,\text{univ.}}^{(n)}  
    =&S_{A,\text{univ.}}^{(n)}+S_{ B,\text{univ.}}^{(n)}+\frac{n+1}{12n}
    \log \left|\frac{\theta_1\left(\frac{w_{X_2}-w_{Y_2}^{\text{New},\alpha}}{2\epsilon}|\tau\right)
    \theta_1\left(\frac{w_{Y_1}^{\text{New},\alpha}-w_{X_1}}{2\epsilon}|\tau\right)
    }{    \theta_1\left(\frac{w_{X_1}-w_{Y_2}^{\text{New},\alpha}}{2\epsilon}|\tau\right)\theta_1\left(\frac{w_{X_2}-w_{Y_1}^{\text{New},\alpha}}{2\epsilon}|\tau\right)}\right|^2,
    \\ \nonumber
    S_{A\cup B,\nu,\text{non-univ.}}^{(n)} 
    =& \frac{1}{1-n}\sum_{k=-\frac{n-1}{2}}^{\frac{n-1}{2}} \log\left|
    \frac{\theta_\nu\left(\frac{k}{n}\frac{w_{X_2}-w_{X_1}+w_{Y_1}^{\text{New},\alpha}-w_{Y_2}^{\text{New},\alpha}}{2\epsilon}\bigg|\tau\right)}{\theta_\nu(0|\tau)}\right|^2,
\end{align}
where the $\log 2\epsilon$ terms come from rescaling the torus coordinates to have periodicities $1$ and $\tau$. Note also that when applying the bosonization formulas in  \cite{Herzog2013}, the coordinates of the twist operators in the different Hilbert spaces are swapped relative to one another as explained in \cite{Nie:2018dfe}.
 
\subsection{Quasiparticle picture}
Suppose that we prepare the systems considered in the thermofield double state, and then evolve them with the Hamiltonians acting on only $\mathcal{H}_1$.
In the infinite temperature limit, the thermofield double state can be written as a product of Bell pairs of quasiparticles as in (\ref{eq:TFD-Bell-pairs}).
The quasiparticles that live on the Hilbert space that is being acted upon by the Hamiltonians move according to inhomogeneous velocity fields $f(x)$ and $-f(x)$ for the right-moving and left-moving quasiparticles, respectively. These quasiparticles describe the dynamics of entanglement in non-chaotic theories. When the Hamiltonian changes as in the case where different unitary operators are composed, the velocity field simply gets replaced by the envelope of the new Hamiltonian that governs the time-evolution.

In the uniform case where $f(x)=1$, the quasiparticles simply propagate with unit speed as explained in \cite{Nie:2018dfe}. In the SSD limit, the speed vanishes at the fixed point $X^1_f$. Therefore, the quasiparticles tend to cluster around the fixed point $X_f^1$ as shown in \cite{Goto:2021sqx}, giving rise to black hole-like excitations.
\subsubsection{System 1}
Let us begin by looking at the case where a single inhomogeneous Hamiltonian acts on the first Hilbert space. Denote the density of the right and left moving quasiparticles at position $x$ at time $t$ by $\rho^{(n)}_R(x,t)$ and $\rho^{(n)}_L(x,t)$. The superscript $n$ denotes the R\'{e}nyi index which determines the density of quasiparticles. Assuming that the quasiparticles are conserved, the corresponding densities have to obey the continuity equation
\begin{equation}\label{ContinuityEquation}
    \frac{\partial \rho^{(n)}_i(x,t)}{\partial t} = \pm \left(f(x)\frac{\partial \rho^{(n)}_i(x,t)}{\partial x}+\rho^{(n)}_i(x,t) \frac{d f(x)}{dx}\right)
\end{equation}
where the $+(-)$ sign is for the $i=L(R)$ chiralities. Since the quasiparticles are moving with a speed $f(x)$, a quasiparticle initially located at $x_0$ at time $t_0$ will be located at position $x$ at a later time $t$ as determined by
\begin{equation}
    dt =\pm \frac{dx}{f(x)} \Rightarrow t-t_0 = \pm \int_{x_0}^x \frac{dx'}{1-\tanh{2\theta}\cos{\frac{2\pi x'}{L}}}
\end{equation}
where "$+$" refers to right-moving quasiparticles while "$-$" refers to the left-moving quasiparticles. The integral is straightforward to perform and yields the trajectories $x_i(t)$ for $i=L,R$. This trajectory can also be inverted to give the initial position of $x_{i,0}(x,t)$ of a quasiparticle that is at position $x$ at time $t$. Since the number of quasiparticles is conserved, the number of particles initially located in the interval $dx_{i,0}$, $\rho^{(n)}(x_{i,0}(x,t),0)dx_{i,0}$, is the same as the number of quasiparticles in $dx$ at time $t$, $\rho^{(n)}(x,t)dx$. Hence, the solution to the continuity equation \eqref{ContinuityEquation} for a constant velocities $\pm f(x)$ is \cite{doi:10.1098/rspa.1978.0190}
\begin{equation}\label{ContinuityEquationSolution}
    \rho^{(n)}_i(x,t) = \rho^{(n)}_i(x_{i,0}(x,t),0) \frac{\partial x_{i,0} (x,t)}{\partial x}
\end{equation}
for $i = L,R$. Since the trajectory, $x_{i,0}(x,t)$, is a periodic function with period $L\cosh{2\theta}$, the corresponding quasiparticle densities also possess the same periodicity.

Now, we turn to the computation of entanglement entropy and mutual information using the quasiparticle picture. In this paper, the unitaries only act on one Hilbert space, so only the quasiparticles in that Hilbert space move while their immobile partners remain fixed at position $x_0$. Each such Bell pair contributes to the correlation between the point $x$ in $\mathcal{H}_1$ and the point $x_0$ in $\mathcal{H}_2$. The methods for computing the mutual information and entanglement entropy in the quasiparticle picture are very similar but not identical so we explain the technique for computing each quantity separately.

\subsubsection*{Entanglement entropy}
The entanglement entropy of a pure state measures the amount of correlation between the subsystem and its complement. Therefore, the entanglement entropy for a subsystem $B$ is proportional to the number of bell pairs shared by subsystem $B$ and its complement. Since the Bell pair partner of any quasiparticle in $B$ lives in the other Hilbert space, any Bell pair with a quasiparticle that winds up in $B$ at a certain time $t$ contributes to the entanglement entropy $S_B(t)$. Therefore, the initial quasiparticle density in \eqref{ContinuityEquationSolution} is a simple constant that can be fixed by equating the quasiparticle prediction for the entanglement entropy to the entanglement entropy in 2d free fermion CFT. This constant turns out to be $\rho_0 = \frac{n+1}{24n} \frac{\pi}{\epsilon}$. For a single interval $B=[Y_2,Y_1]$, the entanglement entropy according to the quasiparticle picture is
\begin{equation}
    S_B(t) = \int_{x\in B} \rho_L(x,t)+ \int_{x\in B}\rho_R(x,t) = \sum_{i=L,R} 
    \text{mod}
    \left[x_{0,i}(Y_1,t)-x_{0,i}(Y_2,t),L\right]
\end{equation}
where the integral was carried out by a simple change of variables from $x$ to the initial position $x_{0,i}(x,t)$ and the modulo operation takes the periodicity of the system into account. This result simply states that the quasiparticles in the interval $[x_{0,i}(X_2,t),x_{0,i}(X_1,t)]$ flow to $[X_2,X_1]$ at time $t$.
\subsubsection*{Mutual Information}
The MI is obtained by the same integral. The only difference comes from the initial quasiparticle density in \eqref{ContinuityEquationSolution}. This is because the MI between subsystems $B$ and $A$ of $\mathcal{H}_1$ and $\mathcal{H}_2$ measures the correlations between subsystems $A$ and $B$ and hence only receives contributions from Bell pairs one quasiparticle in subsystem $A$ and the other in subsystem $B$. Since the quasiparticles in the second Hilbert space are immobile, only the quasiparticles that are initially in subsystem $A$ can potentially contribute to the MI. Therefore, for the computation of mutual information, the initial quasiparticle density is 
\begin{equation}
    \rho_i(x,0) = \rho_0 \theta(x \in A)
\end{equation}
where $\rho_0 = \frac{N+1}{12N}  \frac{\pi}{\epsilon}$ is a constant that is fixed by equating the initial MI for two symmetric intervals $A=B$ with that of the 2d free fermion CFT. If $B$ is the union of $m$ disjoint $[Y_{2j},Y_{2j-1}]$ for $j=1,\ldots,m$, the MI between two subsystems A and B at a fixed time $t$ is given by
\begin{align}
  I_{AB}^{(n)}(t)
  &= \sum_{j=1}^m\sum_{i=L,R} \int_{Y_{2j}}^{Y_{2j-1}} dx \rho_i(x_{0,i}(x,t),0) \frac{\partial x_{0,i}(x,t)}{\partial x}
    \nonumber \\
    &= \rho_0\sum_{j=1}^m \sum_{i=L,R} \text{length of }[x_{0,i}(Y_{2j},t),x_{0,i}(Y_{2j-1},t)] \cap A
\end{align}
The second equality comes from the usual change of variables from $x$ to $x_{0,i}$ where $t$ is held fixed so that $x_{0,i}$ is viewed as a function of a single variable $x$. The final expression has a simple interpretation; the quasiparticles located in $[x_{0,i}(Y_2,t),x_{0,i}(Y_1,t)]$ are the only ones that can be in subsystem $B$ at time $t$. Out of these quasiparticles, only the ones that were also simultaneously in $A$ can contribute to the mutual information between $A$ and $B$. 

\subsubsection{System 2 and 3}
System 2 and 3 correspond to time evolutions where two different unitaries are applied one after the other. The overall time evolution  corresponds to a product of two unitary evolutions for durations $t$ and $T$ that sends a quasiparticle with an initial spacetime position
\begin{equation}
    (0,x_0) \rightarrow (t,x)\rightarrow (t+T,y)
\end{equation}
Under each unitary, the quasiparticle density evolves according to \eqref{ContinuityEquationSolution}, so the final quasiparticle density can be related to the initial density by the chain rule
\begin{equation}
    \rho_i(y,T+t) = \rho_i(x_0(x(y,T),t),0) \frac{\partial x_0}{\partial y}.
\end{equation}
The entanglement entropy is given by
\begin{equation}
    S_B =\sum_{i=L,R} \int_{Y_2}^{Y_1} dy \frac{n+1}{24n} \frac{\pi}{\epsilon} \frac{\partial x_0}{\partial y} =\frac{n+1}{24n} \frac{\pi}{\epsilon}\sum_{i=L,R} \text{mod}\left[x_0(x(Y_1,T),t)-x_0(x(Y_2,T),t),L\right]
\end{equation}
where the final equality comes from the exact same reasoning as in the M\"{o}bius/SSD case. Just as in System 1, this result simply says that the entanglement entropy of a subsystem at a particular instant in time is given by the number of quasiparticles that end up in the subsystem at that time. The mutual information as predicted by quasiparticles is similar to the entanglement entropy except for the initial quasiparticle density. If the subsystem $B$ is a union of $m$ disjoint intervals $[Y_{2j},Y_{2j-1}]$, $j=1,\ldots,m$, the mutual information is
\begin{align}
    I_{AB}^{(n)}(t) =&
    \frac{n+1}{12n} \frac{\pi}{\epsilon} \sum_{i=L,R}\sum_{j=1}^m \int_{Y_{2j}}^{Y_{2j-1}} dy \frac{\partial x_{i,0}}{\partial y} \theta(x_{i,0} \in A)
    \nonumber \\
    =&\frac{n+1}{12n} \frac{\pi}{\epsilon} \sum_{i=L,R}\sum_{j=1}^m \text{length of }[x_0(x(Y_{2j},T),t),x_0(x(Y_{2j-1},T),t)] \cap A
\end{align}
where a change of variables from the final spatial coordinate $y$ to the initial position $x_{i,0}$ was made to carry out the integral. The physical meaning of this result is identical to that in the M\"{o}bius/SSD case.
\subsection{Summary of results for non-chaotic systems \label{App:sum-of-results}}
Using the formulas outlined in the previous subsections, the entanglement entropy and MI can be computed in 2d free fermion CFTs and quasiparticle pictures. For the various subsystems and unitary time evolutions, the entanglement for the two physical spin-structures $\nu=3$ and $\nu=4$ are found to be identical. 
Furthermore, the global TMI for 2d free fermion CFT vanishes in all the cases considered. This is because the entanglement entropy and MI for the 2d free fermion CFT agree with the quasiparticle picture to leading order in $1/\epsilon$ \footnote{As mentioned in \cite{Goto:2021sqx}, the quasiparticle prediction for the operator entanglement entropy deviates away from the 2d free fermion CFT when the entanglement entropy is small, such as in the case where the entanglement entropy approaches the vacuum value.}. 
The agreement with the quasiparticle picture describes the key differences between 2d free fermion and holographic CFTs. Firstly, for finite values of $\theta$, the quasiparticle distributions are periodic with a period of $L\cosh{2\theta}$, so that the MI will possess the same periodicity. Secondly, the MI is separately carried by the right-moving and left-moving quasiparticles which travel independently of one another, as opposed to the holographic theory where the MI is non-zero only when the subsystem contains both the left and the right-moving B.H.-like excitations. Lastly, the TMI is observed to vanish for the 2d free fermion CFTs but that is not always the case for the holographic theories.

\begin{figure}
    \centering
    \includegraphics[scale=0.4]{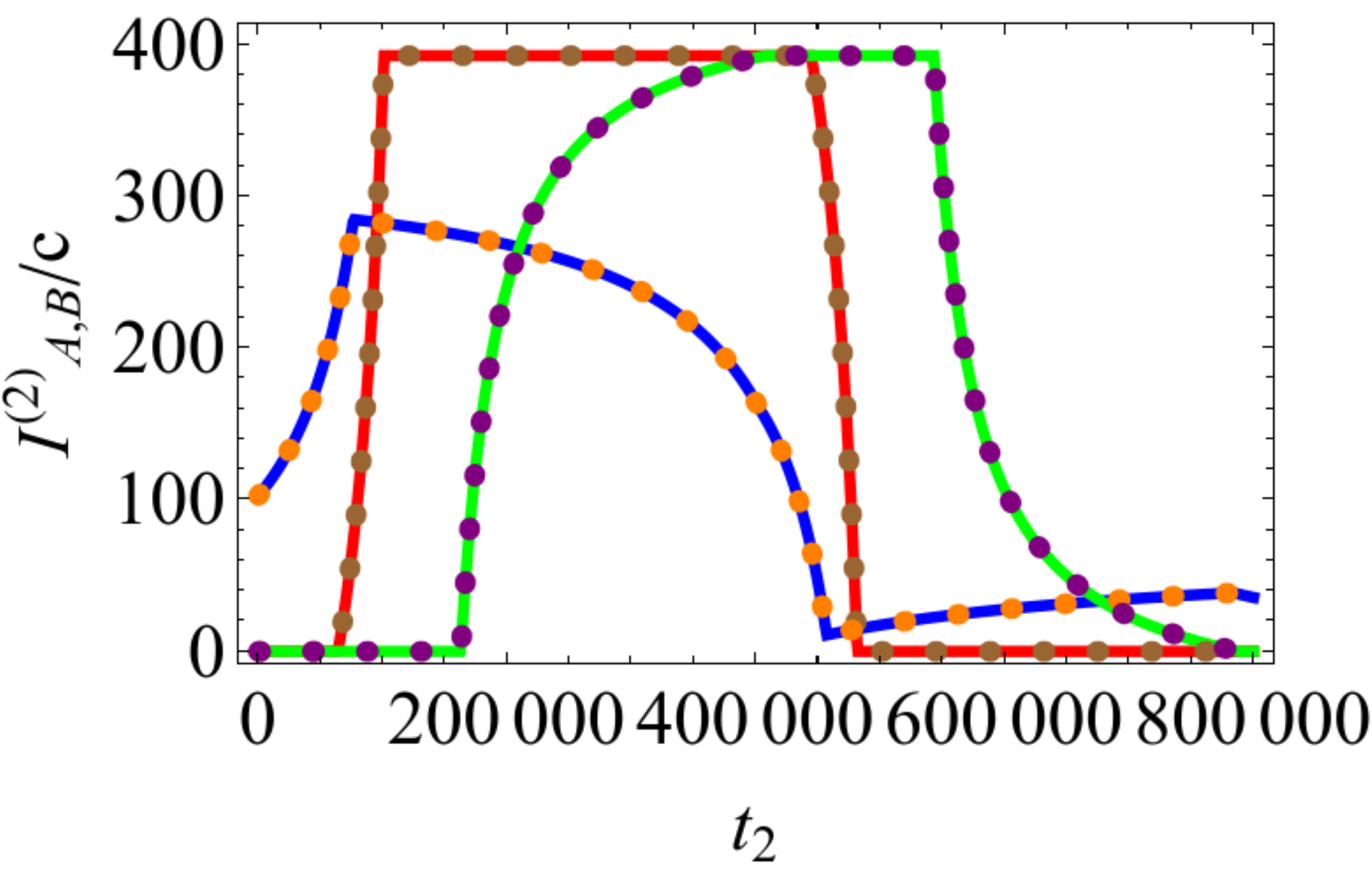}
    \hspace{0.7cm}
    \includegraphics[scale=0.35,trim=0cm -2cm 0cm 0cm  ]{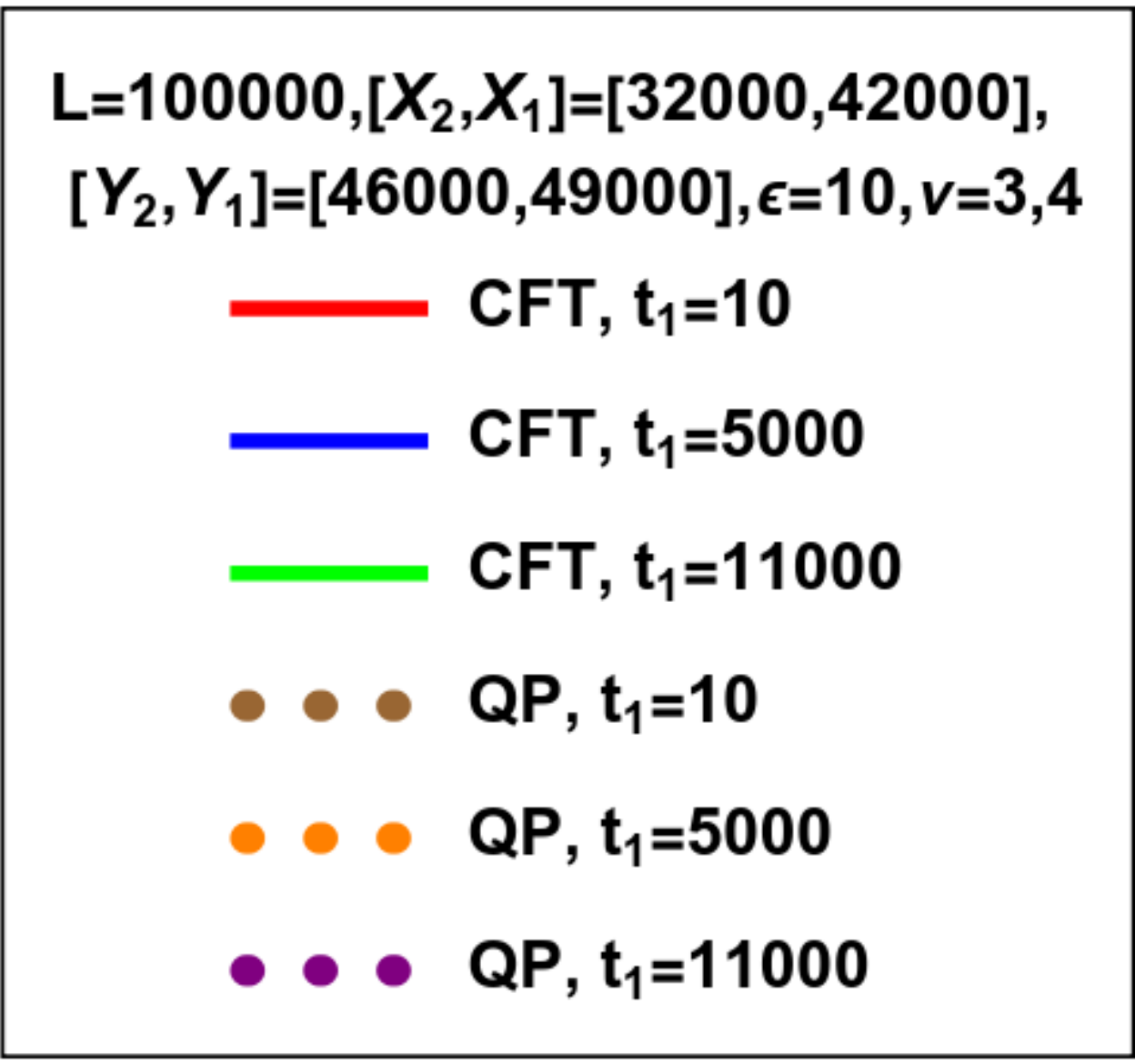}
    \caption{Plots of the operator mutual information when the system is first acted upon by the SSD evolution for a duration of $t_1$ followed by a time evolution of $t_2$ under the CSD Hamiltonian. The solid lines are the 2d free fermion CFT results while the dots are the predictions by the quasiparticle picture.}
    \label{FreeFermionVsQP_SSDThenCSD_NeitherFixedPointsInSystems}
\end{figure}

In Fig. \ref{FreeFermionVsQP_SSDThenCSD_NeitherFixedPointsInSystems}, we show a representative plot comparing the 2d free fermion CFT MI with the quasiparticle prediction. In this setup, we first evolve the system with the SSD Hamiltonian before evolving it with the CSD Hamiltonian which is essentially the SSD Hamiltonian but with the envelope function vanishing at $X_f^2$ instead. 
The holographic results for this kind of evolution are discussed in appendix \ref{App:system4}. The subsystems in Fig. \ref{FreeFermionVsQP_SSDThenCSD_NeitherFixedPointsInSystems} are placed away from both fixed points $X_f^1$ and $X_f^2$.
The quasiparticles will pass through $B$, giving rise to a non-zero BMI. However, since the subsystem does not contain the CSD fixed point, these quasiparticles will eventually leave $B$ although they take a long time to do so for the subsystem in Fig. \ref{FreeFermionVsQP_SSDThenCSD_NeitherFixedPointsInSystems} because $B$ is located close to the CSD fixed point $X_f^2$ where the quasiparticle speed is small. 
This figure also highlights the key difference between the dynamics of BMI in the free fermion CFT as well as the holographic CFTs. 
The BMI vanishes for this choice of subsystems in the latter but not the former for large values of $t_1$. 
This is because BMI is non-zero in the holographic theory when both chiral and anti-chiral B.H.-like excitations are simultaneously present in $B$ which does not occur when $B$ does not contain any of the fixed points and when the SSD evolution time $t_1$ is large which causes the B.H.-like excitation to be sharply peaked. 
By contrast, BMI is separately carried by the left and right moving quasiparticles so as long as either one of them is present in $B$, BMI is non-zero. 
For this choice of subsystems, the left-moving quasiparticles travel leftwards around the spatial circle and approach the CSD fixed point $X_f^2$ from the side opposite to subsystem $B$ and hence do not contribute to the BMI. 
When the SSD quench time is $t_1 = 5000$, there are already right-moving quasiparticles in the output subsystem, so the initial value of BMI is non-zero. Some right-moving quasiparticles start off at $t_2=0$ at positions infinitesimally close to the CSD fixed point $X_f^2$ and take a long time to go around the spatial circle leading to a long tail in the BMI. 
When $t_1=11000$, the right-moving quasiparticles start off at $t_2=0$ to the right of the CSD fixed point at $X_f^2$ and eventually circle around back to subsystem $A$ giving rise to a bump in the BMI.

\section{The gravity dual of the systems\label{sec:gravity-dutal-app}}
Here, we report the gravity dual of the systems considered in this paper.
\subsection{The dual geometries \label{sec:dual-geometries}}
The dual geometries of $\rho_{\mathcal{H}_1}$ considered in this paper are given by
\be \label{eq:geo_i_bh}
\begin{split}
ds_{\alpha=1,2}^2 = L^2\bigg{[}&\f{dr^2}{r^2-r^2_0}+\left(\f{L}{\pi}\right)^4\bigg{\{}\left(r^2f^1_{\alpha=1,2;XX}+r_0^2f^2_{XX}\right)dX^2+\sum_{j=0,1}\left(r^2f^1_{\alpha=1,2;t_jt_j}+r_0^2f^2_{\alpha=1,2;t_jt_j}\right)dt_jdt_j\\
&+2\left(r^2f^1_{t_0t_1}+r_0^2f^2_{t_0t_1}\right)dt_0dt_1+2\sum_{j=0,1}\left(r^2f^1_{\alpha=1,2;Xt_j}+r_0^2f^2_{\alpha=1,2;Xt_j}\right)dXdt_j\bigg{\}}\bigg{]},\\
ds_{\alpha=3}^2 = L^2\bigg{[}&\f{dr^2}{r^2-r^2_0}+\left(\f{L}{\pi}\right)^4\bigg{\{}\left(r^2f^1_{\alpha=3;XX}+r_0^2f^2_{\alpha=3;XX}\right)dX^2+\sum_{j=1,2}\left(r^2f^1_{\alpha=3;t_jt_j}+r_0^2f^2_{\alpha=3;t_jt_j}\right)dt_jdt_j\\
&+2\left(r^2f^1_{\alpha=3;t_1t_2}+r_0^2f^2_{\alpha=3;t_1t_2}\right)dt_1dt_2+2\sum_{j=1,2}\left(r^2f^1_{\alpha=3;Xt_j}+r_0^2f^2_{\alpha=3;Xt_j}\right)dXdt_j\bigg{\}}\bigg{]},
\end{split}
\ee
where the details of metric are summarised in Appendix. \ref{sec:metric-of-geometries}.

In the expression in (\ref{eq:geo_i_bh}), the components such as $dt_jdt_{i\neq j}$ exist.
However, in the time-evolution considered, one of them should be constant.
In the case of the system 1, this system is evolved with $H^1$ from $t_0=0$ to $t_0=t_{0,\text{const.}}$, and then it is evolved with $H^1_{\text{SSD}}$ from $t_1=0$.
Therefore, we should take $t_0$ to be constant and consider the $t_1$-dependence of the geometry.
In this procedure, let us rewrite the radial direction as $r'=\left(\f{L^2}{\pi^2}\right)\sqrt{f^1_{\alpha=1;XX}}r$ that guarantees that the asymptotic geometry near the boundary, $r' \rightarrow \infty$, is given by the SSD AdS$_3$ geometry:
\be
\begin{split}
    ds_{\alpha=1}^2&\approx L^2\left[\f{dr'^2}{r'^2}+r'^2dX^2+r'^2\f{f^{1}_{\alpha=1;t_1t_1}}{f^1_{\alpha=1;XX}}dt_1dt_1\right]\\
    &=L^2\left[\f{dr'^2}{r'^2}+r'^2dX^2-4 \sin ^4\left(\frac{\pi  X}{L}\right)r'^2 dt_1dt_1\right],
\end{split}
\ee
where the time-component of the metric depends on $X$.

In the case of the systems.2 and 3, the system is evolved with $H^1_{\text{SSD}}$ from $t_1=0$ to $t_1=t_{1,\text{const.}}$, and then it is evolved with $H^1$ from $t_0=0$ or $H^1_{CSD}$ from $t_2=0$.
Therefore,  we should take $t_1$ to be constant and consider the geometries.
Rewrite the radial coordinate as $r'_{\alpha=2,3}=\left(\f{L^2}{\pi^2}\right)\sqrt{f^1_{\alpha=2,3;XX}}r$, and then the metric near the boundary, $r'_{i=2,3}\rightarrow \infty$, is given by the global $AdS_3$ for $\alpha=2$ and the CSD $AdS_3$ for $\alpha=3$:
\be
\begin{split}
    ds_{\alpha=3}^2&\approx L^2\left[\f{dr'_{\alpha=3}}{r'^{2}_{\alpha=3}}+r'^{2}_{\alpha=3}dX^2-4 \cos ^4\left(\frac{\pi  X}{L}\right)r'^{2}_{\alpha=3} dt_2dt_2\right].
\end{split}
\ee
As a consequence, the location of the black hole horizon in $r'_{\alpha=2,3}$ coordinate is given by
\be
r'_{\alpha=2,3:\text{Horizon}}=r_0 \left(\f{L^2}{\pi^2}\right)\sqrt{f^1_{\alpha=2,3:XX}}.
\ee
Thus, $r'_{\alpha=2:\text{Horizon}}$ depends on $X$, $t_0$ and $t_1$, while $r'_{\alpha=3:\text{Horizon}}$ depends on $X$, $t_1$, and $t_2$. 

\subsubsection{The temporal and spatial dependence of the inhomogeneous horizon}
Let us focus on the temporal and spatial dependence of the black hole horizon of the black hole geometries dual to the system $2$ and $3$.
\subsubsection*{Asymptotic behavior of horizon with small $t_1$}
Let us begin by looking closely at the temporal and spatial dependence of inhomogeneous black hole horizon in the small $t_1$-region. 
At the second order of the small $t_1$ expansion, the $t_0$-dependence of $r'_{\alpha=2;\text{Horizon}}$ and the $t_2$-dependence of $r'_{\alpha=3;\text{Horizon}}$ are given by
\be
\begin{split}
r'_{\alpha=2:\text{Horizon}} \approx &r_0\left(\f{\pi^2}{L^2}+2\f{\pi^3t_1}{L^3}\sin{\left[\f{2\pi t_0}{L}\right]}\cos{\left[\f{2\pi X}{L}\right]}\right),\\
r'_{\alpha=3:\text{Horizon}} \approx &r_0\bigg{[}\frac{\pi ^2}{\sqrt{L^4+2 \pi ^2 t_2^2 \left(2 \left(L^2+2 \pi ^2 t_2^2\right) \cos \left(\frac{2 \pi  X}{L}\right)+\left(L^2+\pi ^2 t_2^2\right) \cos \left(\frac{4 \pi  X}{L}\right)\right)+2 \pi ^2 L^2 t_2^2+6 \pi ^4 t_2^4}}\\
&+\frac{8 \pi ^4 t_1 t_2\cos ^2\left(\frac{\pi  X}{L}\right) \left(\left(L^2+2 \pi ^2 t_2^2\right) \cos \left(\frac{2 \pi  X}{L}\right)+2 \pi ^2 t_2^2\right)}{\left(L^4+2 \pi ^2 t_2^2 \left(2 \left(L^2+2 \pi ^2 t_2^2\right) \cos \left(\frac{2 \pi  X}{L}\right)+\left(L^2+\pi ^2 t_2^2\right) \cos \left(\frac{4 \pi  X}{L}\right)\right)+2 \pi ^2 L^2 t_2^2+6 \pi ^4 t_2^4\right)^{3/2}}\bigg{]}
\end{split}
\ee
where $r'_{\alpha=2:\text{Horizon}}$ at $X=0, \f{L}{2}$ is independent of $t_0$, and $r'_{\alpha=3:\text{Horizon}}$ at $X=\f{L}{2}$ is independent of $t_1$.
\subsubsection*{Asymptotic behavior of horizon in $t_1 \rightarrow \infty$ \label{sec:Asym-behavior-of-horizon}}
Now, turn to the temporal and spatial dependence of inhomogeneous black hole horizon in the large $t_1$-regime. 
In the large $t_1$-regime excluding $t_0\approx X+n L$ and $t_0\approx -X+n L$, the asymptotic time-dependence of $r'_{i=2,\text{Horizon}}$ is approximated by
\be
r'_{\alpha=2,\text{Horizon}} \approx  \f{r_0}{4t_1^2\left|\sin{\left(\f{\pi (t_0-X)}{L}\right)}\sin{\left(\f{\pi (t_0+X)}{L}\right)}\right|},
\ee
where $n$ is and integer number, and $t_0=\pm X + nL$ are the trajectories of the right- and left-moving B.H.-like excitations.
This suggests the black hole horizon far from the B.H.-like excitations is proportional to $t_1^{-2}$.
\if[0]
Around the the black-hole-like excitations, this large $t_1$-expansion breaks down.
At $t_0= X+n L$ or $t_0= -X+n L$, the profile of black hole horizon is given by
\be \label{blackholehorizon_at_BHE}
r^{'\pm}_{\alpha=2,\text{Horizon}}= \f{\pi^2 r_0}{\sqrt{L^4+2L^2\pi^2t_1^2 (1-\cos{\left(\f{4\pi X}{L}\right)})\mp 2L^3\pi t_1 \sin{\left[\f{4\pi X}{L}\right]}}},
\ee
where $r^{'\pm}_{\alpha=2,\text{Horizon}}$ denotes the black hole horizon at $t_0=\pm X+n L$, respectively.
By using the physical interpretation discussed in section \ref{sec:physical-interpretation}, $X=\pm t_0+nL$ are interpreted as the location of the right- and left-moving B.H.-like excitations, respectively at $t_0$. 
Therefore, $r^{'\pm}_{\alpha=2,\text{Horizon}}$ can be interpreted as the black hole horizon at the points where the right- and left-moving B.H.-like excitations exist at $t_0$.
At $X=0$ or $X=\f{L}{2}$, the black hole horizon is the same as the static BTZ, $r^{'\pm}_{\alpha=2,\text{Horizon}}=\f{\pi^2 r_0}{L^2}$, while At $X=\f{L}{4}$ or $X=\f{3L}{4}$, the black hole horizon is smaller than the static BTZ, and it decreases with $t_1$.
In the spatial regions, $0<X<\f{L}{4}$ and $\f{L}{2}<X<\f{3L}{4}$, $r^{'-}_{\alpha=2,\text{Horizon}}$ is smaller than $\f{\pi^2 r_0}{L^2}$.
In the spatial regions, $\f{L}{4}<X<\f{L}{2}$ and $\f{3L}{4}<X<L$, $r^{'+}_{\alpha=2,\text{Horizon}}$ is smaller than that.
The locations of black hole excitations where $r^{'\pm}_{\alpha=2;\text{Horizon}}$ is maximized is determined by 
\be \label{eq:condtion-for-XRL}
\tan{\left[\f{4\pi X}{L}\right]}=\pm \f{L}{\pi t_1},
\ee
where the equation about $\f{L}{\pi t_1}$ is for the right-mover, while  the equation about $-\f{L}{\pi t_1}$ is for the left-mover.
Let $X_{j=R,L}$ denotes the locations determined by (\ref{eq:condtion-for-XRL}).
In the large $t_1$ limit, $X_{R}$ is approximated by $X_R\approx \f{L^2}{4\pi^2t_1}+\f{mL}{4}$, while $X_L\approx -\f{L^2}{4\pi^2t_1}+\f{(1+m)L}{4}$. 
Here, $m=0,1,2,3$. 
Consequently, the extremal values of $r^{'\pm}_{\alpha=2,\text{Horizon}}$ are given by
\be
\begin{split}
   & r^{'+}_{\alpha=2,\text{Horizon}} \approx r_0\times
   \begin{cases}
    \f{\pi t_1}{L}~\text{for}~ m=0,2\\
    \f{L}{2\pi t_1}~ \text{for}~ m=1,3\\
    \end{cases},
    ~ r^{'-}_{\alpha=2,\text{Horizon}} \approx 
\end{split}
\ee
\fi
\if[0]
Let us consider the asymptotic behavior of $r'_{\text{Horizon}}$ in the limit of $t_1 \rightarrow \infty$.
In this limit, the leading behavior of $r'_{\text{Horizon}}$ is given by
\be
\begin{split}
    \f{r'_{\text{Horizon}}}{r_0} \approx \begin{cases}
    \f{r_0}{4t_1^2 \left|\sin{\left(\f{\pi(t_0-X)}{L}\right)}\sin{\left(\f{\pi(t_0+X)}{L}\right)}\right|} & \text{for}~ \f{\pi(t_0-X)}{L} \neq n \pi, \f{\pi(t_0+X)}{L} \neq m \pi\\
    \text{for}~\f{\pi(t_0+X)}{L} = n \pi &\\
    \f{\pi}{2Lt_1\left|\sin{\left(\f{2\pi X}{L}\right)}\right|} & X\neq \f{L}{2} + n L, ~X\neq m L,\\
    \f{\pi^2}{L^2} &X=\f{L}{2} + n L~\text{or}~X= mL,\\
     \text{for}~\f{\pi(t_0-X)}{L} = n \pi &\\
       \f{\pi}{2Lt_1\left|\sin{\left(\f{2\pi X}{L}\right)}\right|} & X\neq \f{L}{2} + n L, ~X\neq m L,\\
    \f{\pi^2}{L^2} &X=\f{L}{2} + n L~\text{or}~X= mL,\\
    \end{cases},
\end{split}
\ee
where $n$ and $m$ are integers. \textcolor{red}{\bf MN: I am not sure if this behavior is consistent with the time evolution of OEE and BOMI} 
\fi

\subsubsection*{Extremes of $r'_{\alpha=2,\text{Horizon}}$ \label{sec:Asym-behavior-of-ralpha2}}
Let us analyze the spatial extremes of $r'_{\alpha=2,\text{Horizon}}$.
These spatial extremes are determined by $\partial_X r'_{\alpha=2,\text{Horizon}}=0$, and these solutions are given by
\be
\begin{split}
    X=0,\f{L}{2},~\cos{\left(\f{2\pi X_{j=L,R}}{L}\right)}=\f{L^2+2\pi^2t_1^2}{2\pi t_1 (L^2+\pi^2t_1^2)}\left(L\sin{\left(\f{2\pi t_0}{L}\right)}+\pi t_1\cos{\left(\f{2\pi t_0}{L}\right)}\right)
\end{split}
\ee
For $X_{j=L,R}$, $r'_{\alpha=2,\text{Horizon}}$ is given by
\be
r'_{\alpha=2,\text{Horizon}} =\f{r_0\sqrt{L^2+\pi^2t_1^2}}{\left|L \cos{\left(\f{2\pi t_0}{L}\right)}-\pi t_1 \sin{\left(\f{2\pi t_0}{L}\right)}\right|}
\ee
In the large $t_1$-limit, $X_{i=L,R}$ are approximated by
\be
X_{j=L,R} +m L =\pm t_0,
\ee
where $m$ is an integer number.
By using the physical interpretation discussed in section \ref{sec:physical-interpretation}, $X+mL=\pm t_0$ are interpreted as the trajectories at $t_0$ of the right- and left-moving B.H.-like excitations, respectively. 
In other words, the spatial extremes for the large $t_1$ are determined by the trajectories at $t_0$ of the right- and left-moving B.H.-like excitations.
As a consequence, the asymptotic behavior of $r'_{\alpha=2,\text{Horizon}}$ for the large $t_1$ is given by
\be
\begin{split}
    r'_{\alpha=2,\text{Horizon}} \approx \begin{cases}
    \f{r_0}{\left|\sin{\left(\f{2\pi t_0}{L}\right)}\right|} ~\text{for}~t_0\neq \f{nL}{2}\\
    \f{\pi r_0 t_1}{L}~\text{for}~t_0= \f{nL}{2}
     \end{cases},
\end{split}
\ee
where $n$ is an integer number.
Thus, if the B.H.-like excitations are at $X \neq X_{i=1,2}^f$, then $r'_{\alpha=2,\text{Horizon}}$ depends on only $t_0$, while if these excitations are at $X = X_{i=1,2}^f$, then $r'_{\alpha=2,\text{Horizon}}$ depends on only $t_1$, and it linearly increases with $t_1$.
Note that the asymptotic form of the black hole horizon for $t_0\neq \f{nL}{2}$ is invalid in the $t_0$-regimes where $t_0\approx\f{nL}{2}$. 
In these $t_0$-regimes, we need more detailed calculations.
\subsubsection*{Extremes of $r'_{\alpha=3,\text{Horizon}}$ \label{sec:Asym-behavior-of-ralpha3}}

\begin{figure}[tbp]
    \begin{tabular}{ccc}
      \begin{minipage}[t]{0.45\hsize}
        \centering
        \includegraphics[keepaspectratio, scale=0.55]{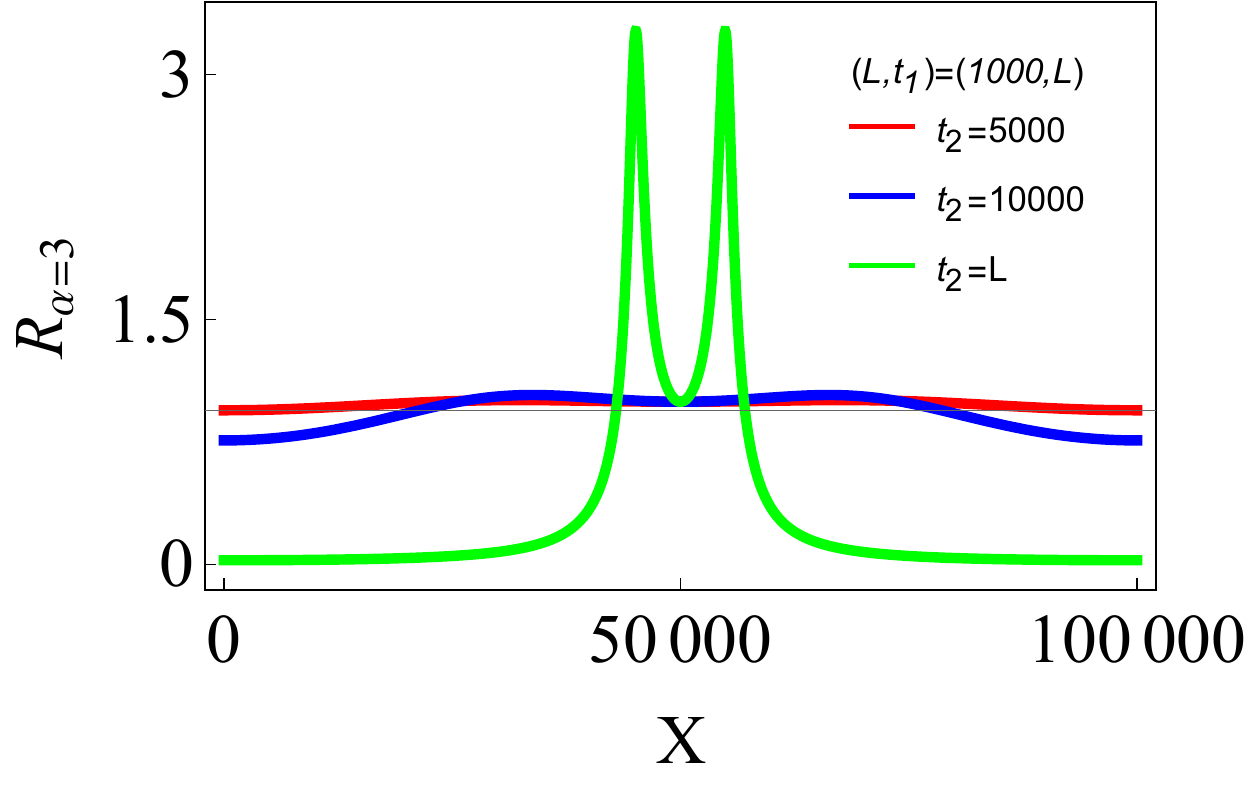}
        
   (a) Small $t_1$-regime.
      \end{minipage} & 
     
     \begin{minipage}[t]{0.45\hsize}
        \centering
        \includegraphics[keepaspectratio, scale=0.55]{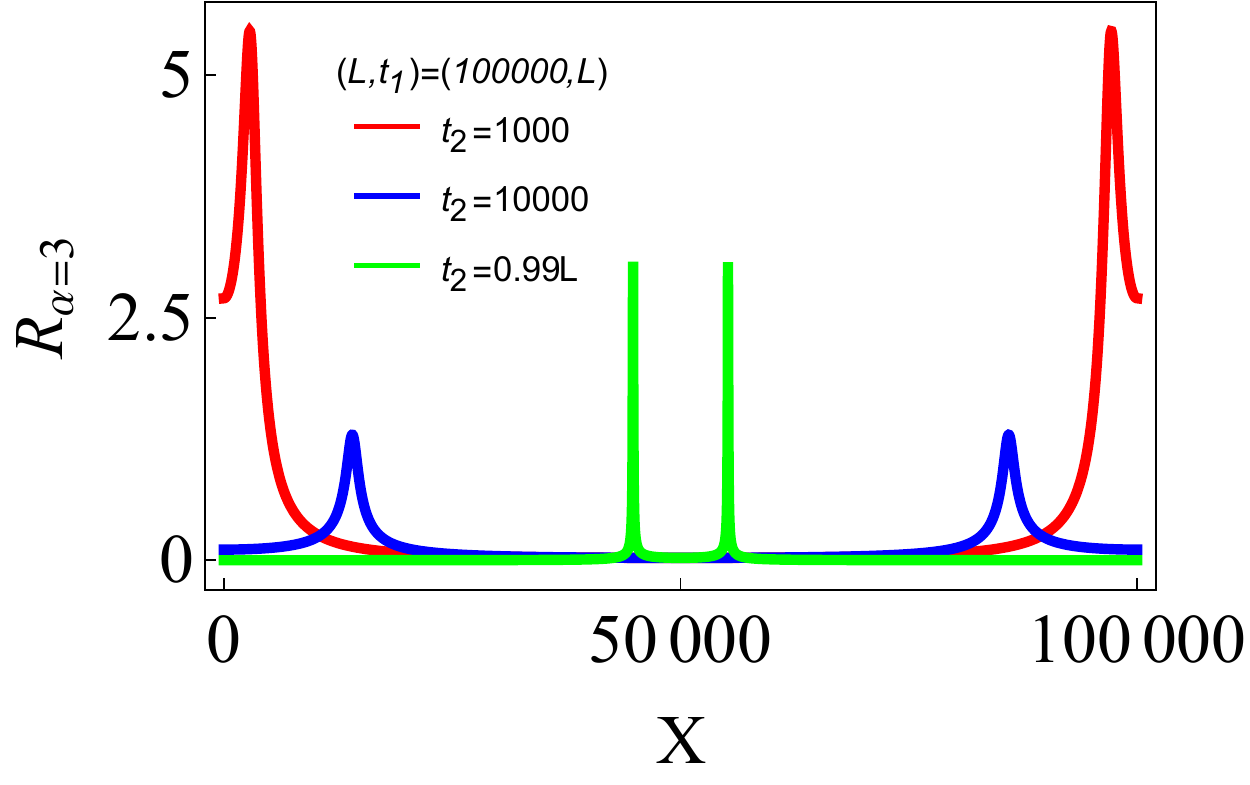}
        
    (b) Large $t_1$-regime.
      \end{minipage} 
     
    \end{tabular}
    \caption{The spatial dependence of the black hole horizon for various $t_1$ and $t_2$ as a function of $X$. Here, $R_{\alpha=3}$ is defined by $r_{\alpha=3,\text{Horizon}}/r_0$. For the large $t_1$, the spatial locations where the peaks of $R_{\alpha=3}$ emerge are approximately equal to the locations of the B.H.-like excitations. }
    \label{Fig:BH_for_alpha-3}
  \end{figure}

Now, let us turn to the analysis of the spatial extremes of $r'_{\alpha=3,\text{Horizon}}$.
The spatial extremes are determined by $\partial_X r'_{\alpha=3,\text{Horizon}}=0$, and the solutions of this equation are given by
\be
\begin{split}
    X=0,~\f{L}{2},~\cos{\left(\f{2\pi X_{j=L,R}}{L}\right)}=-\f{\left(L^4+2\pi^2L^2(t_1-t_2)^2+8\pi^4t_1^2t_2^2\right)\left(4\pi^2t_1^2t_2^2-L^2(t_1-t_2)^2\right)}{2\left(L^4+\pi^2L^2(t_1-t_2)^2+4\pi^4t_1^2t_2^2\right)\left(4\pi^2t_1^2t_2^2+L^2(t_1-t_2)^2\right)}.
\end{split}
\ee
In the large $t_1$-limit, $X_{j=L,R}$ are determined by 
\be
\cos{\left(\f{2\pi X_{j=L,R}}{L}\right)}\approx -\f{(4\pi^2t_2^2-L^2)}{(4\pi^2t_2^2+L^2)}.
\ee
This is the same as the trajectories of the left- and right-moving B.H.-like excitations emerging at $x=X^f_1$ at time $t_2=0$.
Under the evolution by $H^1_{\text{CSD}}$, these excitations moves at the velocities, $v_{j=L,R}=\pm 2\cos^2{\left(\f{\pi x}{L}\right)}$, to the $X^{f}_2$.
The extremes of the black hole horizon for $x=X_{j=L,R}$ are given by
\be
r'_{\alpha=3,\text{Horizon}}=\frac{r_0\sqrt{\left(L^2 (t_1-t_2)^2+4 \pi ^2 t_1^2 t_2^2\right) \left(L^4+\pi ^2 L^2 (t_1-t_2)^2+4 \pi ^4 t_1^2 t_2^2\right)}}{\left|L^3 (t_2-t_1)+4 \pi ^2 L t_1^2 t_2\right|}
\ee
In the large $t_1$-limit for $t_2>0$, the extremes of $r'_{\alpha=3,\text{Horizon}}$ are given by
\be
r'_{\alpha=3,\text{Horizon}}\approx \f{r_0\left(L^2+4\pi^2t_2^2\right)}{4\pi L t_2}.
\ee
Furthermore, take the large $t_2$-limit, and then these extremes are given by
\be
r'_{\alpha=3,\text{Horizon}}\approx \f{\pi r_0 t_2 }{L}.
\ee
Thus, these extremes linear grow with $t_2$.
In Fig.\ \ref{Fig:BH_for_alpha-3}, we depict $r'_{\alpha=3,\text{Horizon}}$ for various $t_1$ and $t_2$ as a function of $X$.
\subsection{Geodesic length in the static BTZ black hole \label{App:GL-BTZ}}
Here, we present the non-universal piece of (\ref{eq:free-energy-correlator}).
It is given by 
\be 
\begin{split}
G(X_1,Y_1) =2 \log{\left(\f{2\epsilon}{\pi}\right)}+\text{Min}\bigg{[}S_1,S_2,S_3\bigg{]},
\end{split}
\ee
where $S_{i=1,2,3}$ are defined as
\be
\begin{split}
S_1&=\log{\left[\left|\sin{\left[\f{\pi}{2\epsilon}\left(w^{\text{New},\alpha}_{Y_1,\epsilon}-w_{X_1}\right)\right]}\right|^2\right]},\\
S_2&=\log{\left[\left|\sin{\left[\f{\pi}{2\epsilon}\left(iL-\left(w^{\text{New},\alpha}_{Y_1,\epsilon}-w_{X_1}\right)\right)\right]}\right|^2\right]},\\
S_3&=\log{\left[\left|\sin{\left[\f{\pi}{2\epsilon}\left(-iL-\left(w^{\text{New},\alpha}_{Y_1,\epsilon}-w_{X_1}\right)\right)\right]}\right|^2\right]}.
\end{split}
\ee

\subsection{The metric of the inhomogeneous black holes\label{sec:metric-of-geometries}}
Here, we present the inhomogeneous black hole geometries.
The dual geometries of $\rho_{\mathcal{H}_1}$ considered in this paper are given
by (\ref{eq:geo_i_bh}) and the components are given
as follows.
For $\alpha=1$,
\be
\begin{split}
  & f^{1}_{\alpha=1;XX}=\frac{\pi ^4}{D_{\alpha=1}},
    \quad
    f^{2}_{\alpha=1;XX}=\frac{4 \pi ^6 L^2 t_1^2 \sin ^2\left(\frac{2 \pi
    X}{L}\right)}{\left(D_{\alpha=1}\right)^2},
    \quad
    f^{1}_{\alpha=1;t_0t_0}=-1,
    \quad f^{2}_{\alpha=1;t_0t_0}=1,\\
   &f^{1}_{\alpha=1;t_1t_1}=-\frac{4 \pi ^4 \sin ^4\left(\frac{\pi X}{L}\right)}{D_{\alpha=1}},
     \quad
   f^{2}_{\alpha=1;t_1t_1}=\frac{4 \pi ^4 \sin ^4\left(\frac{\pi  X}{L}\right) \left(L^2-2 \pi ^2 t_1^2 \cos \left(\frac{2 \pi  X}{L}\right)+2 \pi ^2 t_1^2\right)^2}{\left(D_{\alpha=1}\right)^2},\\
  &f^{1}_{\alpha=1;t_0t_1}=
    f^{2}_{\alpha=1;t_0t_1}=
    -\frac{2 \pi ^2 \sin ^2\left(\frac{\pi  X}{L}\right)
     \left(L^2+4 \pi ^2 t_1^2 \sin ^2\left(\frac{\pi
     X}{L}\right)\right)}{D_{\alpha=1}},
  \\
  &f^{1}_{\alpha=1;Xt_0}=
    -f^{2}_{\alpha=1;Xt_0}
    =-\frac{2 \pi ^3 L t_1 \sin \left(\frac{2 \pi
     X}{L}\right)}{D_{\alpha=1}},
    \quad f^{1}_{\alpha=1;Xt_1}=0,\\
   &f^{2}_{\alpha=1;Xt_1}=\frac{8 \pi ^5 L  t_1 \sin ^3\left(\frac{\pi
     X}{L}\right) \cos \left(\frac{\pi  X}{L}\right) \left(L^2-2 \pi ^2 t_1^2
     \cos \left(\frac{2 \pi  X}{L}\right)+2 \pi ^2
     t_1^2\right)}{\left(D_{\alpha=1}\right)^2},
  \\
   &D_{\alpha=1}=\left(L^2+4 \pi ^2 t_1^2 \sin ^2\left(\frac{\pi  X}{L}\right)\right)^2-4 \pi ^2 L^2 t_1^2 \sin ^2\left(\frac{2 \pi  X}{L}\right)
\end{split}
\ee
For $\alpha=2$,
\be
\begin{split}
  f^1_{\alpha=2;XX}
  &=\frac{\pi ^4}{D_{\alpha=2} },
    \quad
    f^2_{\alpha=2;XX}=\frac{4 \pi ^6t_1^2 \sin ^2\left(\frac{2 \pi  X}{L}\right)
    \left(L \cos \left(\frac{2 \pi  t_0}{L}\right)-\pi  t_1 \sin \left(\frac{2
    \pi  t_0}{L}\right)\right)^2}{(D_{\alpha=2})^2
    },
    \quad
    f^1_{\alpha=2;t_0t_0}=-\frac{\pi ^4}{D_{\alpha=2}},
  \\
  f^2_{\alpha=2;t_0t_0}
  &=\f{\pi^4}{(D_{\alpha=2})^2} \bigg{(}L^2-\pi  t_1 \bigg{(}\pi  t_1 \left(\cos \left(\frac{2 \pi  (t_0-X)}{L}\right)+\cos \left(\frac{2 \pi  (t_0+X)}{L}\right)\right)\\
  &+L \left(\sin \left(\frac{2 \pi  (t_0-X)}{L}\right)+\sin \left(\frac{2 \pi
    (t_0+X)}{L}\right)\right)\bigg{)}+2 \pi ^2 t_1^2\bigg{)}^2,
  \\
  f^1_{\alpha=2;t_1t_1}
  &=-\frac{4 \pi ^4 \sin ^2\left(\frac{\pi  (t_0-X)}{L}\right) \sin^2\left(\frac{\pi  (t_0+X)}{L}\right)}{D_{\alpha=2} },
  \\
  f^2_{\alpha=2;t_1t_1}
  &=\pi ^4 \left[\frac{\sin ^2\left(\frac{\pi  (t_0+X)}{L}\right)}{d_{\alpha=2;p}}+\frac{\sin^2\left(\frac{\pi  (t_0-X)}{L}\right)}{d_{\alpha=2;m}}\right]^2\\
  &+\frac{\sin ^2\left(\frac{\pi
    (t_0-X)}{L}\right)}{\left(d_{\alpha=2;m}\right)^2}\left(\frac{2 \sin
    ^2\left(\frac{\pi  (t_0+X)}{L}\right) d_{\alpha=2;m}+\sin ^2\left(\frac{\pi
    (t_0-X)}{L}\right) d_{\alpha=2;p}}{d_{\alpha=2;p}}\right)\bigg{]},
  \\
  f^1_{\alpha=2;t_0t_1}
  &=-\frac{\pi ^4 \left(\sin ^2\left(\frac{\pi  (t_0-X)}{L}\right)+\sin
    ^2\left(\frac{\pi  (t_0+X)}{L}\right)\right)}{D_{\alpha=2}},
  \\
  f^2_{\alpha=2;t_0t_1}
  &=\frac{1}{2} \pi ^4 \left[\frac{\sin ^2\left(\frac{\pi
    (t_0-X)}{L}\right)}{\left(d_{\alpha=2;m}\right)^2}+\frac{\sin
    ^2\left(\frac{\pi  (t_0-X)}{L}\right)+\sin ^2\left(\frac{\pi
    (t_0+X)}{L}\right)}{D_{\alpha=2}}+\frac{\sin ^2\left(\frac{\pi
    (t_0+X)}{L}\right)}{(d_{\alpha=2;p})^2}\right],
  \\
  D_{\alpha=2}
  &=d_{\alpha=2;p} \times d_{\alpha=2;m},
  \\
  d_{\alpha=2;m}
  &=\left(L^2-2 \pi  t_1 \left(\pi  t_1 \cos \left(\frac{2 \pi
    (t_0-X)}{L}\right)+L \sin \left(\frac{2 \pi  (t_0-X)}{L}\right)\right)+2 \pi
    ^2 t_1^2\right),
  \\
  d_{\alpha=2;p}
  &=\left(L^2-2 \pi  t_1 \left(\pi  t_1 \cos \left(\frac{2 \pi  (t_0+X)}{L}\right)+L \sin \left(\frac{2 \pi  (t_0+X)}{L}\right)\right)+2 \pi ^2 t_1^2\right),\\
  f^1_{\alpha=2;Xt_0}
  &=0,
    \quad f^2_{Xt_0}=\frac{1}{4} \pi ^4
    \left[\frac{1}{\left(d_{\alpha=2;p}\right)^2}-\frac{1}{\left(d_{\alpha=2;m}\right)^2}\right],
    \nonumber \\
  f^1_{\alpha=2;Xt_1}
  &=\frac{\pi ^4 \left(\sin ^2\left(\frac{\pi
    (t_0+X)}{L}\right)-\sin ^2\left(\frac{\pi
    (t_0-X)}{L}\right)\right)}{D_{\alpha=2} },
  \\ 
  f^{2}_{\alpha=2;Xt_1}
  &=-\frac{\pi ^5 t_1}{\left(D_{\alpha=2}\right)^2 }\left[\sin ^2\left(\frac{\pi  (t_0-X)}{L}\right) d_{\alpha=2,p}+\sin ^2\left(\frac{\pi  (t_0+X)}{L}\right) d_{\alpha=2,m}\right] \\
  &\times
    \Big[
    \pi  t_1
    \left(\cos \left(\frac{2 \pi  (t_0-X)}{L}\right) -\cos \left(\frac{2 \pi  (t_0+X)}{L}\right)\right)
    \nonumber \\
  &\quad
    +L \left(\sin \left(\frac{2 \pi  (t_0-X)}{L}\right)
    -\sin \left(\frac{2 \pi  (t_0+X)}{L}\right)\right)\Big],\\
\end{split}
\ee
For $\alpha=3$,
\be
\begin{split}
  f^{1}_{\alpha=3;XX}
  &= \frac{\pi ^4 L^4}{d^1_{XX} },
    \quad f^{2}_{XX}=\frac{4 \pi ^6 L^6  \left(L^2 (t_2-t_1)+4 \pi ^2 t_1^2 t_2\right)^2 \sin ^2\left(\frac{2 \pi  X}{L}\right)}{d^2_{XX}},\\
  f^{1}_{t_1t_1}
  &=-\frac{\pi ^4 \left(-\left(L^2+4 \pi ^2 t_2^2\right) \cos \left(\frac{2 \pi  X}{L}\right)+L^2-4 \pi ^2 t_2^2\right)^2}{d^1_{t_1t_1}},\\
  f^{2}_{\alpha=3;t_1t_1}
  &=\frac{\pi ^4  }{d^2_{t_1t_1}}\Big{(}L^6+\pi ^2 L^4 \left(3 t_1^2+2 t_1 t_2+t_2^2\right)+4 \pi ^4 L^2 t_2^2 \left(-2 t_1^2-6 t_1 t_2+3 t_2^2\right)\\
  &+\pi ^2 \big{(}L^4 \left(t_1^2-2 t_1 t_2+3 t_2^2\right)+4 \pi ^2 L^2 t_2^2 \left(2 t_1^2-2 t_1 t_2+t_2^2\right)+16 \pi ^4 t_1^2 t_2^4\big{)} \cos \left(\frac{4 \pi  X}{L}\right)\\
  &-\left(L^6+4 \pi ^2 L^4 (t_1-t_2) (t_1+t_2)+16 \pi ^4 L^2 t_2^3 (2 t_1-t_2)-64 \pi ^6 t_1^2 t_2^4\right)\cos \left(\frac{2 \pi  X}{L}\right)\\
  &+48 \pi ^6 t_1^2 t_2^4\Big{)}^2,\\
  f^{1}_{\alpha=3;t_2t_2}
  &=-\frac{4 \pi ^4 L^4 \cos ^4\left(\frac{\pi  X}{L}\right)}{d^1_{t_2t_2}},\\
  f^{2}_{\alpha=3;t_2t_2}
  &=\frac{4 \pi ^4 L^4  \cos ^4\left(\frac{\pi  X}{L}\right)}{d^2_{t_2t_2}}\Big{(}L^4+2 \pi ^2 \left(L^2 \left(-t_1^2-2 t_1 t_2+t_2^2\right)+4 \pi ^2 t_1^2 t_2^2\right)\cos \left(\frac{2 \pi  X}{L}\right)\\
  &+2 \pi ^2 L^2 (t_1-t_2)^2+8 \pi ^4 t_1^2 t_2^2\Big{)}^2,\\
  f^{1}_{\alpha=3;Xt_1}
  &=\frac{4 \pi ^5 L^3 t_2 \sin \left(\frac{2 \pi  X}{L}\right)}{d^1_{Xt_1}},\\
  f^{2}_{\alpha=3;Xt_1}
  &=-\frac{2 \pi ^5 L^3 \left(L^2 (t_2-t_1)+4 \pi ^2 t_1^2 t_2\right) \sin \left(\frac{2 \pi  X}{L}\right) }{d^2_{Xt_1}}\Big{(}L^6+\pi ^2 L^4 \left(3 t_1^2+2 t_1 t_2+t_2^2\right)\\
  &+4 \pi ^4 L^2 t_2^2 \left(-2 t_1^2-6 t_1 t_2+3 t_2^2\right)+\pi ^2 \big{(}L^4 \left(t_1^2-2 t_1 t_2+3 t_2^2\right)+4 \pi^2 L^2 t_2^2\\
  &\times\left(2 t_1^2-2 t_1 t_2+t_2^2\right)+16 \pi ^4 t_1^2 t_2^4\big{)} \cos \left(\frac{4 \pi  X}{L}\right)-\big{(}L^6+4 \pi ^2 L^4 (t_1-t_2) (t_1+t_2)\\
  &+16 \pi ^4 L^2 t_2^3 (2 t_1-t_2)-64 \pi ^6 t_1^2 t_2^4\big{)} \cos \left(\frac{2 \pi  X}{L}\right)+48 \pi ^6 t_1^2 t_2^4\Big{)},\\
  f^{1}_{\alpha=3;Xt_2}
  &=0,
    \quad
    f^{2}_{Xt_2}=\frac{1}{2} \pi ^4 L^4 \cos ^2\left(\frac{\pi  X}{L}\right) \left(\frac{1}{d^{2,a}_{Xt_2}}-\frac{1}{d^{2,b}_{Xt_2}}\right),\\
  f^{1}_{\alpha=3;t_1t_2}
  &=-\frac{2 \pi ^4 L^2 \cos ^2\left(\frac{\pi  X}{L}\right) \left(-\left(L^2-4 \pi ^2 t_2^2\right) \cos \left(\frac{2 \pi  X}{L}\right)+L^2+4 \pi ^2 t_2^2\right)}{d^{1}_{t_1t_2}},\\
  f^2_{\alpha=3;t_1t_2}
  &=\frac{N_{t_1t_2}}{d^{2}_{t_1t_2}}
\end{split}
\ee
In the above equations, we used the following notations: 
\be
\begin{split}
  d^1_{XX}
  &=\Big{[}L^4+2 \pi  \big{(}\pi  \left(L^2 \left(-t_1^2-2 t_1 t_2+t_2^2\right)+4 \pi ^2 t_1^2 t_2^2\right) \cos \left(\frac{2 \pi  X}{L}\right)\\
   &+L \left(L^2 (t_1-t_2)-4 \pi ^2 t_1^2 t_2\right) \sin \left(\frac{2 \pi
     X}{L}\right)\big{)}+2 \pi ^2 L^2 (t_1-t_2)^2+8 \pi ^4 t_1^2 t_2^2
     \Big{]}\\
  &\times
    \Big{[}L^4+2 \pi  \big{(}\pi  \left(L^2 \left(-t_1^2-2 t_1 t_2+t_2^2\right)+4 \pi ^2 t_1^2 t_2^2\right) \cos \left(\frac{2 \pi  X}{L}\right)\\
   &+L \left(L^2 (t_2-t_1)+4 \pi ^2 t_1^2 t_2\right) \sin \left(\frac{2 \pi
     X}{L}\right)\big{)}+2 \pi ^2 L^2 (t_1-t_2)^2+8 \pi ^4 t_1^2 t_2^2
     \Big{]},\\
  d^2_{XX}
  &=\Big{[}L^4+2 \pi  \big{(}\pi  \left(L^2 \left(-t_1^2-2 t_1 t_2+t_2^2\right)+4 \pi ^2 t_1^2 t_2^2\right) \cos \left(\frac{2 \pi  X}{L}\right)\\
   &+L \left(L^2 (t_1-t_2)-4 \pi ^2 t_1^2 t_2\right) \sin \left(\frac{2 \pi
     X}{L}\right)\big{)}+2 \pi ^2 L^2 (t_1-t_2)^2+8 \pi ^4 t_1^2 t_2^2
     \Big{]}^2 \\
   &\times\Big{[}L^4+2 \pi  \big{(}\pi  \left(L^2 \left(-t_1^2-2 t_1 t_2+t_2^2\right)+4 \pi ^2 t_1^2 t_2^2\right) \cos \left(\frac{2 \pi  X}{L}\right)\\
   &+L \left(L^2 (t_2-t_1)+4 \pi ^2 t_1^2 t_2\right) \sin \left(\frac{2 \pi
     X}{L}\right)\big{)}+2 \pi ^2 L^2 (t_1-t_2)^2+8 \pi ^4 t_1^2 t_2^2
     \Big{]}^2,\\
  d^1_{t_1t_1}
  &=\Big{[}L^4+2 \pi  \big{(}\pi  \left(L^2 \left(-t_1^2-2 t_1 t_2+t_2^2\right)+4 \pi ^2 t_1^2 t_2^2\right) \cos \left(\frac{2 \pi  X}{L}\right)\\
   &+L \left(L^2 (t_1-t_2)-4 \pi ^2 t_1^2 t_2\right) \sin \left(\frac{2 \pi
     X}{L}\right)\big{)}+2 \pi ^2 L^2 (t_1-t_2)^2+8 \pi ^4 t_1^2 t_2^2
     \Big{]}\\
  &\times
    \Big{[}L^4+2 \pi  \big{(}\pi  \left(L^2 \left(-t_1^2-2 t_1 t_2+t_2^2\right)+4 \pi ^2 t_1^2 t_2^2\right) \cos \left(\frac{2 \pi  X}{L}\right)\\
   &+L \left(L^2 (t_2-t_1)+4 \pi ^2 t_1^2 t_2\right) \sin \left(\frac{2 \pi
     X}{L}\right)\big{)}+2 \pi ^2 L^2 (t_1-t_2)^2+8 \pi ^4 t_1^2 t_2^2
     \Big{]},\\
  d^2_{t_1t_1}
  &=\Big{[}L^4+2 \pi  \big{(}\pi  \left(L^2 \left(-t_1^2-2 t_1 t_2+t_2^2\right)+4 \pi ^2 t_1^2 t_2^2\right) \cos \left(\frac{2 \pi  X}{L}\right)\\
   &+L \left(L^2 (t_1-t_2)-4 \pi ^2 t_1^2 t_2\right) \sin \left(\frac{2 \pi
     X}{L}\right)\big{)}+2 \pi ^2 L^2 (t_1-t_2)^2+8 \pi ^4 t_1^2 t_2^2
     \Big{]}^2\\
   &\times\Big{[}L^4+2 \pi  \big{(}\pi  \left(L^2 \left(-t_1^2-2 t_1 t_2+t_2^2\right)+4 \pi ^2 t_1^2 t_2^2\right) \cos \left(\frac{2 \pi  X}{L}\right)\\
   &+L \left(L^2 (t_2-t_1)+4 \pi ^2 t_1^2 t_2\right) \sin \left(\frac{2 \pi
     X}{L}\right)\big{)}+2 \pi ^2 L^2 (t_1-t_2)^2+8 \pi ^4 t_1^2 t_2^2
     \Big{]}^2,\\
   \end{split}
\ee
\be
\begin{split}
  d^1_{t_2t_2}
  &=\Big{[}
    L^4+2 \pi  \big{(}\pi  \left(L^2 \left(-t_1^2-2 t_1 t_2+t_2^2\right)+4 \pi ^2 t_1^2 t_2^2\right) \cos \left(\frac{2 \pi  X}{L}\right)\\
  &+L \left(L^2 (t_1-t_2)-4 \pi ^2 t_1^2 t_2\right) \sin \left(\frac{2 \pi
    X}{L}\right)\big{)}+2 \pi ^2 L^2 (t_1-t_2)^2+8 \pi ^4 t_1^2 t_2^2
    \Big{]}\\
  &\times
    \Big{[}
    L^4+2 \pi  \big{(}\pi  \left(L^2 \left(-t_1^2-2 t_1 t_2+t_2^2\right)+4 \pi ^2 t_1^2 t_2^2\right) \cos \left(\frac{2 \pi  X}{L}\right)\\
  &+L \left(L^2 (t_2-t_1)+4 \pi ^2 t_1^2 t_2\right) \sin \left(\frac{2 \pi
    X}{L}\right)\big{)}+2 \pi ^2 L^2 (t_1-t_2)^2+8 \pi ^4 t_1^2 t_2^2
    \Big{]},\\
  d^2_{t_2t_2}
  &=\Big{[}L^4+2 \pi  \big{(}\pi  \left(L^2 \left(-t_1^2-2 t_1 t_2+t_2^2\right)+4 \pi ^2 t_1^2 t_2^2\right) \cos \left(\frac{2 \pi  X}{L}\right)\\
  &+L \left(L^2 (t_1-t_2)-4 \pi ^2 t_1^2 t_2\right) \sin \left(\frac{2 \pi
    X}{L}\right)\big{)}+2 \pi ^2 L^2 (t_1-t_2)^2+8 \pi ^4 t_1^2 t_2^2
    \Big{]}^2 \\
  &\times
    \Big{[}L^4+2 \pi  \big{(}\pi  \left(L^2 \left(-t_1^2-2 t_1 t_2+t_2^2\right)+4 \pi ^2 t_1^2 t_2^2\right) \cos \left(\frac{2 \pi  X}{L}\right)\\
  &+L \left(L^2 (t_2-t_1)+4 \pi ^2 t_1^2 t_2\right) \sin \left(\frac{2 \pi
    X}{L}\right)\big{)}+2 \pi ^2 L^2 (t_1-t_2)^2+8 \pi ^4 t_1^2 t_2^2
    \Big{]}^2,\\
  d^{1}_{Xt_1}
  &=\Big{[}
    L^4+2 \pi  \big{(}\pi  \left(L^2 \left(-t_1^2-2 t_1 t_2+t_2^2\right)+4 \pi ^2 t_1^2 t_2^2\right) \cos \left(\frac{2 \pi  X}{L}\right)\\
  &+L \left(L^2 (t_1-t_2)-4 \pi ^2 t_1^2 t_2\right) \sin \left(\frac{2 \pi
    X}{L}\right)\big{)}+2 \pi ^2 L^2 (t_1-t_2)^2+8 \pi ^4 t_1^2 t_2^2
    \Big{]} \\
  & \times
    \Big{[}
    L^4+2 \pi  \big{(}\pi  \left(L^2 \left(-t_1^2-2 t_1 t_2+t_2^2\right)+4 \pi ^2 t_1^2 t_2^2\right) \cos \left(\frac{2 \pi  X}{L}\right)\\
  &+L \left(L^2 (t_2-t_1)+4 \pi ^2 t_1^2 t_2\right) \sin \left(\frac{2 \pi
    X}{L}\right)\big{)}+2 \pi ^2 L^2 (t_1-t_2)^2+8 \pi ^4 t_1^2 t_2^2
    \Big{]},\\
  d^{2}_{Xt_1}
  &=\Big{[}
    L^4+2 \pi  \big{(}\pi  \left(L^2 \left(-t_1^2-2 t_1 t_2+t_2^2\right)+4 \pi ^2 t_1^2 t_2^2\right) \cos \left(\frac{2 \pi  X}{L}\right)\\
  &+L \left(L^2 (t_1-t_2)-4 \pi ^2 t_1^2 t_2\right) \sin \left(\frac{2 \pi
    X}{L}\right)\big{)}+2 \pi ^2 L^2 (t_1-t_2)^2+8 \pi ^4 t_1^2 t_2^2
    \Big{]}^2 \\
  &\times
    \Big{[}L^4+2 \pi  \big{(}\pi  \left(L^2 \left(-t_1^2-2 t_1 t_2+t_2^2\right)+4 \pi ^2 t_1^2 t_2^2\right) \cos \left(\frac{2 \pi  X}{L}\right)\\
  &+L \left(L^2 (t_2-t_1)+4 \pi ^2 t_1^2 t_2\right) \sin \left(\frac{2 \pi
    X}{L}\right)\big{)}+2 \pi ^2 L^2 (t_1-t_2)^2+8 \pi ^4 t_1^2 t_2^2
    \Big{]}^2,
\end{split}
\ee
\be
\begin{split}
  d^{2,a}_{Xt_2}
  &=\Big{[}L^4+2 \pi  \big{(}\pi  \left(L^2 \left(-t_1^2-2 t_1 t_2+t_2^2\right)+4 \pi ^2 t_1^2 t_2^2\right) \cos \left(\frac{2 \pi  X}{L}\right)\\
  &\quad +L \left(L^2 (t_2-t_1)+4 \pi ^2 t_1^2 t_2\right) \sin \left(\frac{2 \pi
    X}{L}\right)\big{)}
    +2 \pi ^2 L^2 (t_1-t_2)^2+8 \pi ^4 t_1^2 t_2^2\Big{]}^2,
  \\
  d^{2,b}_{Xt_2}
  &=\Big{[}L^4+2 \pi  \big{(}\pi  \left(L^2 \left(-t_1^2-2 t_1 t_2+t_2^2\right)+4 \pi ^2 t_1^2 t_2^2\right) \cos \left(\frac{2 \pi  X}{L}\right)\\
  &\quad
    +L \left(L^2 (t_1-t_2)-4 \pi ^2 t_1^2 t_2\right) \sin \left(\frac{2 \pi
    X}{L}\right)\big{)}+2 \pi ^2 L^2 (t_1-t_2)^2
    +8 \pi ^4 t_1^2 t_2^2\Big{]}^2,
  \\
  d^{1}_{t_1t_2}
  &=\Big{[}
    L^4
    +2 \pi ^2 \left(L^2 \left(-t_1^2-2 t_1 t_2+t_2^2\right)
    +4 \pi ^2 t_1^2 t_2^2\right) \cos \left(\frac{2 \pi  X}{L}\right)
  \\
  &\quad
    -2 \pi  L( L^2 (t_2-t_1)
    +4 \pi ^2 t_1^2 t_2)\times\sin \left(\frac{2 \pi  X}{L}\right)
  \\
  &\quad
    +2 \pi ^2 L^2 t_1^2-4 \pi ^2 L^2 t_1 t_2+2 \pi ^2 L^2 t_2^2+8 \pi ^4 t_1^2 t_2^2\Big{]}\\
  &\quad
    \times\Big{[}L^4+2 \pi ^2 \left(L^2 \left(-t_1^2-2 t_1 t_2+t_2^2\right)+4
    \pi ^2 t_1^2 t_2^2\right) \cos \left(\frac{2 \pi  X}{L}\right)
  \\
  &\quad +2 \pi  L \big{(}L^2 (t_2-t_1)
    +4 \pi ^2 t_1^2 t_2\big{)} \sin \left(\frac{2 \pi  X}{L}\right)
  \\
  &\quad +2 \pi ^2 L^2 t_1^2-4 \pi ^2 L^2 t_1 t_2+2 \pi ^2 L^2 t_2^2+8 \pi ^4 t_1^2 t_2^2\Big{]},\\
  d^{2}_{t_1t_2}
  &=2 \Bigg{[}L^4+2 \pi ^2 (t_1-t_2)^2 L^2+8 \pi ^4 t_1^2 t_2^2
  \\
  &\quad
    +2 \pi  \Big{(}\pi  \left(\left(-t_1^2-2 t_2 t_1+t_2^2\right) L^2
    +4 \pi ^2 t_1^2 t_2^2\right) \cos \left(\frac{2 \pi  X}{L}\right)
    \nonumber \\
  &\quad
    +L \left(L^2 (t_1-t_2)-4 \pi ^2 t_1^2 t_2\right) \sin \left(\frac{2 \pi  X}{L}\right)\Big{)}\Bigg{]}^2 \\
  &\quad
    \times\Bigg{[}L^4+2 \pi ^2 (t_1-t_2)^2 L^2+8 \pi ^4 t_1^2 t_2^2
  \\
  &\quad +2 \pi  \Big{(}\pi  \left(\left(-t_1^2-2 t_2 t_1+t_2^2\right) L^2
    +4 \pi ^2 t_1^2 t_2^2\right) 
    \cos \left(\frac{2 \pi  X}{L}\right)
    \nonumber \\
  &
    \quad
    +L \left((t_2-t_1) L^2+4 \pi ^2 t_1^2
    t_2\right) \sin \left(\frac{2 \pi  X}{L}\right)\Big{)}
    \Bigg{]}^2
\end{split}
\ee
\be
\begin{split}
  N_{t_1t_2}
  &=
  \frac{1}{L^2 \pi^4}
  \left[
  N^{(0)}_{t_1t_2}
  + \cos \frac{2\pi X}{L} N^{(1)}_{t_1t_2}
  + \cos \frac{4\pi X}{L} N^{(2)}_{t_1t_2}
  + \cos \frac{6\pi X}{L} N^{(3)}_{t_1t_2}
  + \cos \frac{8\pi X}{L} N^{(4)}_{t_1t_2}
  \right],
    \\
  N^{(0)}_{t_1t_2}
  &=
    L^{10}+4 \pi ^2 \left(t_1^2+2 t_2^2\right) L^8+\pi ^4 \left(5 t_1^4-26 t_2^2 t_1^2-120 t_2^3 t_1+65 t_2^4\right) L^6
  \nonumber \\
  &\quad +4 \pi ^6 t_2^2 \left(-3 t_1^4+20 t_2 t_1^3+200 t_2^2 t_1^2-140 t_2^3 t_1+35 t_2^4\right) L^4
    \nonumber \\
  &\quad -80 \pi ^8 t_1^2 t_2^4 \left(t_1^2+28 t_2 t_1-14 t_2^2\right) L^2 +2240 \pi ^{10} t_1^4 t_2^6,
  \nonumber \\
  N^{(1)}_{t_1t_2}
  &=
  2 \pi ^2 \big{[}-\left(t_1^2-7 t_2^2\right) L^8
    -2 \pi ^2 \left(t_1^4+4 t_2^2 t_1^2+52 t_2^3 t_1-27 t_2^4\right) L^6
  \nonumber \\
  &\qquad
    +16 \pi ^4 t_2^3 \left(2 t_1^3+41 t_2 t_1^2-28 t_2^2 t_1+7 t_2^3\right) L^4
  \nonumber \\
   &\qquad -32 \pi ^6 t_1^2 t_2^4 \left(t_1^2+56 t_2 t_1-28 t_2^2\right) L^2+1792 \pi
     ^8 t_1^4 t_2^6\big{]},
  \nonumber \\
  N^{(2)}_{t_1t_2}
  &=
    -L^{10}
   -4 \pi ^2 \left(t_1^2-2 t_2^2\right) L^8-4 \pi ^4 \left(t_1^4-6 t_2^2
    t_1^2+32 t_2^3 t_1-15 t_2^4\right) L^6
  \nonumber \\
  &\qquad +16 \pi ^6 t_2^2 (t_1^4-4 t_2 t_1^3+44 t_2^2 t_1^2-28 t_2^3 t_1+7
    t_2^4) L^4
  \nonumber \\
  & \qquad +64 \pi ^8 t_1^2 t_2^4 \left(t_1^2-28 t_2 t_1+14 t_2^2\right) L^2+1792 \pi ^{10} t_1^4 t_2^6,
  \nonumber \\
  N^{(3)}_{t_1t_2}
  &=
    2 \pi ^2 \big{[}\left(t_1^2+t_2^2\right) L^8+2 \pi ^2 (t_1^4+4 t_2^2 t_1^2-12 t_2^3 t_1+5 t_2^4) L^6+16 \pi ^4 t_2^3 (-2 t_1^3+7 t_2 t_1^2-4 t_2^2 t_1
     +t_2^3) L^4
  \nonumber \\
  &\qquad +32 \pi ^6 t_1^2 t_2^4 \left(t_1^2-8 t_2 t_1+4 t_2^2\right)
    L^2+256 \pi ^8 t_1^4 t_2^6\big{]},
  \nonumber \\
    N^{(4)}_{t_1t_2}
    &=
    \pi ^4 \big{[}(-t_1^2-2 t_2 t_1
    +t_2^2) L^2+4 \pi ^2 t_1^2 t_2^2\big{)}
  \nonumber \\
  &\qquad
    \times \left(\left(t_1^2-2 t_2 t_1+3
    t_2^2\right) L^4+4 \pi ^2 t_2^2 \left(2 t_1^2-2 t_2 t_1+t_2^2\right) L^2+16
    \pi ^4 t_1^2 t_2^4\right)
  \big{]}.
    \end{split}
\ee

\bibliographystyle{ieeetr}
\bibliography{reference}
\end{document}